\documentstyle[aps,amssymb,12pt,a4,epsfig]{revtex}

\widetext

\begin{document}
\tightenlines

\title{Electromagnetic transition form factors \\
and
dilepton decay rates of nucleon resonances}
\author{M.I. Krivoruchenko$^{a,b)}$, B.V. Martemyanov$^{a,b)}$, Amand
Faessler$^{a)}$, C. Fuchs$^{a)}$\vspace{0.5 cm}{} \\
$^{a)}${\small Institut f\"{u}r Theoretische Physik, Universit\"{a}t
T\"{u}bingen, Auf der Morgenstelle 14}\\
{\small D-72076 T\"{u}bingen, Germany}\\
{\small \ }$^{b)}${\small Institute for Theoretical and Experimental
Physics, B. Cheremushkinskaya 25}\\
{\small 117259 Moscow, Russia}}
\maketitle

\begin{abstract}
Relativistic, kinematically complete phenomenological expressions for the
dilepton decay rates of nucleon resonances with arbitrary spin and parity
are derived in terms of the magnetic, electric, and Coulomb transition form
factors. The dilepton decay rates of the nucleon resonances with masses
below $2$ GeV are estimated using the extended vector meson dominance (VMD)
model for the transition form factors. The model provides a unified
description of the photo- and electroproduction data, $\gamma(\gamma ^{*})N
\rightarrow N^{*}, $ the vector meson decays, $N^{*}\rightarrow
N\rho(\omega) $, and the dilepton decays, $N^{*}\rightarrow N\ell ^{+}\ell
^{-}$. The constraints on the transition form factors from the quark
counting rules are taken into account. The parameters of the model are fixed
by fitting the available photo- and electroproduction data and using results
of the multichannel partial-wave analysis of the $\pi N$ scattering. Where
experimental data are not available, predictions of the non-relativistic
quark models are used as an input. The vector meson coupling constants of
the magnetic, electric, and Coulomb types are determined. The dilepton
widths and the dilepton spectra from decays of nucleon resonances with
masses below $2$ GeV are calculated.

\vspace{0.5cm}{\bf keywords}: nucleon resonances, electromagnetic transition
form factors, dileptons

\end{abstract}

\textheight=61pc

\newpage{}

\section{Introduction}
\setcounter{equation}{0}
$\;$
\vspace{-0.5cm}

Particle properties in the medium are generally known to be different from 
particle properties in the vacuum. The
change of the nucleon mass in the nuclear matter was studied within the
Walecka model \cite{Walecka:1974qa,Chin:1977iz} already in the 1970's. During
the last decade, the problem of the description of hadrons in dense and hot
nuclear matter received new attention \cite
{Drukarev:1988ib,Tsushima:1991fe,Brown:1991kk,Adami:1993tp}. The goal of the
current investigations is to determine mass shifts and broadening of the
resonances in nuclear matter. The best probe for measuring the in-medium
modifications of the vector mesons masses and widths are dileptons which, being produced,
leave the reaction zone essentially undistorted by the final-state
interactions.

The data on the total
photoabsorption cross section on heavy nuclei \cite{Bianchi:1993nh} give an
evidence for a broadening of nucleon resonances in nuclear medium \cite
{Kondratyuk:1994ah}. The physics behind this effect is the same as in the
collision broadening of the atomic spectral lines in hot and dense gases,
discussed by Weisskopf \cite{Wei} in early 1930's. 

Dilepton spectra from heavy-ion collisions have been measured by the CERES and
HELIOS-3 Collaborations at SPS \cite{Agakishiev:1995xb,Masera:1995ck} (a few
hundreds GeV per nucleon) and by the DLS Collaboration at the BEVALAC \cite
{Porter:1997rc} (a few GeV per nucleon). The collision broadening
makes the peaks associated with the $V\rightarrow e^{+}e^{-}$ vector
meson decays in the CERES and HELIOS-3 experiments unobservable, but the 
spectra can be described theoretically, although the origin of the enhanced
dilepton yields in the low mass region i.e. below the $\rho$-meson peak is
still a matter of current debate.

However, concerning the DLS experiment, the measured dilepton spectra do not 
match with the theoretical estimates, even when possible
reduction of the $\rho $-meson mass and the $\rho $-meson broadening are
taken into account \cite{Bratkovskaya:1999pr}. The future HADES
experiment at GSI will study the dilepton spectra in the same energy range in
greater details \cite{Friese:1999qm}.

The dilepton modes of nucleon resonances are important sources of the
dilepton production in proton-proton and heavy-ion collisions. In this
paper, we derive kinematically complete phenomenological expressions for the
dilepton decays of nucleon resonances with arbitrary spin and parity,
parameterized in terms of the magnetic, electric, and Coulomb transition form
factors, and give numerical estimates for the dilepton spectra and dilepton
widths of the nucleon resonances with masses below $2$ GeV.

In order to calculate the dilepton decays, one needs to know the
electromagnetic transition form factors of nucleon resonances in the
time-like region. The standard vector meson dominance (VMD) with the
ground-state $\rho $-$,$ $\omega $-$,$ and $\phi $-mesons predicts monopole
form factors with $1/q^{2}$ asymptotics at $q^{2}\rightarrow \infty $.
Such asymptotics are, according to the quark counting rules \cite{Matveev:1973ra}, 
valid for the electromagnetic pion form factor. However,
already in the case of the nucleon form factors, radially excited vector
mesons $\rho ^{\prime },$ $\rho ^{\prime \prime }$ $...$ etc. should be
added into the VMD model in order to provide a dipole behavior for the Sachs
form factors and describe the experimental data \cite
{Hohler:1976ax,Krivoruchenko:1994qb,Mergell:1996bf,Dubnicka:1996sp}. It was
pointed out \cite{Faessler:2000md} that the standard VMD model overestimates
the photon branching ratios of the nucleon resonances if decay widths of the
nucleon resonances are used as an input. It disagrees also
with the quark counting rules for the transition form factors. The higher
powers of $1/q^{2} $ in the asymptotics imply a destructive interference
between contributions of vector mesons at small $q^{2}$. This effect
decreases the photon branching ratios. It can help to describe 
the experimental data for the vector meson and the photon decays of the
nucleon resonances within the VMD model framework.

We use the extended VMD model for the description of the transition form factors
of the nucleon resonances. The model provides a unified description of
photo- and electroproduction data and the vector meson decays of the nucleon
resonances. Its parameters are fixed by fitting the available data. The
form factors are finally used for the calculation of the dilepton decays of
nucleon resonances. 

The same philosophy has been used in Ref. \cite{Faessler:2000de} 
in order to predict unknown meson decay channels to dileptons. Recent
measurements e.g. for $\phi \rightarrow \eta e^+ e^-$, 
$\eta \rightarrow \pi^+ \pi^- e^+ e^-$ by the CMD-2 Collaboration \cite{Akhmetshin:2001bw}
are in excellent agreement with predictions \cite{Faessler:2000de}. This gives a support for the
extended VMD model in general, and also concerning the reliability of the 
present investigations of the dilepton decay channels of the nucleon resonances.

The outline of this paper is as follows: In the next Sect., we describe
the general framework for the description of the higher-spin resonances,
electromagnetic vertexes in terms of the covariant form factors, $%
F_{k}(q^{2})$, which are free from kinematical singularities, and calculate
helicity $\gamma ^{*}N\rightarrow N^{*}$ amplitudes in terms of the covariant form
factors $F_{k}(q^{2})$.

The relations between the magnetic, electric, and Coulomb transition form
factors and the covariant form factors $F_{k}(q^{2})$ have been established for the $\Delta
(1232)$-resonance by Jones and Scadron \cite{JS} and for arbitrary spin
resonances by Devenish, Eisenschitz and K\"orner \cite{DEK}. 
We find it worthwhile to rederive for methodical purposes 
in Sect. 3 these relations in view of the
controversy existing in the literature concerning the simple
$\Delta (1232)$ radiative and dilepton decays (for a discussion see
Ref. \cite{Krivoruchenko:2001hs}). First, we transform
amplitudes with the fixed total angular momentum of the photon and orbital momentum of the 
$\gamma ^{*}N$ system to the magnetic, electric, and Coulomb amplitudes,
and, second, transform amplitudes with fixed total angular momentum and
the orbital momentum to the helicity basis.

The kinematically complete phenomenological expressions for the $%
N^{*}\rightarrow N\gamma ^{*}$ decay rates, where $\gamma ^{*}$ is a virtual
massive photon, and the $N^{*}\rightarrow Ne^{+}e^{-}$ decay rates are then
obtained in terms of the the magnetic, electric, and Coulomb transition form
factors. The experimental data on the photo- and electroproduction of the
nucleon resonances are quoted for the helicity amplitudes and/or magnetic,
electric, and Coulomb transition form factors. The results of Sects. 2 and 3
are sufficient for the description of these data.

The data for the vector meson decays of the nucleon resonances are quoted usually in
the partial-wave basis for a fixed total spin and a fixed orbital momentum of 
the $NV$ system. In Sect. 4, we establish the connection between the partial-wave 
basis and the helicity basis of the $VN\rightarrow N^{*}$ amplitudes. The results 
of Sect. 4 are sufficient to fit the data for the vector meson decays of the 
nucleon resonances.

In Sect. 5, we establish a general representation for the transition form factors 
in the no-width vector meson limit 
within the extended VMD model, consistent with the quark counting rules. The 
quark counting rules reduce the number of the phenomenological parameters which 
otherwise cannot be determined. 
The overall sign of the vector meson decay amplitudes is not fixed experimentally
with respect to the photo- and electroproduction amplitudes. We use
the non-relativistic quark model to fix this sign.

In Sect. 6, numerical results are presented. We perform a fit to the
amplitudes of the photo- and electroproduction of the nucleon resonances and
to the vector meson decay amplitudes and determine free parameters of the
extended VMD model. The minimal
extension is found to be sufficient to describe the available data. We use
for the vector meson decay amplitudes the data from PDG \cite{Groom:2000in}.
When these data are not available, the results of the multichannel $\pi N$
partial-wave analysis by the Manley and Saleski \cite{Manley:1992yb} and
Longacre and Dolbeau \cite{Longacre:1977ja} are used. In other cases, we use
the quark model predictions by Koniuk \cite{Koniuk:1982ej} and Capstick and
Roberts \cite{Capstick:1994kb}. We give the coupling constants
of the nucleon resonances with the $\rho $-, and $\omega $-mesons,
determined form the fit, the total widths of the dilepton decays
of the nucleon resonances, and their dilepton spectra.

\section{The $\gamma ^{*}N\rightarrow N^{*}$ helicity amplitudes}
\setcounter{equation}{0}

$\;$
\vspace{-0.5cm}

The electromagnetic transition current between the nucleon and a
spin-$J$ nucleon resonance has the form 
\begin{equation}
J_{\mu }(p_{*},\lambda _{*},p,\lambda )=e\overline{u}_{\beta _{1}...\beta
_{l}}(p_{*},\lambda _{*}){\sf \Gamma }_{\beta _{1}...\beta _{l}\mu }^{(\pm
)}u(p,\lambda )
\end{equation}
where $m_{*}$ and $m$ are masses, $p_{*}$ and $p$ are momenta, $\lambda
_{*}\ $and $\lambda $ are helicities of the resonance and the nucleon, $e=-%
\sqrt{4\pi \alpha }$ is the electron charge, $\alpha =1/137$. In the
resonance rest frame, $p_{*}=(m_{*},0,0,0),$ $p=(E,0,0,-k).$

The spinor $u_{\beta _{1}...\beta _{l}}(p_{*},\lambda _{*})$ is the
generalized Rarita-Schwinger spinor (see e.g. \cite{BW,LAN}) that describes
fermions with $J=l+\frac{1}{2}$ $\geq \frac{3}{2}$. It is symmetric with
respect to the indices $\beta _{1}...\beta _{l}$ and traceless. The spinors
are normalized by 
\begin{eqnarray}
\overline{u}(p,\lambda)u(p,\lambda) &=&2m,  \nonumber \\
(-)^{l}\overline{u}_{\beta _{1}...\beta _{l}}(p_{*},\lambda_{*})u_{\beta
_{1}...\beta _{l}}(p_{*},\lambda_{*}) &=&2m_{*}  \label{NORM}
\end{eqnarray}
(there is a misprint in Eq.(15.7) of Ref. \cite{LAN}). 
The matrices ${\sf \Gamma }_{\beta _{1}...\beta _{l}\mu }^{(\pm )}$ stand
for the normal- and abnormal parity resonances, $J^{P}=\frac{1}{2}^{-},\frac{%
3}{2}^{+},\frac{5}{2}^{-},\;...$ (the upper sign) and $J^{P}=\frac{1}{2}^{+},%
\frac{3}{2}^{-},\frac{5}{2}^{+},...$ (the lower sign).

The photon polarization vectors have the form 
\begin{eqnarray}
\epsilon _{\mu }^{(\pm 1)}(q) &=&\frac{1}{\sqrt{2}}(0,\mp 1,-i,0),  \nonumber
\\
\epsilon _{\mu }^{(0)}(q) &=&\frac{1}{M}(k,0,0,\omega ),  \label{PH}
\end{eqnarray}
where $q=p_{*}-p=(\omega ,0,0,k)$, $q^{2}=M^{2}.$ These vectors are
transversal, $q_{\mu }\epsilon _{\mu }^{(\lambda )}(q)=0,$ and normalized by 
\begin{equation}
\epsilon _{\mu }^{(\lambda )}(q)^{*}\epsilon _{\mu }^{(\lambda ^{\prime
})}(q)=-\delta _{\lambda \lambda ^{\prime }}.
\end{equation}

In the limit $M\rightarrow 0$, $\epsilon _{\mu }^{(0)}(q)=q_{\mu }/M+O(M)$.
Due to the current conservation $q_{\mu }J_{\mu }=0,$ the longitudinal
component of the vector current equals $\epsilon _{\mu }^{(0)}(q)J_{\mu
}=O(M),$ so it vanishes for physical photons at $M=0$.

\subsection{The $\gamma ^{*}N\rightarrow N^{*}$ vertexes}
$\;$
\vspace{-0.5cm}

{\it Spin }$J\geq \frac{3}{2}${\it \ resonances.} The resonances with
arbitrary spin have three independent helicity amplitudes in the $\gamma
^{*}N\rightarrow N^{*}$ transitions. It means that there are three
independent scalar functions to fix the vertexes. The most general
decomposition of the vertex ${\sf \Gamma }_{\beta _{1}...\beta _{l}\mu
}^{(\pm )}$ over the Lorentz vectors and the Dirac gamma matrices has the
form \cite{JS,DEK,Trueman:1969wn} 
\begin{equation}
{\sf \Gamma }_{\beta _{1}...\beta _{l}\mu }^{(\pm )}=q_{\beta
_{1}}...q_{\beta _{l-1}}{\sf \Gamma }_{\beta _{l}\mu }^{(\pm )}  \label{dec}
\end{equation}
where 
\begin{equation}
{\sf \Gamma }_{\beta \mu }^{(\pm )}=\sum_{k}{\sf \Gamma }_{\beta \mu }^{(\pm
)k}F_{k}^{(\pm )}.  \label{dec_F}
\end{equation}
In Eq.(\ref{dec}), the symmetrization over the indices $\beta _{1},...,\beta
_{l}$ is assumed. The values $F_{k}^{(\pm)}$ are scalar functions of $q^2$ and
are called covariant form factors in the following. In this representation,
the Dirac structure of the transition amplitudes is fully separated off and
expressed by the ${\sf \Gamma }_{\beta \mu }^{(\pm)k}$ matrices.

For the normal-parity case, the matrices ${\sf \Gamma }_{\beta \mu }^{(+)i}$
($i=1,2,3$)\ have the form 
\begin{eqnarray}
{\sf \Gamma }_{\beta \mu }^{(+)1} &=&m_{*}(q_{\beta }\gamma _{\mu }-\not%
{q}g_{\beta \mu })\gamma _{5},  \label{K1} \\
{\sf \Gamma }_{\beta \mu }^{(+)2} &=&(q_{\beta }P_{\mu }-q\cdot Pg_{\beta
\mu })\gamma _{5},  \label{K2} \\
{\sf \Gamma }_{\beta \mu }^{(+)3} &=&(q_{\beta }q_{\mu }-q^{2}g_{\beta \mu
})\gamma _{5}  \label{K3}
\end{eqnarray}
where $\gamma _{5}=i\gamma ^{0}\gamma ^{1}\gamma ^{2}\gamma ^{3},$ $P=\frac{1%
}{2}(p_{*}+p)$. For the abnormal-parity case, the matrices ${\sf \Gamma }%
_{\beta \mu }^{(-)i}$ ($i=1,2,3$)\ can be taken to be 
\begin{equation}
{\sf \Gamma }_{\beta \mu }^{(-)k}={\sf \Gamma }_{\beta \mu }^{(+)k}\gamma
_{5}.  \label{K4}
\end{equation}
The sets (\ref{K1})-(\ref{K4}) are simply related to the sets used in Refs. 
\cite{JS,DEK}.

{\it Spin }$J=\frac{1}{2}${\it \ resonances. }The vertex ${\sf \Gamma }_{\mu
}^{(\pm )}$ ($l=0$)\ can also be expanded like in Eq.(\ref{dec_F}). There
are two matrices ${\sf \Gamma }_{\mu }^{(+)i}$ ($i=1,2$)\ for the
normal-parity case $J^{P}=\frac{1}{2}^{-}$, 
\begin{eqnarray}
{\sf \Gamma }_{\mu }^{(+)1} &=&(q^{2}\gamma _{\mu }-\not{q}q_{\mu })\gamma
_{5}, \\
{\sf \Gamma }_{\mu }^{(+)2} &=&(P\cdot q\gamma _{\mu }-P_{\mu }{q})\gamma
_{5},  \label{K5}
\end{eqnarray}
and two matrices for the abnormal-parity case $J^{P}=\frac{1}{2}^{+},$%
\begin{equation}
{\sf \Gamma }_{\mu }^{(-)k}={\sf \Gamma }_{\mu }^{(+)k}\gamma _{5}.
\label{K6}
\end{equation}
The sets (\ref{K5})-(\ref{K6}) are identical to the sets used in Ref. \cite
{DEK}. The vertex dimensions are ${\sf \Gamma }_{\beta _{1}...\beta _{l}\mu
}^{(\pm )}\backsim 1,$ ${\sf \Gamma }_{\beta \mu }^{(\pm )}\backsim 1/
m_{*}^{l-1}$, and $F_{k}^{(\pm )}\backsim 1/m_{*}^{l+1}.$

\subsection{The $N^{*}\rightarrow N\gamma ^{*}$ decay width in terms of
helicity amplitudes}
$\;$
\vspace{-0.5cm}

The photo- and electroproduction $T$-matrix elements ($S=1+iT$), 
\begin{equation}
<JJ_{z}|{\it T}|\lambda \lambda _{\gamma }{\bf n}>,
\end{equation}
depend on the resonance spin, $J$, its projection on the $z$-axis, $J_{z}$,
the nucleon and photon helicities, $\lambda $ and $\lambda _{\gamma }$, and
on the unit vector, ${\bf n}$, in the direction of the photon momentum. The $%
N^{*}\rightarrow N\gamma ^{*}$ width has the form 
\begin{equation}
\Gamma (N^{*}\rightarrow N\gamma ^{*})=\frac{k}{32\pi ^{2}m^{*2}}\int
d\Omega _{{\bf n}}\sum_{\lambda \lambda _{\gamma }}|<\lambda \lambda
_{\gamma }{\bf n}|{\it T}|JJ_{z}>|^{2}~.  \label{gammanng}
\end{equation}

The angular dependence of the matrix element $<JJ_{z}|{\it T}|\lambda
\lambda _{\gamma }{\bf n}>$ is a universal function (see e.g. \cite{Rose})
determined by the resonance total spin and its spin projections: 
\begin{equation}
<JJ_{z}|{\it T}|\lambda \lambda _{\gamma }{\bf n}>=D_{\lambda _{*}J_{z}}^{J}(%
{\bf n})^{*}<J\lambda _{*}{\bf n}|{\it T}|\lambda \lambda _{\gamma }{\bf n}>
\end{equation}
where $\lambda _{*}=-\lambda +\lambda _{\gamma }$ is the resonance helicity.
The rotation matrices 
\begin{equation}
D_{\lambda _{*}J_{z}}^{J}({\bf n})=<J\lambda _{*}{\bf n}|JJ_{z}>  \label{D}
\end{equation}
are the amplitudes of finding the resonance with the spin projection $%
\lambda _{*}$ on the unit vector ${\bf n}$ in a state with the spin
projection $J_{z}$ on the $z$-axis. The helicity amplitudes $<J\lambda _{*}%
{\bf n}|{\it T}|\lambda \lambda _{\gamma }{\bf n}>$ do not depend on the
vector ${\bf n}$, so the symbol ${\bf n}$ can be suppressed. There exist six
helicity amplitudes, three ones with positive $\lambda _{*}$'s and three
ones with negative $\lambda _{*}$'s. The $P$-invariance of the
electromagnetic interactions gives a symmetry relation for the amplitudes
with opposite signs of the helicities (see e.g. \cite{LAN}, Eq.(70.13)): 
\begin{equation}
<J-\lambda _{*}|{\it T}|-\lambda -\lambda _{\gamma }>=\mp <J\lambda _{*}|%
{\it T}|\lambda \lambda _{\gamma }>.
\end{equation}

The functions $D_{\lambda _{*}J_{z}}^{J}({\bf n})$ are the unitary matrices
with respect to the indices $\lambda _{*}$ and $J_{z}$. The normalization
condition reads
\begin{equation}
\int d\Omega _{{\bf n}}D_{\lambda _{*}J_{z}}^{J}({\bf n})^{*}D_{\lambda
_{*}^{\prime }J_{z}^{\prime }}^{J^{\prime }}({\bf n})=\frac{4\pi }{2J+1}%
\delta ^{JJ^{\prime }}\delta ^{\lambda _{*}\lambda _{*}^{\prime }}\delta
^{J_{z}J_{z}^{\prime }}.~  \label{O}
\end{equation}
Using the properties of the helicity amplitudes and of the $D_{\lambda
_{*}J_{z}}^{J}({\bf n})$ matrices, one obtains the $N^{*}\rightarrow N\gamma
^{*}$ decay width in terms of the three helicity amplitudes: 
\begin{equation}
\Gamma (N^{*}\rightarrow N\gamma ^{*})=\frac{k}{32\pi ^{2}m_{*}^{2}}\frac{%
8\pi }{2J+1}\sum_{\lambda _{*}=-\lambda +\lambda _{\gamma }>0}|<\lambda
\lambda _{\gamma }|{\it T}|J\lambda _{*}>|^{2}~.  \label{gammanng1}
\end{equation}

\subsection{Helicity amplitudes in terms of the covariant 
form factors F$_{k}^{(\pm )}$}
$\;$
\vspace{-0.5cm}

Since the amplitudes $<\lambda \lambda _{\gamma }|{\it T}|J\lambda _{*}>$ do
not depend on the vector ${\bf n,}$ it is convenient to choose it in the
direction of the $z$-axis. The helicity amplitudes can then be calculated in
terms of the covariant form factors $F_{k}^{(\pm )}\ $from equation 
\begin{equation}
\left\langle J\lambda _{*}^{(\pm )}|{\it T}|\lambda \lambda _{\gamma
}\right\rangle =-e\overline{u}_{\beta _{1}...\beta _{l}}(p_{*},\lambda _{*})%
{\sf \Gamma }_{\beta _{1}...\beta _{l}\mu }^{(\pm )}u(p,\lambda )\epsilon
_{\mu }^{(\lambda _{\gamma })}(q).  \label{amplitude}
\end{equation}
The sign $\pm$ refers to the natural- and abnormal-parity resonances. We use 
the following notations for these amplitudes: 
\begin{eqnarray}
{\frak F}_{\frac{3}{2}}^{(\pm )} &=&\left\langle J\frac{3}{2}^{(\pm )}\left|
T\right| -\frac{1}{2}1\right\rangle ,  \nonumber \\
{\frak F}_{\frac{1}{2}}^{(\pm )} &=&\left\langle J\frac{1}{2}^{(\pm )}\left|
T\right| +\frac{1}{2}1\right\rangle ,  \nonumber \\
\frac{M}{m_{*}}{\frak C}_{\frac{1}{2}}^{(\pm )} &=&\left\langle J\frac{1}{2}%
^{(\pm )}\left| T\right| -\frac{1}{2}0\right\rangle .  \label{AAS}
\end{eqnarray}
These amplitudes describe, respectively, the double-spin-flip, no-spin-flip,
and single-spin-flip transitions. For $l=0$, the amplitude ${\frak F}_{\frac{%
3}{2}}^{(\pm )}$ should be set equal to zero.

The experimental data for the helicity amplitudes (\ref{AAS}) are quoted by
PDG in the non-relativistic normalization for the fermions, including a
factor of $1/\sqrt{2\omega _{0}}$ from the photon wave function, and also a
sign of the $\pi N\rightarrow N^{*}$ amplitude and an additional sign $\mp $
for nucleon and $\Delta $-resonances. The value $\omega
_{0}=(m_{*}^{2}-m^{2})/(2m_{*})$ is the real-photon energy.

{\it Spin }$J\geq \frac{3}{2}${\it \ resonances.} The direct calculations
give 
\begin{eqnarray}
\left( 
\begin{array}{l}
\pm {\frak F}_{\frac{3}{2}}^{(\pm )} \\ 
-{\frak F}_{\frac{1}{2}}^{(\pm )} \\ 
\pm {\frak C}_{\frac{1}{2}}^{(\pm )}
\end{array}
\right) &=&\lambda _{l}^{(\pm )}\frac{2(\pm m)}{3m_{\pm }}\times  \nonumber
\\
&&\left( 
\begin{array}{lll}
\sqrt{\frac{l+2}{2l}}2m_{\pm }m_{*} & \sqrt{\frac{l+2}{2l}}m_{+}m_{-} & 
\sqrt{\frac{l+2}{2l}}2M^{2} \\ 
\sqrt{\frac{1}{2}}2(m_{\pm }(\mp m)+M^{2}) & \sqrt{\frac{1}{2}}m_{+}m_{-} & 
\sqrt{\frac{1}{2}}2M^{2} \\ 
2m_{*}^{2} & 2m_{*}^{2}-\frac{1}{2}\Delta _{0}^{2} & \Delta _{0}^{2}
\end{array}
\right) \left( 
\begin{array}{l}
F_{1}^{(\pm )} \\ 
F_{2}^{(\pm )} \\ 
F_{3}^{(\pm )}
\end{array}
\right)  \label{HEL3FFF}
\end{eqnarray}
where 
\begin{eqnarray*}
m_{\pm } &=&m_{*}\pm m, \\
\Delta _{0}^{2} &=&m_{+}m_{-}+M^{2}.
\end{eqnarray*}
The coefficients $\lambda _{l}^{(\pm )}$ are defined as
\begin{equation}
\lambda _{l}^{(\pm )}=e\frac{3m_{\pm }}{4(\pm m)}\sqrt{m_{\mp }^{2}-M^{2}}%
k^{l-1}\sqrt{\frac{2^{l}(l!)^{2}(l+1)}{(2l+1)!}}  \label{lamb}
\end{equation}
with $J = l + \frac{1}{2}$. Notice that 
\begin{eqnarray}
{\frak F}_{\frac{3}{2}}^{(-)}(m_{*},m) &=&-{\frak F}_{\frac{3}{2}%
}^{(+)}(m_{*},-m),  \nonumber \\
{\frak F}_{\frac{1}{2}}^{(-)}(m_{*},m) &=&+{\frak F}_{\frac{1}{2}%
}^{(+)}(m_{*},-m),  \nonumber \\
{\frak C}_{\frac{1}{2}}^{(-)}(m_{*},m) &=&-{\frak C}_{\frac{1}{2}%
}^{(+)}(m_{*},-m).
\end{eqnarray}

{\it Spin }$J=\frac{1}{2}${\it \ resonances. }The direct calculation gives
the following expression for the helicity amplitudes$:$

\begin{equation}
\left( 
\begin{array}{c}
{\frak F}_{\frac{1}{2}}^{(\pm )} \\ 
\pm \sqrt{2}{\frak C}_{\frac{1}{2}}^{(\pm )}
\end{array}
\right) =\frac{\lambda _{0}^{(\pm )}}{m_{*}}\left( 
\begin{array}{ll}
2M^{2} & m_{+}m_{-} \\ 
-2m_{*}m_{\mp } & -m_{*}m_{\pm }
\end{array}
\right) \left( 
\begin{array}{l}
F_{1}^{(\pm )} \\ 
F_{2}^{(\pm )}
\end{array}
\right) .  \label{HEL2FFF}
\end{equation}
The parameters $\lambda _{0}^{(\pm )}$ are defined by 
\begin{equation}
\lambda _{0}^{(\pm )}=e\frac{m_{*}}{\sqrt{2}}\sqrt{m_{\pm }^{2}-M^{2}}.
\end{equation}

\section{Magnetic, electric, and Coulomb transition form factors}
\setcounter{equation}{0}
$\;$
\vspace{-0.5cm}

The photo- and electroproduction of nucleon resonances$\ $can be described
in terms of the amplitudes ${\frak A}_{J_{\gamma },\ell }^{(\pm )}$ with the
definite total angular momentum $J_{\gamma }=J+\frac{1}{2}$, $J-\frac{1}{2}$
of the photon and the definite orbital momentum $\ell =J_{\gamma }+1,$ $%
J_{\gamma },$ $J_{\gamma }-1$ of the $N\gamma ^{*}$ system ($\ell$
should not be mixed with $l$ in Eq.(II.1)). There are three
nonvanishing amplitudes: 
\begin{equation}
{\frak A}_{J_{\gamma }^{\prime },J_{\gamma }^{\prime }}^{(\pm )},\;{\frak A}%
_{J_{\gamma },J_{\gamma }+1}^{(\pm )},\;{\frak A}_{J_{\gamma },J_{\gamma
}-1}^{(\pm )}  \label{JLM}
\end{equation}
where $J_{\gamma }^{\prime }=J\mp \frac{1}{2}$ and $J_{\gamma }=J\pm \frac{1%
}{2}.\;$In the spin $J=\frac{1}{2}$ case, the photo- and electroproduction
processes are described by two amplitudes ${\frak A}_{1,2}^{(+)}$ and$\;%
{\frak A}_{1,0}^{(+)}$ for negative parity resonances and ${\frak A}%
_{0,1}^{(-)}$ and ${\frak A}_{1,1}^{(-)}$ for positive parity resonances.

We establish first a connection of the magnetic, electric, and Coulomb
amplitudes to the amplitudes ${\frak A}_{J_{\gamma },\ell }^{(\pm )}$ and
find then a connection of the amplitudes ${\frak A}_{J_{\gamma },\ell
}^{(\pm )}$ to the helicity amplitudes. In this way, using Eqs. (\ref
{HEL3FFF}) and (\ref{HEL2FFF}), we express the magnetic, electric, and
Coulomb form factors in terms of the covariant form factors $F_{k}$.

\subsection{Relation between $J_{\gamma }\ell $ amplitudes and monopole amplitudes
}
$\;$
\vspace{-0.5cm}

In the momentum space, the photon vector potentials with the definite total
angular momentum, $J_{\gamma },$ its projection, $m$, on the $z$-axis, and
orbital momentum, $\ell ,$ have the form 
\begin{equation}
{\bf A}_{J_{\gamma }m\ell }({\bf n})=\sum_{\lambda _{\gamma }}C_{\ell
m-\lambda _{\gamma }1\lambda _{\gamma }}^{J_{\gamma }m}Y_{\ell m-\lambda
_{\gamma }}({\bf n}){\bf \varepsilon }^{(\lambda _{\gamma })}
\end{equation}
where ${\bf \varepsilon }^{(\lambda _{\gamma })}$ is the space-like part of
the polarization vector (\ref{PH}) with the vanishing momentum $k=0$. The
spherical coordinates of the vector potential have the form 
\begin{equation}
({\bf A}_{J_{\gamma }m\ell }({\bf n}))_{\lambda _{\gamma }}\equiv -{\bf %
\varepsilon }^{(\lambda _{\gamma })}\cdot {\bf A}_{J_{\gamma }m\ell }({\bf n}%
)=(-)^{1+\lambda _{\gamma }}C_{\ell m+\lambda _{\gamma }1-\lambda _{\gamma
}}^{J_{\gamma }m}Y_{\ell m+\lambda _{\gamma }}({\bf n}).
\end{equation}

In momentum space, the magnetic, electric and Coulomb potentials equal  
\cite{LAN}, 
\begin{equation}
{\bf A}_{J_{\gamma }m}^{T}({\bf n})={\bf a}^{T}Y_{J_{\gamma }m}({\bf n})
\label{MULT}
\end{equation}
where $T=M,E,C$ and 
\begin{eqnarray}
{\bf a}^{M} &=&\frac{{\bf l}}{\sqrt{J_{\gamma }(J_{\gamma }+1)}}, \nonumber \\
{\bf a}^{E} &=&\frac{{\bf \nabla }_{n}}{\sqrt{J_{\gamma }(J_{\gamma }+1)}},
\nonumber \\
{\bf a}^{C} &=&{\bf n}.
\end{eqnarray}
Here, ${\bf l}=-i{\bf k}\times $ $\partial /\partial {\bf k}$ is the orbital
momentum operator and ${\bf \nabla }_{n}=k\partial /\partial {\bf k.}$ Using
the Wigner-Eckart theorem (see e.g. \cite{Rose}), 
\begin{equation}
<\ell m^{\prime }|({\bf a}^{T})_{\lambda _{\gamma }}|J_{\gamma
}m>=C_{J_{\gamma }m1\lambda _{\gamma }}^{\ell m^{\prime }}<\ell ||{\bf a}%
^{T}||J_{\gamma }>,
\end{equation}
the spherical coordinates of the vector potentials ${\bf A}_{J_{\gamma
}m}^{T}({\bf n})$\ can be found to be 
\begin{equation}
({\bf A}_{J_{\gamma }m}^{T}({\bf n}))_{\lambda _{\gamma }}=\sum_{\ell
}(-)^{\ell +1-J_{\gamma }}\sqrt{\frac{2\ell +1}{2J_{\gamma }+1}}<\ell ||{\bf %
a}^{T}||J_{\gamma }>({\bf A}_{J_{\gamma }m\ell }({\bf n}))_{\lambda _{\gamma
}}.  \label{MUJL}
\end{equation}
Notice that $\ell =J_{\gamma }\pm 1$ for $T=M,E$ and $\ell =J_{\gamma }$ for 
$T=C.$ Eq.(\ref{MUJL}) gives a connection between the $J_{\gamma }\ell $\
and $J_{\gamma }T$ representations for the photon wave functions and,
respectively, for the $\gamma ^{*}N\rightarrow N^{*}$ transition amplitudes.

The linear combinations, ${\frak R}_{J_{\gamma },T}^{(\pm )}$, of the
amplitudes ${\frak A}_{J_{\gamma },\ell }^{(\pm )}$ with the coefficients of
Eq.(\ref{MUJL}) describe absorption of photons of the magnetic, electric,
and Coulomb types. In the matrix form, 
\begin{equation}
\left( 
\begin{array}{l}
{\frak R}_{J_{\gamma }^{\prime },M}^{(\pm )} \\ 
{\frak R}_{J_{\gamma },E}^{(\pm )} \\ 
{\frak R}_{J_{\gamma },C}^{(\pm )}
\end{array}
\right) =\left( 
\begin{array}{lll}
1 & 0 & 0 \\ 
0 & +\sqrt{\frac{J_{\gamma }}{2J_{\gamma }+1}} & +\sqrt{\frac{J_{\gamma }+1}{%
2J_{\gamma }+1}} \\ 
0 & -\sqrt{\frac{J_{\gamma }+1}{2J_{\gamma }+1}} & +\sqrt{\frac{J_{\gamma }}{%
2J_{\gamma }+1}}
\end{array}
\right) \left( 
\begin{array}{l}
{\frak A}_{J_{\gamma }^{\prime },J_{\gamma }^{\prime }}^{(\pm )} \\ 
{\frak A}_{J_{\gamma },J_{\gamma }+1}^{(\pm )} \\ 
{\frak A}_{J_{\gamma },J_{\gamma }-1}^{(\pm )}
\end{array}
\right) .  \label{MEC}
\end{equation}

\subsection{Relation between $J_{\gamma }\ell$ amplitudes and helicity amplitudes}
$\;$
\vspace{-0.5cm}

The amplitudes ${\frak A}_{J_{\gamma },\ell }^{(\pm )}$ can be expressed in
terms of the helicity amplitudes ${\frak F}_{\frac{3}{2}}^{(\pm )},$ ${\frak %
F}_{\frac{1}{2}}^{(\pm )},$ and ${\frak C}_{\frac{1}{2}}^{(\pm )}$. Let us
consider $T$-matrix elements 
\begin{equation}
<JJ_{z}|{\it T}|JJ_{z}J_{\gamma }\ell >,  \label{JgL}
\end{equation}
with $|JJ_{z}J_{\gamma }\ell >$ given by

\begin{equation}
|JJ_{z}J_{\gamma }\ell >=\sum_{{s_{z}}J_{\gamma z}}\sum_{{s_{\gamma z}}%
m}C_{J_{\gamma }J_{\gamma z}\frac{1}{2}{s_{z}}}^{JJ_{z}}C_{\ell m1{s_{\gamma
z}}}^{J_{\gamma }J_{\gamma z}}|\ell m>|\frac{1}{2}{s_{z}}>|1{s_{\gamma z}}>.
\label{DECO}
\end{equation}
The states $|JJ_{z}J_{\gamma }\ell >$ are eigenstates of the total angular
momentum $J,$ its projection on the $z$- axis $J_{z},\;$the total photon
angular momentum $J_{\gamma }$, and of the orbital angular momentum $%
\ell $ of the $N\gamma ^{*}$ system. The values $C_{J_{\gamma }J_{\gamma z}%
\frac{1}{2}{s_{z}}}^{JJ_{z}}$ and $C_{\ell m1{s_{\gamma z}}}^{J_{\gamma
}J_{\gamma z}}$ are the usual Clebsh-Gordon coefficients (CGC's) with the
phase conventions of PDG \cite{Groom:2000in}. In Eq.(\ref{DECO}) and below,
the standard rules for combining the angular momenta and spins are
used (see e.g. \cite{Manley:1984jz}), according to which baryons in CGC's
appear before mesons, the orbital momentum in CGC's comes before the
intrinsic spin, and the angles in the spherical harmonic $Y_{lm}({\bf n})$
are measured with respect to the first particle in the corresponding isospin
CGC's.

Using the set $|\lambda \lambda _{\gamma }{\bf n}>$ of the helicity states,
the $T$-matrix elements $<JJ_{z}|{\it T }$ $|JJ_{z}J_{\gamma }\ell >$ 
can be transformed as follows: 
\begin{equation}
<JJ_{z}|{\it T}|JJ_{z}J_{\gamma }\ell >=\int d\Omega _{{\bf n}}\sum_{\lambda
\lambda _{\gamma }}<JJ_{z}|{\it T}|\lambda \lambda _{\gamma }{\bf n}%
><\lambda \lambda _{\gamma }{\bf n}|JJ_{z}J_{\gamma }\ell >.  \label{QQQ}
\end{equation}
The state $|JJ_{z}J_{\gamma }\ell >$ is a superposition of eigenstates of
the helicity $\lambda _{*}$ of the resonance: 
\[
|JJ_{z}J_{\gamma }\ell >=\sum_{\lambda _{*}}D_{\lambda _{*}J_{z}}^{J}({\bf n}%
)|J\lambda _{*}J_{\gamma }\ell >. 
\]
Using the decomposition of Eq.(\ref{DECO}) for the states $|J\lambda
_{*}J_{\gamma }\ell >$, the amplitudes $<\lambda \lambda _{\gamma }{\bf n}%
|JJ_{z}J_{\gamma }\ell >$ can be found to be 
\begin{equation}
<\lambda \lambda _{\gamma }{\bf n}|JJ_{z}J_{\gamma }\ell >=\sqrt{\frac{2\ell
+1}{4\pi }}D_{\lambda _{*}J_{z}}^{J}({\bf n})C_{J_{\gamma }\lambda _{\gamma }%
\frac{1}{2}-{\lambda }}^{J\lambda _{*}}C_{\ell 01\lambda _{\gamma
}}^{J_{\gamma }\lambda _{\gamma }}.
\end{equation}
The CGC's entering Eq.(III.2) originate from the decomposition of Eq.(\ref{DECO}), where one
should set $J_{z}=\lambda _{*}$, $J_{\gamma z}={s_{\gamma z}=}\lambda
_{\gamma }$,\ and $m=J_{\gamma z}-{s_{\gamma z}=0.}$\ The coefficient $\sqrt{%
\frac{2\ell +1}{4\pi }}$ arises from the spherical harmonic $<{\bf n}|\ell
m>=Y_{lm}({\bf n})$ evaluated at $m=0$ and ${\bf n}=(0,0,1)$.

The integration in Eq.(\ref{QQQ}) removes the $D$-matrices. The $T$-matrix
elements $<JJ_{z}|{\it T}$ $|JJ_{z}J_{\gamma }\ell >$ 
are expressed in terms of the $T$-matrix elements $%
<J\lambda _{*}|{\it T}$ $|\lambda \lambda_{\gamma }>$ as follows 
\begin{equation}
<JJ_{z}|{\it T}|JJ_{z}J_{\gamma }\ell >=\frac{\sqrt{4\pi (2\ell +1)}}{2J+1}%
\sum_{\lambda \lambda _{\gamma }}C_{J_{\gamma }\lambda _{\gamma }\frac{1}{2}-%
{\lambda }}^{J\lambda _{*}}C_{\ell 01\lambda _{\gamma }}^{J_{\gamma }\lambda
_{\gamma }}<J\lambda _{*}|{\it T}|\lambda \lambda _{\gamma }>.  \label{AAA}
\end{equation}
The amplitudes ${\frak A}_{J_{\gamma },\ell }^{(\pm )}$ can be defined by

\begin{equation}
{\frak A}_{J_{\gamma },\ell }^{(\pm )}=\sqrt{\frac{2J+1}{8\pi }}<JJ_{z}|{\it %
T}|JJ_{z}J_{\gamma }\ell >.
\end{equation}
The symmetry of the helicity amplitudes under the $P$-transformation can be
used to reduce the summation in Eq.(\ref{AAA}) to positive values of $%
\lambda _{*}=-\lambda +\lambda _{\gamma }>0$, which yields
\begin{equation}
{\frak A}_{J_{\gamma },\ell }^{(\pm )}=\frac{1\pm (-)^{-l+\ell }}{\sqrt{2}}%
\sqrt{\frac{2\ell +1}{2J+1}}\sum_{\lambda _{*}=-\lambda +\lambda _{\gamma
}>0}C_{J_{\gamma }\lambda _{\gamma }\frac{1}{2}-{\lambda }}^{J\lambda
_{*}}C_{\ell 01\lambda _{\gamma }}^{J_{\gamma }\lambda _{\gamma }}<J\lambda
_{*}|{\it T}|\lambda \lambda _{\gamma }>.  \label{UJgL}
\end{equation}

\subsection{Relation between magnetic, electric, and Coulomb transition form
factors and helicity amplitudes}
$\;$
\vspace{-0.5cm}

{\it Spin }$J\geq \frac{3}{2}${\it \ resonances. }The electric, magnetic and
Coulomb transition form factors are defined by 
\begin{eqnarray}
{\frak R}_{J-\frac{1}{2},M/E}^{(\pm )} &=&\mp \lambda _{l}^{(\pm )}\sqrt{%
\frac{l+1}{l}}G_{M/E}^{(\pm )},  \nonumber \\
{\frak R}_{J+\frac{1}{2},E/M}^{(\pm )} &=&\mp \lambda _{l}^{(\pm )}\sqrt{%
(l+1)(l+2)}G_{E/M}^{(\pm )},  \nonumber \\
{\frak R}_{J\pm \frac{1}{2},C}^{(\pm )}\;\;\; &=&\mp \lambda _{l}^{(\pm )}%
\frac{M}{m_{*}}G_{C}^{(\pm )}.
\end{eqnarray}
Here, the $G_{M/E}^{(\pm )}$ stands for $G_{M}^{(+)}$ or $G_{E}^{(-)}$ and
the $G_{E/M}^{(\pm )}$ stands for $G_{E}^{(+)}$ or $G_{M}^{(-)}$ with the 
coefficients $\lambda _{l}^{(\pm )}$ given by Eq.(2.24). Using Eqs.(%
\ref{MEC}) and (\ref{AAS})\ and substituting the coefficients $C{}_{\ell
01\lambda _{\gamma }}^{J_{\gamma }\lambda _{\gamma }}$ and $C_{J_{\gamma
}\lambda _{\gamma }\frac{1}{2}\lambda }^{J\lambda _{*}}$ in Eq.(\ref{UJgL}),
we obtain the following relation between the electric, magnetic, and Coulomb
form factors and the helicity amplitudes:

\begin{equation}
\lambda _{l}^{(\pm )}\left( 
\begin{array}{l}
\sqrt{\frac{l+1}{l}}G_{M/E}^{(\pm )} \\ 
\sqrt{(l+1)(l+2)}G_{E/M}^{(\pm )} \\ 
G_{C}^{(\pm )}
\end{array}
\right) =\left( 
\begin{array}{lll}
+\sqrt{\frac{l+2}{2(l+1)}} & +\sqrt{\frac{l}{2(l+1)}} & 0 \\ 
+\sqrt{\frac{l}{2(l+1)}} & -\sqrt{\frac{l+2}{2(l+1)}} & 0 \\ 
0 & 0 & +1
\end{array}
\right) \left( 
\begin{array}{l}
{\frak F}_{\frac{3}{2}}^{(\pm )} \\ 
{\frak F}_{\frac{1}{2}}^{(\pm )} \\ 
{\frak C}_{\frac{1}{2}}^{(\pm )}
\end{array}
\right) .  \label{EMC3HEL}
\end{equation}
The transformation matrix is an orthogonal matrix.

The helicity amplitudes are expressed in terms of the covariant form factors $F_{k}$
in Eq.(\ref{HEL3FFF}). Eqs.(\ref{HEL3FFF}) and (\ref{EMC3HEL}) can be
combined to give a linear relation between the magnetic, electric, and
Coulomb form factors and the covariant form factors $F_{k}$.

The monopole form factors are expressed in terms of
the covariant form factors $F_{k}^{(+)}$ as follows:

\begin{eqnarray}
\left( 
\begin{array}{l}
G_{M}^{(+)} \\ 
G_{E}^{(+)} \\ 
G_{C}^{(+)}
\end{array}
\right) &=&\frac{2m}{3m_{{\large +}}}\left( 
\begin{array}{lll}
\frac{\Delta _{l+1}^{2}}{l+1} & \frac{m_{+}m_{-}}{l+1} & \frac{2M^{2}}{l+1}
\\ 
\frac{\Delta _{0}^{2}}{l+1} & \frac{m_{+}m_{-}}{l+1} & \frac{2M^{2}}{l+1} \\ 
2m_{*}^{2} & 2m_{*}^{2}-\frac{1}{2}\Delta _{0}^{2} & \Delta _{0}^{2}
\end{array}
\right) \left( 
\begin{array}{l}
F_{1}^{(+)} \\ 
F_{2}^{(+)} \\ 
F_{3}^{(+)}
\end{array}
\right) ,  \label{FG_lP} \\
\left( 
\begin{array}{l}
G_{M}^{(-)} \\ 
G_{E}^{(-)} \\ 
G_{C}^{(-)}
\end{array}
\right) &=&\frac{2m}{3m_{{\large -}}}\left( 
\begin{array}{lll}
\frac{\sigma _{-}^{2}}{l+1} & 0 & 0 \\ 
\frac{\sigma _{-}^{2}}{l+1}+\Delta _{0}^{2} & m_{+}m_{-} & 2M^{2} \\ 
2m_{*}^{2} & 2m_{*}^{2}-\frac{1}{2}\Delta _{0}^{2} & \Delta _{0}^{2}
\end{array}
\right) \left( 
\begin{array}{l}
F_{1}^{(-)} \\ 
F_{2}^{(-)} \\ 
F_{3}^{(-)}
\end{array}
\right)  \label{FG_lM}
\end{eqnarray}
where 
\begin{eqnarray*}
m_{\pm } &=&m_{*}\pm m, \\
\Delta _{l}^{2} &=&m_{+}m_{-}+M^{2}+l\sigma _{+}^{2}, \\
\sigma _{\pm }^{2} &=&m_{\pm }^{2}-M^{2}.
\end{eqnarray*}
The photon momentum appearing in Eq.(\ref{PH}) can be written as $k=\sqrt{%
\sigma _{-}^{2}\sigma _{+}^{2}}/(2m_{*}).$ The inverse transformations have
the form 
\begin{eqnarray}
\left( 
\begin{array}{l}
F_{1}^{(+)} \\ 
F_{2}^{(+)} \\ 
F_{3}^{(+)}
\end{array}
\right) &=&\frac{3m_{+}}{2m\sigma _{+}^{2}\sigma _{{\large -}}^{2}}\times 
\nonumber \\
&&\left( 
\begin{array}{lll}
\;\sigma _{{\large -}}^{2} & -\sigma _{{\large -}}^{2} & 0 \\ 
-\sigma _{{\large -}}^{2} & 2m_{*}m_{-}+l\Delta _{0}^{2} & -2M^{2} \\ 
-\frac{1}{2}\sigma _{{\large -}}^{2} & -m_{*}m_{+}-\frac{1}{2}%
l(4m_{*}^{2}-\Delta _{0}^{2}) & m_{+}m_{-}
\end{array}
\right) \left( 
\begin{array}{l}
G_{M}^{(+)} \\ 
G_{E}^{(+)} \\ 
G_{C}^{(+)}
\end{array}
\right) ,  \label{FGPLUS3} \\
\left( 
\begin{array}{l}
F_{1}^{(-)} \\ 
F_{2}^{(-)} \\ 
F_{3}^{(-)}
\end{array}
\right) &=&\frac{3m_{-}}{2m\sigma _{+}^{2}\sigma _{{\large -}}^{2}}\times 
\nonumber \\
&&\left( 
\begin{array}{lll}
\;(l+1)\sigma _{{\large +}}^{2} & 0 & 0 \\ 
-2m_{*}m_{+}-l\sigma _{{\large +}}^{2} & \Delta _{0}^{2} & -2M^{2} \\ 
m_{*}m_{-}-\frac{1}{2}l\sigma _{{\large +}}^{2} & -2m_{*}^{2}+\frac{1}{2}%
\Delta _{0}^{2} & m_{+}m_{-}
\end{array}
\right) \left( 
\begin{array}{l}
G_{M}^{(-)} \\ 
G_{E}^{(-)} \\ 
G_{C}^{(-)}
\end{array}
\right) .  \label{FGMINU3}
\end{eqnarray}

In terms of the magnetic, electric, and Coulomb form factors, the resonance
decay widths equal 
\begin{eqnarray}
\Gamma (N_{(\pm )}^{*} &\rightarrow &N\gamma ^{*})=\frac{9\alpha }{16}\frac{%
(l!)^{2}}{2^{l}(2l+1)!}\frac{m_{\pm }^{2}(m_{\mp }^{2}-M^{2})^{l+1/2}(m_{\pm
}^{2}-M^{2})^{l-1/2}}{m_{*}^{2l+1}m^{2}}  \nonumber \\
&&\left( \frac{l+1}{l}\left| G_{M/E}^{(\pm )}\right| ^{2}+(l+1)(l+2)\left|
G_{E/M}^{(\pm )}\right| ^{2}+\frac{M^{2}}{m_{*}^{2}}\left| G_{C}^{(\pm
)}\right| ^{2}\right) .  \label{GAMMA_l}
\end{eqnarray}

There is a symmetry between expressions for decay widths of the normal- and
abnormal-parity resonances: $m_{+}\leftrightarrow m_{-},$ $%
G_{M}^{(+)}\leftrightarrow G_{E}^{(-)},\;G_{E}^{(+)}\leftrightarrow
G_{M}^{(-)},$ $G_{C}^{(+)}\leftrightarrow G_{C}^{(-)}.$ For $l=1$, we recover
the result of Ref.\cite{Krivoruchenko:2001hs}.

{\it Spin }$J=\frac{1}{2}${\it \ resonances. }In the lowest spin case, the
electric and magnetic form factors are defined by 
\begin{eqnarray*}
{\frak R}_{J-\frac{1}{2},M/E}^{(\pm )} &=&G_{M/E}^{(\pm )}\equiv 0, \\
{\frak R}_{J+\frac{1}{2},E/M}^{(\pm )} &=&\mp \sqrt{2}\lambda _{0}^{(\pm
)}G_{E/M}^{(\pm )}, \\
{\frak R}_{J\pm \frac{1}{2},C\;}^{(\pm )}\; &=&\mp \lambda _{0}^{(\pm )}%
\frac{M}{m_{*}}G_{C}^{(\pm )}.
\end{eqnarray*}
The helicity amplitudes are simply connected to the electric (magnetic) and
Coulomb amplitudes: 
\begin{equation}
\lambda _{0}^{(\pm )}\left( 
\begin{array}{l}
\sqrt{2}G_{E/M}^{(\pm )} \\ 
G_{C}^{(\pm )}
\end{array}
\right) =\left( 
\begin{array}{ll}
-1 & 0 \\ 
0 & 1
\end{array}
\right) \left( 
\begin{array}{l}
{\frak F}_{\frac{1}{2}}^{(\pm )} \\ 
{\frak C}_{\frac{1}{2}}^{(\pm )}
\end{array}
\right)  \label{EMC2HEL}
\end{equation}

The Eqs. (\ref{HEL2FFF}) and (\ref{EMC2HEL}) can be combined to give a
relation between the magnetic, electric, and Coulomb form factors and the
form factors $F_{k}$. In the spin-$\frac{1}{2}$ case, these form factors
have the form

\begin{equation}
\left( 
\begin{array}{l}
G_{E/M}^{(\pm )} \\ 
\pm G_{C}^{(\pm )}
\end{array}
\right) =-\frac{1}{\sqrt{2}m_{*}}\left( 
\begin{array}{ll}
\ 2M^{2} & \ m_{+}m_{-} \\ 
\ 2m_{*}m_{\mp } & \ m_{*}m_{\pm }
\end{array}
\right) \left( 
\begin{array}{l}
F_{1}^{(\pm )} \\ 
F_{2}^{(\pm )}
\end{array}
\right) .  \label{FG2}
\end{equation}
The inverse relations are as follows 
\begin{equation}
\left( 
\begin{array}{l}
F_{1}^{(\pm )} \\ 
F_{2}^{(\pm )}
\end{array}
\right) =\frac{1}{\sqrt{2}m_{\pm }\sigma _{\mp }^{2}}\left( 
\begin{array}{ll}
m_{*}m_{\pm } & \ -m_{+}m_{-} \\ 
\ -2m_{*}m_{\mp } & \ \ 2M^{2}
\end{array}
\right) \left( 
\begin{array}{l}
G_{E/M}^{(\pm )} \\ 
\pm G_{C}^{(\pm )}
\end{array}
\right) .
\end{equation}

The resonance decay width can be found to be 
\begin{eqnarray}
\Gamma (N_{(\pm )}^{*} &\rightarrow &N\gamma ^{*})=\frac{\alpha }{8m_{*}}%
(m_{\pm }^{2}-M^{2})^{3/2}(m_{\mp }^{2}-M^{2})^{1/2}  \nonumber \\
&&\left( 2\left| G_{E/M}^{(\pm )}\right| ^{2}+\frac{M^{2}}{m_{*}^{2}}\left|
G_{C}^{(\pm )}\right| ^{2}\right) .  \label{GAMMA_0}
\end{eqnarray}

We use the normalization for the monopole form factors identical to Ref. \cite
{DEK}. The $\Delta (1232)$-resonance form factors of Refs. \cite
{JS,Krivoruchenko:2001hs} contain an additional factor of $\sqrt{\frac{2}{3}}%
.$ In the limit $M\rightarrow 0,$ we recover Eqs.(2.59) and (2.60) of Ref. \cite{DEK}.
Eqs. (\ref{GAMMA_l}) and (\ref{GAMMA_0}) are the main results of this Sect.

\section{The $N^{*}\rightarrow N\gamma ^{*}$ partial-wave amplitudes}
\setcounter{equation}{0}
$\;$
\vspace{-0.5cm}

The decay width of the $N^{*}\rightarrow N\gamma ^{*}$ transition can be
calculated in terms of the amplitudes ${\frak H}_{S,\ell }^{(\pm )}$ with
the definite total spin, $S$, of the $N\gamma ^{*}$ system and the definite
orbital momentum, $\ell ,$ of the $N\gamma ^{*}$ system\ ($\ell =J\mp \frac{1%
}{2},J\pm \frac{3}{2}$ for $J^{P}=\frac{3}{2}^{\pm },$ $\frac{5}{2}^{\mp },$ 
$...$ ). The nucleon decays are described by three independent amplitudes 
\begin{equation}
{\frak H}_{\frac{1}{2},J\mp \frac{1}{2}}^{(\pm )},\;{\frak H}_{\frac{3}{2}%
,J\mp \frac{1}{2}}^{(\pm )},\;{\frak H}_{\frac{3}{2},J\pm \frac{3}{2}}^{(\pm
)},
\end{equation}
while in the case $J^{P}=\frac{1}{2}^{\mp },$ the decays are described by
two independent amplitudes ${\frak H}_{\frac{1}{2},0}^{(+)},$ ${\frak H}_{%
\frac{3}{2},2}^{(+)}$ and ${\frak H}_{\frac{1}{2},1}^{(-)},$ ${\frak H}_{%
\frac{3}{2},1}^{(-)}$. The amplitudes ${\frak H}_{S,\ell }^{(\pm )}$ can be
connected to the helicity amplitudes ${\frak F}_{\frac{3}{2}}^{(\pm )},$ $%
{\frak F}_{\frac{1}{2}}^{(\pm )},$ and ${\frak C}_{\frac{1}{2}}^{(\pm )}$.

\subsection{Relation between partial-wave and helicity amplitudes}
$\;$
\vspace{-0.5cm}

Let us consider the $T$-matrix elements 
\begin{equation}
<JJ_{z}S\ell |{\it T}|JJ_{z}>.
\end{equation}
The states $|JJ_{z}S\ell >$ are defined by 
\begin{equation}
|JJ_{z}S\ell >=\sum_{S_{z}m}C_{\ell mSS_{z}}^{JJ_{z}}|\ell m>|SS_{z}>.
\label{SL}
\end{equation}
They are eigenstates of the total spin, $S,$ of the nucleon and a virtual
photon, of the orbital momentum, $\ell $, and of the total angular momentum, 
$J$, and its projection on $z$- axis$,J_{z},$ of the decaying resonance $%
N^{*}$. Using the set of the helicity states $|\lambda \lambda _{\gamma }%
{\bf n}>$, one can relate the $T$-matrix elements $<JJ_{z}S\ell |{\it T}%
|JJ_{z}>$ to the $T$-matrix elements $<\lambda \lambda _{\gamma }{\bf n}|%
{\it T}|JJ_{z}>$: 
\begin{equation}
<JJ_{z}S\ell |{\it T}|JJ_{z}>=\int d\Omega _{{\bf n}}\sum_{\lambda \lambda
_{\gamma }}<JJ_{z}S\ell |\lambda \lambda _{\gamma }{\bf n}><\lambda \lambda
_{\gamma }{\bf n}|{\it T}|JJ_{z}>~.  \label{SLJJ}
\end{equation}

The state $<JJ_{z}S\ell |$ can be rotated to give a superposition 
\begin{equation}
<JJ_{z}S\ell |=\sum_{\lambda _{*}}D_{J_{z}\lambda _{*}}^{J}({\bf n}%
)^{*}<J\lambda _{*}S\ell |.
\end{equation}
Using the decomposition of Eq.(%
\ref{SL}) with $J_{z}=S_{z}=\lambda _{*}$ and $m=0$ 
for the rotated state $<J\lambda _{*}S\ell |$, we obtain 
\begin{equation}
<JJ_{z}S\ell |\lambda \lambda _{\gamma }{\bf n}>=\sqrt{\frac{2\ell +1}{4\pi }%
}C_{\ell 0S\lambda _{*}}^{J\lambda _{*}}C_{\frac{1}{2}-\lambda 1\lambda
_{\gamma }}^{S\lambda _{*}}D_{J_{z}\lambda _{*}}^{J}({\bf n})^{*}.
\end{equation}
The factor $\sqrt{\frac{2\ell +1}{4\pi }}$ originates from the product $%
<\ell m|{\bf n}>=Y_{lm}^{*}({\bf n})$ evaluated at $m=0$ and ${\bf n}%
=(0,0,1) $. The angular dependence in the amplitudes $<\lambda \lambda
_{\gamma }{\bf n}|{\it T}|JJ_{z}>$ is factorized with the help of Eq.(\ref
{D}). The $D$-matrices are then removed by the angular integration in Eq.(\ref{SLJJ}%
). The $T$-matrix elements in the partial-wave representation, 
$<JJ_{z}S\ell |{\it T}|JJ_{z}>$, become \cite{JW}

\begin{equation}
<JJ_{z}S\ell |{\it T}|JJ_{z}>=\frac{\sqrt{4\pi (2\ell +1)}}{2J+1}%
\sum_{\lambda \lambda _{\gamma }}C_{\ell 0S\lambda _{*}}^{J\lambda _{*}}C_{%
\frac{1}{2}-\lambda 1\lambda _{\gamma }}^{S\lambda _{*}}<\lambda \lambda
_{\gamma }|{\it T}|J\lambda _{*}>.  \label{SLJJOK}
\end{equation}

We extract from the amplitude the kinematical factors $\lambda _{l}^{(\pm )}$
and define the amplitude ${\frak H}_{S,\ell }^{(\pm )}$ as follows 
\begin{equation}
\lambda _{l}^{(\pm )}{\frak H}_{S,\ell }^{(\pm )}=\sqrt{\frac{2J+1}{8\pi }}%
<S\ell |{\it T}|JJ_{z}>.
\end{equation}
The symmetry of the helicity amplitudes under the $P$-transformation can be
used to remove in Eq.(\ref{SLJJOK}) the summation over the negative values
of $\lambda _{*}.$ We thus obtain 
\begin{equation}
\lambda _{l}^{(\pm )}{\frak H}_{S,\ell }^{(\pm )}=\frac{1\pm (-)^{-l+\ell }}{%
\sqrt{2}}\sqrt{\frac{2\ell +1}{2J+1}}\sum_{\lambda _{*}=-\lambda +\lambda
_{\gamma }>0}C_{\ell 0S\lambda _{*}}^{J\lambda _{*}}C_{\frac{1}{2}-\lambda
1\lambda _{\gamma }}^{S\lambda _{*}}<\lambda \lambda _{\gamma }|{\it T}%
|J\lambda _{*}>.
\end{equation}

{\it Spin }$J\geq \frac{3}{2}${\it \ resonances. }The partial-wave
amplitudes are linear combinations of the helicity amplitudes:

\begin{eqnarray}
\lambda _{l}^{(+)}\left( 
\begin{array}{c}
{\frak H}_{\frac{1}{2},J-\frac{1}{2}}^{(+)} \\ 
{\frak H}_{\frac{3}{2},J-\frac{1}{2}}^{(+)} \\ 
{\frak H}_{\frac{3}{2},J+\frac{3}{2}}^{(+)}
\end{array}
\right) &=&\left( 
\begin{array}{ccc}
0 & -\sqrt{\frac{2}{3}} & +\sqrt{\frac{1}{3}} \\ 
-\sqrt{\frac{3(l+2)}{2(2l+3)}} & -\sqrt{\frac{l}{6(2l+3)}} & -\sqrt{\frac{l}{%
3(2l+3)}} \\ 
-\sqrt{\frac{l}{2(2l+3)}} & +\sqrt{\frac{l+2}{2(2l+3)}} & +\sqrt{\frac{l+2}{%
2l+3}}
\end{array}
\right) \left( 
\begin{array}{r}
{\frak F}_{\frac{3}{2}}^{(+)} \\ 
{\frak F}_{\frac{1}{2}}^{(+)} \\ 
\frac{M}{m_{*}}{\frak C}_{\frac{1}{2}}^{(+)}
\end{array}
\right),  \\
\lambda _{l}^{(-)}\left( 
\begin{array}{c}
{\frak H}_{\frac{1}{2},J+\frac{1}{2}}^{(-)} \\ 
{\frak H}_{\frac{3}{2},J+\frac{1}{2}}^{(-)} \\ 
{\frak H}_{\frac{3}{2},J-\frac{3}{2}}^{(-)}
\end{array}
\right) &=&\left( 
\begin{array}{ccc}
0 & +\sqrt{\frac{2}{3}} & -\sqrt{\frac{1}{3}} \\ 
+\sqrt{\frac{3l}{2(2l+1)}} & -\sqrt{\frac{l+2}{6(2l+1)}} & -\sqrt{\frac{l+2}{%
3(2l+1)}} \\ 
+\sqrt{\frac{l+2}{2(2l+1)}} & +\sqrt{\frac{l}{2(2l+1)}} & +\sqrt{\frac{l}{%
2l+1}}
\end{array}
\right) \left( 
\begin{array}{r}
{\frak F}_{\frac{3}{2}}^{(-)} \\ 
{\frak F}_{\frac{1}{2}}^{(-)} \\ 
\frac{M}{m_{*}}{\frak C}_{\frac{1}{2}}^{(-)}
\end{array}
\right).
\end{eqnarray}
The matrices entering these equations are orthogonal ones.

{\it Spin }$J=\frac{1}{2}${\it \ resonances. }The transformation from the
helicity basis to the partial-wave basis is given by 
\begin{eqnarray}
\lambda _{0}^{(+)}\left( 
\begin{array}{c}
{\frak H}_{\frac{1}{2},J-\frac{1}{2}}^{(+)} \\ 
{\frak H}_{\frac{3}{2},J+\frac{3}{2}}^{(+)}
\end{array}
\right) &=&\left( 
\begin{array}{cc}
-\sqrt{\frac{2}{3}} & +\sqrt{\frac{1}{3}} \\ 
+\sqrt{\frac{1}{3}} & +\sqrt{\frac{2}{3}}
\end{array}
\right) \left( 
\begin{array}{r}
{\frak F}_{\frac{1}{2}}^{(+)} \\ 
\frac{M}{m_{*}}{\frak C}_{\frac{1}{2}}^{(+)}
\end{array}
\right) , \\
\lambda _{0}^{(-)}\left( 
\begin{array}{c}
{\frak H}_{\frac{1}{2},J+\frac{1}{2}}^{(-)} \\ 
{\frak H}_{\frac{3}{2},J+\frac{1}{2}}^{(-)}
\end{array}
\right) &=&\left( 
\begin{array}{cc}
+\sqrt{\frac{2}{3}} & -\sqrt{\frac{1}{3}} \\ 
-\sqrt{\frac{1}{3}} & -\sqrt{\frac{2}{3}}
\end{array}
\right) \left( 
\begin{array}{r}
{\frak F}_{\frac{1}{2}}^{(-)} \\ 
\frac{M}{m_{*}}{\frak C}_{\frac{1}{2}}^{(-)}
\end{array}
\right) .
\end{eqnarray}
The inverse transformations are given by the transposed matrices. 
Eqs.(IV.10) - (IV.13) are in agreement with Ref. \cite
{Koniuk:1982ej} where the transformation
matrices are given for $l=0,1,2,3$.

\subsection{ Dilepton decay widths}
$\;$
\vspace{-0.5cm}

The $N^{*}\rightarrow N\gamma ^{*}$ decay width has many equivalent
representations that can be obtained from Eqs.(\ref{gammanng}) and (\ref
{gammanng1}) making use of the completeness of sets of the $N\gamma ^{*}$ states: 
\begin{eqnarray}
\frac{32\pi ^{2}m^{*2}}{k}\Gamma (N^{*} &\rightarrow &N\gamma ^{*})= 
\nonumber \\
&=&\int d\Omega _{{\bf n}}\sum_{\lambda \lambda _{\gamma }}|<\lambda \lambda
_{\gamma }{\bf n}|{\it T}|JJ_{z}>|^{2}~  \nonumber \\
&=&\frac{8\pi }{2J+1}\sum_{\lambda _{*}=-\lambda +\lambda _{\gamma
}>0}|<\lambda \lambda _{\gamma }|{\it T}|J\lambda _{*}>|^{2}  \nonumber \\
&=&\sum_{J_{\gamma }\ell }|<JJ_{z}J_{\gamma }\ell |{\it T}%
|JJ_{z}>|^{2}=\sum_{S\ell }|<JJ_{z}S\ell |{\it T}|JJ_{z}>|^{2}  \nonumber \\
&=&\frac{8\pi }{2J+1}\sum_{J_{\gamma }\ell }|{\frak A}_{J_{\gamma },\ell
}^{(\pm )}|^{2}=\frac{8\pi }{2J+1}\sum_{J_{\gamma }T}|{\frak R}_{J_{\gamma
},T}^{(\pm )}|^{2}  \nonumber \\
&=&\frac{8\pi }{2J+1}\sum_{S\ell }(\lambda_l^{(\pm)})^2|{\frak H}_{S,\ell }^{(\pm )}|^{2}.
\label{OK}
\end{eqnarray}
One can add here also expressions (\ref{GAMMA_l}) and (\ref{GAMMA_0}) which
calculate the decay widths using the monopole
transition form factors. These form factors are the most frequently used ones
both in analyzing experimental data and in theoretical works.

If the width $\Gamma (N^{*}\rightarrow N\gamma ^{*})$ is known, the
factorization prescription (see {\it e.g.} \cite{Faessler:2000de}) can be
used to find the dilepton decay rate:

\begin{equation}
d\Gamma (N^{*}\rightarrow Ne^{+}e^{-})=\Gamma (N^{*}\rightarrow N\gamma
^{*})M\Gamma (\gamma ^{*}\rightarrow e^{+}e^{-})\frac{dM^{2}}{\pi M^{4}},
\label{OK!}
\end{equation}
where 
\begin{equation}
M\Gamma (\gamma ^{*}\rightarrow e^{+}e^{-})=\frac{\alpha }{3}%
(M^{2}+2m_{e}^{2})\sqrt{1-\frac{4m_{e}^{2}}{M^{2}}}  \label{OK!!}
\end{equation}
is the decay width of a virtual photon $\gamma ^{*}$ into the dilepton
pair with invariant mass $M$.

The physical $N^{*}\rightarrow N\gamma $ decay rate is given by Eqs.(\ref{OK}%
) in the limit of $M=0$. Eqs.(\ref{OK})-(\ref{OK!!}) or (\ref{GAMMA_l}), (%
\ref{GAMMA_0}), (\ref{OK!}), and (\ref{OK!!}) being combined give the $%
N^{*}\rightarrow Ne^{+}e^{-}$ decay rates.

\section{Transition form factors and vector meson dominance}
\setcounter{equation}{0}
$\;$
\vspace{-0.5cm}

The non-relativistic quark models are successful in the description of static
hadron properties and hadron decays. These models, however, are not well suited
for the calculation of the dilepton emission, since the electromagnetic form factors
should be interpolated into the time-like region. In the time-like region, the 
vector meson dominance comes into play, whereas the
non-relativistic quark models have direct photon-quark couplings. They
predict form factors which behave like $exp(q^{2}/\varkappa ),$ whereas the
VMD requires a Breit-Wigner shape of the spectral functions for the form factors, 
centered around the vector meson masses with the physical vector meson widths.
In a finite interval of the space-like region, the experimental data can be fitted with an
exponential formula, whereas at high negative $q^{2}$ the non-relativistic
quark models contradict to the quark counting rules. They contradict also to 
the Frazer-Fulco unitarity relations for the nucleon form factors \cite
{Frazer:1959gy,LLME,SSM,Hohler:1975ht,HBOOK,FUR,Krivoruchenko:1995cv} in the 
time-like region. The
unitarity relations, from the other side, justify the VMD model. The distinction
between expressions given by the exact solution of the unitarity relations
for the isovector nucleon form factors and the VMD expression is not
strong. As to the pion form factor, the deviation of the naive VMD
expression (see Eq.(\ref{PIF}))\ from the FFGS expression \cite
{Frazer:1959gy,Gounaris:1968mw}, obtained by solving the unitarity relations
for the pion form factor (see e.g. \cite{BD}), is quite small. The same is true for the
isovector kaon form factor \cite{Krivoruchenko:1993vd}.

Using the VMD model, we satisfy the quark counting rules, analyticity in the 
complex $q^{2}$-plane, and take into account (approximately) the unitarity 
relations. It is important also that the VMD model gives simple expressions 
for the covariant form factors which can easily be embedded into the heavy-ion codes, 
if residues of the covariant form factors at the vector meson poles are determined. 

\subsection{Extended VMD model}
$\;$
\vspace{-0.5cm}

In terms of the vector meson fields, $V_{\mu }$, the electromagnetic current
has the form \cite{SAK} 
\begin{equation}
J_{\mu }^{em}=-e\sum_{V}\frac{m_{V}^{2}}{g_{V}}V_{\mu }  \label{A2}
\end{equation}
where $m_{V}$ are the vector meson masses. The $SU(3)$ predictions for the
coupling constants, $g_{\rho }:g_{\omega }:g_{\phi }=1:3:\frac{-3}{\sqrt{2}}%
, $ are in good agreement with the values $g_{\rho }=5.03,$ $g_{\omega
}=17.1$, and $g_{\phi }=-12.9$ extracted from the $V\rightarrow e^{+}e^{-}$
decays of the $\rho $-$,$ $\omega $-$,$ and $\phi $-mesons. The expression (%
\ref{A2}) determines the vector meson couplings with the photon.

The VMD model describes well the electromagnetic pion form factor:
\begin{equation}
F_{\pi }(q^{2})=\frac{f_{\rho \pi \pi }}{g_{\rho }}\frac{m_{\rho }^{2}}{%
m_{\rho }^{2}-q^{2}},  \label{PIF}
\end{equation}
with $f_{\rho \pi \pi }$ being the coupling constant of the effective Lagrangian 
\begin{eqnarray}
{\cal L}_{\rho \pi \pi } &=&-\frac{1}{2}f_{\rho \pi \pi }
\epsilon _{\alpha \beta \gamma }\rho _{\mu }^{\alpha }
(\pi ^{\beta }{\partial }_{\mu }\pi ^{\gamma }) \nonumber \\
&=&-f_{\rho \pi \pi }(
\rho _{\mu }^{0}\pi ^{-}i{\partial }_{\mu }\pi ^{+}+
\rho _{\mu }^{+}\pi ^{0}i{\partial }_{\mu }\pi ^{-}+
\rho _{\mu }^{-}\pi ^{+}i{\partial }_{\mu }\pi ^{0})  \label{RHOPIPI}
\end{eqnarray}
where ${\partial }_{\mu } = \overrightarrow{\partial }_{\mu } - \overleftarrow{\partial }_{\mu }$. 
The normalization $F_{\pi }(0)=1$ implies 
\begin{equation}
f_{\rho \pi \pi }/g_{\rho }=1.  \label{SAC}
\end{equation}

The quark counting rules \cite{Matveev:1973ra} show that the pion form
factor decreases like $F_{\pi }(q^{2})\backsim 1/q^{2}$ as $q^{2} \rightarrow \infty .$
The VMD predicts therefore the correct asymptotics.

The electromagnetic nucleon form factors demonstrate experimentally a dipole
behavior. The quark counting rules for the Sachs form factors predict $%
G_{E}(q^{2})\backsim G_{M}(q^{2})\backsim 1/q^{4}$ at $q^{2}\rightarrow
\infty .$ The VMD model with the ground-state $\rho $-$,$ $\omega $-$,$ and $%
\phi $-mesons cannot describe the nucleon form factors at low values of $%
q^{2}$ (the isovector charge radius is underestimated) and gives in contrast to the pion
incorrect asymptotic behavior. It was proposed \cite
{Hohler:1976ax,Krivoruchenko:1994qb,Mergell:1996bf,Dubnicka:1996sp} to
include in the current (\ref{A2}) excited states of the vector mesons $\rho
^{\prime },$ $\rho ^{\prime \prime },$ ... etc. The VMD model extended in
this way allows to reproduce the low- and intermediate-energy experimental
data and yields for the nucleon form factors the correct asymptotic behavior.
The minimal extension of the VMD model improves the description of the $\rho
\pi \gamma$ transition form factor that falls off asymptotically as $1/q^{4}
$ \cite{Faessler:2000de}.

The vector meson couplings with the
nucleon resonances are defined by the $T$-matrix element of the $%
VN\rightarrow N^{*}$ process 
\begin{equation}
<J\lambda _{*}|{\it T}|\lambda \lambda _{V}>=\sum_{k}f_{VNN^{*},k}^{(\pm )}%
\overline{u}_{\beta _{1}...\beta _{l}}(p_{*},\lambda _{*})q_{\beta
_{1}}...q_{\beta _{l-1}}{\sf \Gamma }_{\beta _{l}\mu }^{(\pm )k}u(p,\lambda
)\epsilon _{\mu }^{(\lambda _{V})}(q)  \label{A1}
\end{equation}
where the vertexes ${\sf \Gamma }_{\beta \mu }^{(\pm )k}$ are the same as
for the photon, and $\epsilon _{\mu }^{(\lambda _{V})}(k)$ is the
polarization vector of the vector meson $V$ with momentum $q$ and helicity $%
\lambda _{V}$.

The combination of Eqs.(\ref{A1}) and (\ref{A2}) allows to calculate the photo-
and electroproduction amplitudes 
\begin{eqnarray}
&<&J\lambda _{*}|{\it T}|\lambda \lambda _{\gamma
}>=\sum_{k}\sum_{V}f_{VNN^{*},k}^{(\pm )}\frac{em_{V}^{2}}{g_{V}}\frac{1}{%
q^{2}-m_{V}^{2}}\times  \nonumber \\
&&\overline{u}_{\beta _{1}...\beta _{l}}(p_{*},\lambda _{*})q_{\beta
_{1}}...q_{\beta _{l-1}}{\sf \Gamma }_{\beta _{l}\mu }^{(\pm )k}u(p,\lambda
)\epsilon _{\mu }^{(\lambda _{\gamma })}(q)  \label{A3}
\end{eqnarray}
The comparison with Eq.(\ref{amplitude}) shows that the covariant form factors
have the form 
\begin{equation}
F_{k}^{(\pm )}(M^{2})=\sum_{V}\frac{f_{VNN^{*},k}^{(\pm )}}{g_{V}}\frac{1}{%
1-M^{2}/m_{V}^{2}}.  \label{VMDF}
\end{equation}
The $\Delta $-resonance form factors have only contributions from the $\rho $%
-meson family. If the covariant form factors $F_{k}^{(\pm )}(M^{2})$ are known,
the coupling constants $f_{\rho ^{0}NN^{*},k}^{(\pm )}$ for the $\Delta $%
-resonances can be found from equation 
\begin{equation}
f_{\rho ^{0}NN^{*},k}^{(\pm )}=-\frac{g_{\rho }}{m_{\rho }^{2}}{\normalsize %
res}\left\{ F_{k}^{(\pm )}(M^{2}=m_{\rho }^{2})\right\} .  \label{RESD}
\end{equation}
The nucleon resonances receive contributions from the $\rho $-and $\omega $%
-mesons. The couplings with the nucleon resonances are calculated as
residues of a superposition for isospin projections $I_{3}=+\frac{1}{2}$ and 
$I_{3}=-\frac{1}{2}:$%
\begin{equation}
f_{VNN^{*},k}^{(\pm )}=-\frac{g_{V}}{2m_{V}^{2}}{\normalsize res}\left\{
F_{k}^{(\pm )}(M^{2}=m_{V}^{2})^{I_{3}=+\frac{1}{2}}\mp F_{k}^{(\pm
)}(M^{2}=m_{V}^{2})^{I_{3}=-\frac{1}{2}}\right\}  \label{RESN}
\end{equation}
where $V=\rho ^{0}(\omega )$ with the corresponding upper (lower) sign between
the two isospin form factors.

The quark counting rules predict the following asymptotics for the helicity
amplitudes 
\begin{eqnarray}
{\frak F}_{\frac{3}{2}}^{(\pm )} &=&O(\frac{1}{(-M^{2})^{5/2}}),
\nonumber \\
{\frak F}_{\frac{1}{2}}^{(\pm )} &=&O(\frac{1}{(-M^{2})^{3/2}}),
\nonumber \\
{\frak C}_{\frac{1}{2}}^{(\pm )} &=&O(\frac{1}{(-M^{2})^{5/2}}).
\label{QCR1}
\end{eqnarray}
These constraints can be used to reduce the number of free parameters of the
model. 

The transition form factors of nucleon resonances with high spins decrease 
stronger than the diagonal nucleon form factors.

{\it Spin }$J\geq \frac{3}{2}${\it \ resonances. }Now, taking into account
that $\lambda _{l}^{(\pm )}=O((-M^{2})^{l-1/2})\ $at $M^{2}\rightarrow
-\infty ,$ we get the asymptotics of the covariant form factors $F_{k}^{(\pm )}(M^{2})$
at $M^{2}\rightarrow -\infty $: 
\begin{eqnarray}
F_{1}^{(\pm )}(M^{2}) &=&O(\frac{1}{(-M^{2})^{l+2}}), \nonumber \\
F_{2}^{(\pm )}(M^{2}) &=&O(\frac{1}{(-M^{2})^{l+3}}), \nonumber \\
F_{3}^{(\pm )}(M^{2}) &=&O(\frac{1}{(-M^{2})^{l+3}}).
\end{eqnarray}

These constraints can be resolved to give

\begin{eqnarray}
F_{1}^{(\pm )}(M^{2}) &=&\frac{{\sum }_{j=0}^{n+1}C_{1j}^{(\pm )}{M^{2}}^{j}%
}{{\prod }_{i=1}^{l+3+n}(1-M^{2}/m_{i}^{2})},  \nonumber \\
F_{2}^{(\pm )}(M^{2}) &=&\frac{{\sum }_{j=0}^{n}C_{2j}^{(\pm )}{M^{2}}^{j}}{{%
\prod }_{i=1}^{l+3+n}(1-M^{2}/m_{i}^{2})},  \nonumber \\
F_{3}^{(\pm )}(M^{2}) &=&\frac{{\sum }_{j=0}^{n}C_{3j}^{(\pm )}{M^{2}}^{j}}{{%
\prod }_{i=1}^{l+3+n}(1-M^{2}/m_{i}^{2})}.  \label{FF_l}
\end{eqnarray}
Here, $C_{kj}^{(\pm )}$ are free parameters of the extended VMD model, ${%
l+3+n}$ is the total number of the vector mesons. For each form factor, the
quark counting rules reduce the number of free parameters from ${l+3+n}$ to $%
{n+2}$ for $k=1$ and to ${n+1}${\ for} $k=2,3$. In the simplest case $n=0,$
the knowledge of the four parameters $C_{10}^{(\pm )}$, $C_{11}^{(\pm )}$, 
$C_{20}^{(\pm )}$, and $C_{30}^{(\pm )}$ is sufficient to fix $F_{k}^{(\pm
)}(M^{2})$. In the zero-width limit, the multiplicative representation (\ref
{FF_l}) is completely equivalent to an additive representation of Eq.(\ref
{VMDF}).

The similar multiplicative representation motivated by the Regge theory
is used in Ref. \cite{DEK}, Eqs.(3.7). The asymptotic dominance of the transverse 
covariant form factors, used in that work as an assumption, 
does not agree with the quark counting rules. 

{\it Spin }$J=\frac{1}{2}${\it \ resonances. }In terms of the amplitudes $%
{\frak F}_{\frac{1}{2}}^{(\pm )}$ and ${\frak C}_{\frac{1}{2}}^{(\pm )}$,
the constraints to the asymptotics have the form of Eqs.(%
\ref{QCR1}). Taking into account that $\lambda _{0}^{(\pm
)}=O((-M^{2})^{1/2})\ $at $M^{2}\rightarrow -\infty ,$ we get 
\begin{equation}
F_{1,2}^{(\pm )}(M^{2})=O(\frac{1}{(-M^{2})^{3}}).
\end{equation}
The general representation for the covariant form factors in the spin-$\frac{1}{2}$
case has the form 
\begin{equation}
F_{k}^{(\pm )}(M^{2})=\frac{{\sum }_{j=0}^{n}C_{kj}^{(\pm )}{M^{2}}^{j}}{{%
\prod }_{i=1}^{3+n}(1-M^{2}/m_{i}^{2})}.  \label{FF_0}
\end{equation}

\subsection{Relative sign of the photo- and electroproduction amplitudes 
and amplitudes for the nucleon resonance decays into the vector mesons}
$\;$
\vspace{-0.5cm}

The experimental data for the photo- and electroproduction amplitudes are
quoted by PDG for the amplitudes 
\begin{eqnarray}
A_{\frac{3}{2}} &=&\frac{\xi _{I}\xi }{\sqrt{8m_{*}m\omega _{0}}}{\frak F}_{%
\frac{3}{2}}^{(\pm )},  \label{A32} \\
A_{\frac{1}{2}} &=&\frac{\xi _{I}\xi }{\sqrt{8m_{*}m\omega _{0}}}{\frak F}_{%
\frac{1}{2}}^{(\pm )},  \label{A12} \\
S_{\frac{1}{2}} &=&\frac{\xi _{I}\xi }{\sqrt{8m_{*}m\omega _{0}}}{\frak C}_{%
\frac{1}{2}}^{(\pm )},  \label{S12}
\end{eqnarray}
which include the phase factor of the $N^{*}\rightarrow N\pi $ decay, $\xi
=A(N^{*}\rightarrow N\pi )/|A(N^{*}\rightarrow N\pi )|$ and an isospin factor $%
\xi _{I}=-1$ for nucleon resonances and $\xi _{I}=+1$ for $\Delta $%
-resonances. The phase factor $\xi $ appears, since the amplitude $\gamma
^{*}N\rightarrow N^{*}$ is accompanied in the photoproduction experiments by
the subsequent pion decay\ of the nucleon resonance.

The experimental data for the vector meson decays are quoted for the values $%
\sigma \sqrt{\Gamma _{N^{*}\rightarrow NV}}$ where $\sigma $ is a sign of
the amplitudes $N^{*}\rightarrow N\rho ^{0}$ or $N^{*}\rightarrow N\omega $
multiplied by $\xi _{V}\xi ^{*}$, where $\xi $ is the pion decay phase and $%
\xi _{V}=A(V\rightarrow $pions$)$/ $|A(V\rightarrow $pions$)|.$ The
additional factor $\xi _{V}$ appears, since the vector meson production is
accompanied by the vector meson decays into pions. The isospin symmetry
implies $\Gamma _{\Delta ^{*}\rightarrow N \rho }=\frac{3}{2}\Gamma
_{\Delta ^{*}\rightarrow N \rho ^{0}}$ and $\Gamma _{N^{*}\rightarrow
N\rho }=3\Gamma _{N^{*}\rightarrow N\rho ^{0}}$.

The quark models are very successful in the description of static properties of
hadrons and hadron decays. The models \cite
{Koniuk:1980vy,Warns:1990ic,Li:1990qu,Bijker:1994yr,Capstick:1992uc} give
predictions for the photo- and electroproduction amplitudes (\ref{A32}) - (%
\ref{S12}). The $^{3}P_{0}$ quark-pair creation model by Yaouanc et al. \cite
{LeYaouanc:1975mr} gives the vector-meson decay amplitudes inclusive of the
phase of $\xi _{V}\xi ^{*}.$ The non-relativistic quark model by Koniuk \cite
{Koniuk:1982ej} gives these amplitudes without the factor $\xi _{V}$, so its
predictions are valid up to an overall sign. The quark-pair creation models
of Refs. \cite{Capstick:1994kb,Stassart:1990zt,Stancu:1993xz} do not include
the factor $\xi _{V}$ either. The multichannel $\pi N$-scattering partial-wave
analysis of Manley and Saleski \cite{Manley:1992yb} has an overall sign
ambiguity of the $\pi \pi N$ amplitudes with respect to the $\pi N$
amplitudes. The $\pi \pi N$ amplitudes interfere with the $N\rho $
amplitudes due to the $\rho \rightarrow \pi \pi $ decay, so the overall $%
N\rho $ phase is not fixed.

The extended VMD model, from the other side, gives a unified description of the photo- and
electroproduction data and of the resonance decays into the vector mesons,
including the signs of the amplitudes. We wish to use the data \cite
{Groom:2000in,Manley:1992yb,Koniuk:1982ej,Capstick:1994kb} to fix the transition form
factors. The overall phase of the $N^{*}\rightarrow NV$ decays with respect 
to the $\gamma^{*}N\rightarrow N^{*}$ amplitudes is not fixed, however, neither 
experimentally, nor theoretically.

Yaouanc and co-authors \cite{LeYaouanc:1975mr}
calculate the quantity $\xi _{\rho }.$ It depends on the sign of the coupling
constant. The product $\xi _{V}\xi ^{*}$ is negative, being an even function
of the coupling. They do not analyze, however, the photo- and
electroproduction amplitudes. The photo- and electroproduction and vector
meson decays of the nucleon resonances are calculated in Refs. \cite
{Bijker:1994yr,Bijker:1997tr}. The authors give, however, only relative
signs of the $N^{*}\rightarrow NV$ amplitudes. The overall sign correlated
with the photo- and electroproduction amplitudes is, in principle, provided
by the quark models of Refs. \cite{Koniuk:1982ej,Koniuk:1980vy} and \cite
{Capstick:1994kb,Capstick:1992uc}. The $N^{*}\rightarrow NV$ amplitudes of
Ref. \cite{Koniuk:1982ej} depend on the quark coupling with the vector
mesons, $g$, whose sign is a matter of convention. The quantity $\xi _{\rho }
$ is, however, proportional to the same coupling, $g,$ so the quantity $%
\sigma \sqrt{\Gamma _{N^{*}\rightarrow NV}}\ $is an even function of $g$ and
therefore well defined. We thus propose a solution of the "sign ambiguity" problem
within the non-relativistic Isgur-Koniuk \cite{Koniuk:1982ej,Koniuk:1980vy} 
quark model framework. (The similar analysis can probably be made also in 
the quark-pair creation model of Ref. \cite
{Capstick:1994kb,Capstick:1992uc}.) 

The value of $\xi $ is not calculated in our model. It enters as a common
factor to the experimentally measured photo- and electroproduction
amplitudes and the vector meson decay amplitudes and can be absorbed by the
covariant form factors $F_{k}^{(\pm )}$. The isospin factor $\xi _{I}$ will also be
absorbed by the form factors. The common phase of the form factors does not
influence the dilepton decay rates.

The value of $\xi _{\rho }$ can easily be found. Using the effective
Lagrangian (\ref{RHOPIPI}), the $P$-wave amplitude of the $\rho
^{0}\rightarrow \pi ^{+}\pi ^{-}$ decay can be found to be $<\pi \pi |{\it T}%
|\rho ^{0}>=-f_{\rho \pi \pi }2k_{\pi }$ where $k_{\pi }$ is the absolute
value of the pion momentum. Eq.(\ref{RHOPIPI}) shows that the coupling $%
f_{\rho \pi \pi }$ has a meaning of the $\pi ^{+}$ charge with respect to
the massive vector field $\rho _{\mu }^{0}.$ In the quark model by Koniuk 
\cite{Koniuk:1982ej}, \thinspace the $\rho ^{0}$ static charge of the up-quarks
is assumed to be positive \cite{note}. The value of $f_{\rho
\pi \pi }$ is therefore positive, and so $\xi _{\rho }=-1$.

The vector meson decay amplitudes of the nucleon resonances are proportional
to the product $f_{\rho ^{0}NN^{*},k}^{(\pm )}f_{\rho \pi \pi }$. They are
invariant with respect to the sign change of the vector meson coupling,
since $f_{\rho ^{0}NN^{*},k}^{(\pm )}\backsim g$ and $f_{\rho \pi \pi
}\backsim g$. The VMD in the electromagnetic pion form factor gives rise to
Eq.(\ref{SAC}), so the coupling $g_{\rho }$ is an odd function of $g$. The
sign convention for the coupling $g$ does not affect the isovector parts of
the photo- and electroproduction amplitudes, since they are proportional to
the ratios $f_{\rho ^{0}NN^{*},k}^{(\pm )}/g_{\rho }$. The model we discuss
thus provides a consistent description of the isovector part of the
processes $\gamma ^{*}N\rightarrow N^{*}$ and of the $N^{*}\rightarrow N\rho
^{0}$ decays.

Similar arguments could be applied to the effective Lagrangian ${\cal L}%
_{\omega \rho \pi }$ to determine the $\omega \rightarrow 3\pi $ amplitude
through a two-step mechanism $\omega \rightarrow \rho \pi ,$ $\rho
\rightarrow \pi \pi $. The $\omega \rho \pi $ coupling constant can,
however, be calculated up to an arbitrary phase only. It is proportional to
the $\rho \pi $ transition magnetic moment (with respect to a static
magnetic $\omega $-meson field) which is not a diagonal one. The $\omega
\rho \pi $ coupling is proportional to an arbitrary phase difference of the $%
\rho $- and $\pi $-meson wave functions (as distinct from the $\rho \pi \pi $
diagonal transition where such a phase is identically zero). The relative
phase of the quantities $\xi _{\rho }$ and $\xi _{\omega }$ is well defined
for identical final states, e.g. for the $\rho \rightarrow $ $\pi \gamma $
and $\omega \rightarrow \pi \gamma $ decays, in which case $\xi _{\rho }/\xi
_{\omega }=+1,$ while the $\omega \rightarrow 3\pi $ phase cannot be fixed
with respect to the $\rho \rightarrow \pi \pi $ phase.

Godfrey and Isgur \cite{Godfrey:1985xj} calculate meson decay amplitudes by
considering the pseudoscalar mesons as elementary fields. The overall sign
of these amplitudes is not correlated with the amplitudes calculated by
considering the vector mesons as elementary fields. It can be seen e.g. from
Table V of Ref. \cite{Godfrey:1985xj}, according to which the $\rho
\rightarrow \pi \pi $ amplitude is imaginary, whereas we get above for the $%
\rho \rightarrow \pi \pi $ decay a real negative amplitude. The reason stems
from the fact that the amplitudes of Ref. \cite{Godfrey:1985xj} are
proportional to the coupling constant of the quarks with the pions, whereas
the amplitudes of Ref. \cite{Koniuk:1982ej} are proportional to the coupling
constant of the quarks with the vector mesons. The relative signs (phases)
of these coupling constants are not correlated.

The model by Koniuk \cite{Koniuk:1982ej} has, however, a $SU(3)$ symmetric
vertex for quarks interacting with the vector mesons. In this model, $%
1/g_{\rho }\backsim tr\tau ^{3}Q=1$ and $1/g_{\omega }\backsim trQ=1/3$
where $\tau ^{3}$ is the isospin Pauli matrix and $Q=$ diag$(2/3,-1/3)$ is
the quark charge matrix. The relative sign of the coupling constants $%
g_{\rho }$ and $g_{\omega }$ is apparently known. The same relative sign ($+$%
) is given by Godfrey and Isgur \cite{Godfrey:1985xj}, which is independent on
the quark-pion coupling constant. 

This knowledge is
sufficient to compare predictions of the extended VMD model with the
amplitudes of Refs. \cite{Koniuk:1982ej}, which do not include the phases $%
\xi _{V}$.

Manley and Saleski \cite{Manley:1992yb} found that the $\pi N$ partial-wave
analysis is in good agreement with the results by Koniuk \cite{Koniuk:1982ej}
taken with the opposite sign. We got $\xi _{\rho }=-1$, using the model by
Koniuk, so the results by Manley and Saleski for the $\pi \pi N$ waves have
apparently the overall sign correlated correctly with the photo- and
electroproduction amplitudes, as calculated by Koniuk and Isgur \cite
{Koniuk:1980vy}.

We thus perform a fit to the amplitudes for the photo- and electroproduction
of the nucleon resonances with the standard phase conventions of PDG \cite
{Groom:2000in} and Koniuk and Isgur \cite{Koniuk:1980vy}. We fit further the
vector meson decay {\it amplitudes}. The overall sign of these amplitudes is
chosen such as to reproduce in the extended VMD model both the photo- and
electroproduction amplitudes and the vector meson amplitudes of the nucleon
resonance decays, calculated in the non-relativistic quark model 
by Koniuk and Isgur \cite{Koniuk:1980vy} and by Koniuk \cite
{Koniuk:1982ej}. The transition form factors of nucleon resonances
determined this way do not depend on the sign convention adopted in the
quark model \cite{Koniuk:1982ej} for the coupling constant, $g$, of quarks
with the vector mesons:

In the Koniuk quark model, the amplitudes of the vector meson decays of the
nucleon resonances are proportional to the coupling constant $g$. In the
extended VMD model, these amplitudes are proportional to the coupling
constants $f_{VNN^{*},k}^{(\pm )},$ according to Eq.(\ref{A1}). We found
earlier $f_{\rho \pi \pi }\backsim g$. The coupling constant $g_{\rho }\ $is
also proportional to $g,$ according to Eq.(\ref{SAC}). The $SU(3)$ symmetry
implies $g_{\omega }\backsim g$. The $g$-dependence drops out from the
transition form factors (\ref{VMDF}). In agreement with the quark
model, Eqs.(\ref{RESD}) and (\ref{RESN}) give
afterwards $f_{VNN^{*},k}^{(\pm )}\backsim g$. This completes the consistency check.

\section{Numerical results}
\setcounter{equation}{0}
$\;$
\vspace{-0.5cm}

The parameters $C_{kj}^{(\pm )}$ of the extended VMD model, entering
Eqs.(\ref{FF_l}) and (\ref{FF_0}), are determined from the fit to the photo-
and electroproduction data \cite
{Groom:2000in,Stein:1975yy,Bartel:1968tw,Batzner,Frolov:1999pw} and the
vector meson decay amplitudes of the nucleon resonances \cite
{Groom:2000in,Manley:1992yb,Longacre:1977ja,Koniuk:1982ej,Capstick:1994kb}.
We use the minimal $n=0$ extension of the VMD model for all resonances.

The number of the vector mesons required for each isotopic channel to
ensure the correct asymptotic behavior depends on the total spin of the 
nucleon resonance. For spin-$J$ resonances, we need $3 + l$ ($=3 + J - \frac{1}{2}$) 
excited vector mesons with the same quantum numbers for 
the minimal extension of the VMD. The nucleon resonances we consider have
spins $J$ from $\frac{1}{2}$ to $\frac{7}{2}$. It means that we need at
the most 6 excited vector mesons for each isotopic channel. The following
masses are used: 0.769, 1.250, 1.450, 1.720, 2.150, 2.350 (in GeV). The
numbers appearing on the 1 and 3 - 5 positions are masses of
the physical $\rho$-mesons according to PDG \cite{Groom:2000in}. 
The possible existence of vector mesons with
masses around 1.250 GeV is discussed for a long time. The phenomenology of
the nucleon form factors and, in particularly, the scaling laws for the Sachs 
form factors make the existence of an enhancement in the spectral function of the
nucleon form factors at 1.250 GeV very plausible (for details see \cite{Faessler:2000de}). 
The results on the dilepton emission do not depend strongly
on the exact numerical values of the masses of the excited vector mesons, since the
dilepton energy spectrum extends only slightly 
above 1 GeV for the nucleon resonances with masses about 2 GeV. In the region of 
the invariant masses $M<1$ GeV, the form factors are smooth functions of the masses 
of the excited vector mesons. The last mass is set equal to 2.350 GeV for an
estimate. We assumed further a degeneracy between the $\rho$
and $\omega$ families. The strange $\phi$ mesons are decoupled in our model from the nucleons
due to the OZI rule. The opposite assumption is used in 
Refs. \cite{Hohler:1976ax,Mergell:1996bf}. The widths of the mesons 2 - 6 is set
equal to zero, the widths of the ground-state $\rho$- and $\omega$-mesons are taken
from PDG.

For the nucleon resonance decays into the vector mesons, we use the data from PDG \cite
{Groom:2000in}. When these data are not available (quite often), we use the
Manley and Saleski results (MS) of the multichannel $\pi N$
partial-wave analysis \cite{Manley:1992yb}.  In other cases, we use the quark model
predictions by Koniuk  \cite{Koniuk:1982ej} with $50\%$ errors and $0.05$ MeV$%
^{1/2}$ errors if the values are close to zero. In a few cases, the
results \cite{Longacre:1977ja} of the multichannel $\pi N$ partial-wave
analysis of Longacre and Dolbeau (LD) with $50\%$ errors and of Capstick and
Roberts (CR) quark model predictions \cite{Capstick:1994kb} are used, when
other results do not agree with the most recent PDG constraints to the total
vector meson decay widths. The PDG and MS data are included to the $\chi ^{2}
$ with greater weights. Below we give details of our fitting procedure:

$N(1535)\frac{1}{2}^{-}$: The experimental values for $A_{1/2}$ are from Ref.\cite{Burk}.
The $N\rho$ mode $s_{1/2}$ is taken from PDG. The $N\rho$ mode $%
d_{3/2}$ is taken from MS. The $\omega$-meson mode $d_{3/2}$ is set equal to zero. 

$N(1650)\frac{1}{2}^{-}$: The $N\rho $ mode $s_{1/2}$ is taken from PDG with
the negative sign as predicted by K. The MS sign is not well fixed. The mode $d_{3/2}$ is taken from PDG.
The $N\omega $ modes are from the Koniuk paper (K) with $0.05$ MeV$^{1/2}$ errors.

$N(1520)\frac{3}{2}^{-}$: The experimental values for $A_{1/2}$ and $A_{3/2}$ are 
from Ref.\cite{Burk}. The modes $d_{1/2}$ and $d_{3/2}$ are taken from K
with a $0.05$ MeV$^{1/2}$ error. The mode $s_{3/2}$ is taken from PDG.

$N(1700){\frac{3}{2}}^{-}$: The experimental values for $A_{1/2}$ and $A_{3/2}$ are 
from Ref.\cite{Burk}. The $N\rho $ modes $d_{1/2}$ and $d_{3/2}$ are
taken from K with a $0.05$ MeV$^{1/2}$ error. The mode $s_{3/2}$ is taken
from PDG with the negative sign as predicted by MS. The $N\omega $ modes are
from K with $0.05$ MeV$^{1/2}$ errors.

$N(1675){\frac{5}{2}}^{-}$: The $N\rho $ mode $d_{1/2}$ is taken from MS,
the mode $d_{3/2}$ is taken from PDG. The $g_{3/2}$ modes are from K with a
$0.05$ MeV$^{1/2}$ error. The $N\omega$ mode $g_{1/2}$ is set equal to zero.


$N(1440){\frac{1}{2}}^{+}$: The experimental values for $A_{1/2}$ are from Ref.\cite{Burk}.
The mode $p_{1/2}$ is taken from PDG. The sign
of the mode which is given by PDG in the absolute value is taken to be
positive as predicted by Koniuk. The value of the mode $p_{3/2}$ is taken
from K. We set a $50\%$ error for the fit. The coordinate quark wave function of the
resonance is known to be symmetric, so the neutron charge
radius does vanish. It is proportional to the second derivative of the charge density
with respect to the momentum $k$ at $k=0$ or, equivalently, to the first derivative 
of the helicity amplitude $S_{1/2}$ at
pseudothreshold $M^{2}=(m_{*}-m)^{2}$, so we set $%
2m_{+}F_{1}^{(-)}+m_{-}F_{2}^{(-)}=0$ at $M^{2}=(m_{*}-m)^{2}$.


$N(1710){\frac{1}{2}}^{+}$: The $N\rho $ mode $p_{1/2}$ is taken from PDG.
The sign of the mode is taken to be positive, as predicted by MS. The value of the 
mode $p_{3/2}$ from LD results to $\sqrt{\Gamma
_{N\rho }}=8.9$ MeV$^{1/2}$. This value is too high as compared to the
square root of the PDG total $N\rho $ width of $4.3\pm 1.9$ MeV$^{1/2}$. We
thus take for the mode $p_{3/2}$ the three times smaller Koniuk quark model
value of $\sqrt{B_{N\pi }B_{N\rho }}=0.09$ with a $50\%$ error. The $N\omega 
$ modes from K are included to the fit with $0.05$ MeV$^{1/2}$ errors.


$N(1720){\frac{3}{2}}^{+}$: The $N\rho $ mode $p_{1/2}$ from MS and the mode 
$p_{3/2}$ from LD seems to be overestimated, in view of the PDG value $%
\sqrt{\Gamma _{N\rho }^{tot}}=11\pm 2$ MeV$^{1/2}$ for the total $N\rho$ width. The
modes $p_{1/2}$, $p_{3/2}$ and $f_{3/2}$ are taken from K with $50\%$
errors. The $N\omega $ modes from K are included to the fit with $0.05$ MeV$%
^{1/2}$ errors.


$N(1900){\frac{3}{2}}^{+}$: The $N\rho $ mode $p_{1/2}$ is from MS. The
modes $p_{3/2}$ and $f_{3/2}$ are taken from K with $0.05$ MeV$%
^{1/2}$ errors. The $N\omega $ modes are from K with $0.05$ MeV$^{1/2}$
errors.


$N(1680){\frac{5}{2}}^{+}$: The experimental values for $A_{1/2}$ and $A_{3/2}$ are 
from Ref.\cite{Burk}. The $N\rho $ mode $f_{1/2}$ is from K with a $%
50\%$ error. The modes $f_{3/2}$ and $p_{3/2}$ are from PDG. The $N\omega $
modes are from K with $0.05$ MeV$^{1/2}$ errors.


$N(2000){\frac{5}{2}}^{+}$: The $N\rho $ mode $f_{1/2}$ is from K with a $%
50\%$ error. The modes $f_{3/2}$ and $p_{3/2}$ are from MS. The $N\omega $
modes are from K with $0.05$ MeV$^{1/2}$ errors.


$N(1990){\frac{7}{2}}^{+}$: The $N\rho $ modes $f_{1/2}$, $f_{3/2}$, and $%
h_{3/2}$ are from K with $0.05$ MeV$^{1/2}$ errors. The $N\omega $ modes are
from K with $0.05$ MeV$^{1/2}$ errors.


$\Delta (1620){\frac{1}{2}}^{-}$: PDG values are used.


$\Delta (1900){\frac{1}{2}}^{-}$: MS values are used.


$\Delta (1700){\frac{3}{2}}^{-}$: K values are used for the modes $d_{1/2}$
and $d_{3/2}$ with $0.05$ MeV$^{1/2}$ errors. The PDG absolute value is used
for the $s_{3/2}$ mode with MS sign.


$\Delta (1940){\frac{3}{2}}^{-}$: CR values are used for the modes $d_{1/2}$
and $d_{3/2}$. MS value is used for the $s_{3/2}$ mode.


$\Delta (1930){\frac{5}{2}}^{-}$: CR values are used for the modes $d_{3/2}$
and $g_{3/2}$. MS value is used for the $d_{1/2}$ mode.


$\Delta (1750){\frac{1}{2}}^{+}$: K values are used with $50\%$ errors.


$\Delta (1910){\frac{1}{2}}^{+}$: K values are used with $50\%$ errors.


$\Delta (1232){\frac{3}{2}}^{+}$: 
The data in the space-like region on the magnetic transition form factor are from
Refs.\cite{Bartel:1968tw,Stein:1975yy,Batzner}.  We include into the
fit the experimental results of Refs.\cite{Burkert,Siddle:1971ug,Beck:1997ew,Frolov:1999pw,AS,ALDER} 
for the ratio $G_C/G_M$ and of Refs.
\cite{Frolov:1999pw,AS,Siddle:1971ug,ALDER}
for the ratio $G_E/G_M$. The results of Refs.\cite{Frolov:1999pw} and \cite{AS} for 
the ratio $G_E/G_M$ disagree already
in sign. We fit well the data  \cite{Frolov:1999pw}.
The amplitudes $A_{3/2}$ and $A_{1/2}$ at $M=0$ are
given by PDG. The $\Delta(1232)$ magnetic
form factor in the space-like region is a smooth, well measured function. It is well reproduced 
in the VMD model. Using the knowledge of the $G_M$, we translate on Fig. 19, when data are sufficient, 
the experimental points
from plots for the monopole form factors to plots for the helicity amplitudes, and {\it v. v.} It is seen that
the experimental data for the  form factors $G_E$ and $G_C$ are not stable yet. 
The vector meson decay channels are closed. 


$\Delta (1600){\frac{3}{2}}^{+}$: LD values give too high
coupling constants $\Delta N\rho $. We use CR predictions which are 5 to 10
times smaller.


$\Delta (1920){\frac{3}{2}}^{+}$: K values are used with $50\%$ errors.


$\Delta (1905){\frac{5}{2}}^{+}$: CR values are used for the $f$-modes and
PDG for the $p_{3/2}$-mode.


$\Delta (2000){\frac{5}{2}}^{+}$: K values for the $f$ modes are used with 
$50\%$ errors. MS value is used for the $p_{3/2}$ mode.


$\Delta (1950){\frac{7}{2}}^{+}$: PDG gives an upper limit of $6$ MeV$^{1/2}$
for the total $N\rho $ width. MS and K results are above this limit, so we
use the estimates of CR.


The $S_{1/2}$ amplitudes which we used as an input 
are from Refs. \cite{foster,Gerhardt:1980yg}.

In Table I, we show the parameters $C_{kj}^{(\pm )}$ of the extended VMD
model for nucleon resonances with masses below $2$ GeV. 

In Table II, the
coupling constants $g_{T}^{V}\equiv g_{T}^{V}(m_{V})$ of the vector mesons
of the magnetic, electric, and Coulomb types with the nucleon resonances are
shown. The coupling constants $g_{T}^{V}\ $are defined as in Eqs.(\ref{RESD}%
) and (\ref{RESN}) with the selfevident replacements $f_{VNN^{*},k}^{(\pm
)}\rightarrow g_{T}^{V}$ and $F_{k}^{(\pm )}\rightarrow G_{T}^{(\pm )}.$ The
coupling constants $f_{VNN^{*},k}^{(\pm )}$ and $g_{T}^{V}$ are related by
the same transformation as the covariant form factors $F_{k}^{(\pm )}$ and $%
G_{T}^{(\pm )}:$%
\begin{equation}
g_{T}^{V}=\sum_{k}{\rm M}_{Tk}(m_{V}^{2})f_{VNN^{*},k}^{(\pm )}.
\end{equation}
The matrices ${\rm M}_{Tk}(M^{2})$ are defined by Eqs.(%
\ref{FG_lP}), (\ref{FG_lM}), and (\ref{FG_lM}), respectively, for the
normal- and abnormal-parity $l>0$ resonances, and normal- and
abnormal-parity $l=0$ resonances.

The quality of the fit is generally good. It can be controlled with the help
of Tables III - VI and Figs. 1 - 25. 
The minimal $n=0$ extension of the VMD model appears to
be sufficient to fit the data. The exception is only the $N^{*}(1520)$%
-resonance. It has a sharp $M^{2}$-dependence of the helicity amplitude $%
A_{3/2}^{p}(M^{2})$ at the interval $-4<M^{2}<0$ GeV$^{2}$ (see Fig. 3). The
minimal $n=0$ model cannot reproduce it.

In Tables III - VI, we compare the VMD model results for the vector meson decay
amplitudes of the nucleon resonances with the results of the $\pi N$
multichannel partial-wave analysis \cite
{Groom:2000in,Manley:1992yb,Longacre:1977ja} and the quark models \cite
{Koniuk:1982ej,Capstick:1994kb,LeYaouanc:1975mr,Stassart:1990zt,Stancu:1993xz,Bijker:1997tr}%
. The extended VMD model results should not normally be treated as predictions, since
for every nucleon resonance the number of the parameters in the fit is
comparable with (being always less than or equal to) the number of points
available from the experiment and the quark models. The evident exceptions
are the $\Delta (1232)$-resonance, where many data points exist in the
space-like region, and a few others. It is seen from the Tables III - VI that the
minimal extension $n=0$ of the VMD model describes the vector meson
amplitudes in the different partial waves fairly well, with respect to
both, the signs and the values. 

The vector meson decay widths of the nucleon resonances are calculated using
Eqs.(\ref{GAMMA_l}) and (\ref{GAMMA_0}) after substituting $\alpha
\rightarrow 1/(4\pi )$ and $G_{T}(M^{2})\rightarrow g_{T}^{V}(M^{2}).$ The
''running'' coupling constant $g_{T}^{V}(M^{2})$ is given by 
\begin{equation}
g_{T}^{V}(M^{2})=\sum_{kT^{\prime }}{\rm M}_{Tk}(M^{2}){\rm M}_{kT^{\prime
}}^{-1}(m_{V}^{2})g_{T^{\prime }}^{V}(m_{V}^{2})  \label{RUN}
\end{equation}
The values ${\rm M}_{kT^{\prime }}^{-1}(m_{V}^{2})$ are the inverse transformation 
matrices
given in Sect. 3. The parameters $C_{kj}^{(\pm )}$ and the coupling
constants $f_{VNN^{*}}^{(\pm )}$ of Eqs.(\ref{RESD}) and (\ref{RESN}) are
considered as being independent of $M^{2}$, whereas in the multipole basis
one should take into account the $M^{2}$-dependence of the transformation
matrices ${\rm M}_{Tk}(M^{2})$.

The signs of the vector meson decay amplitudes are defined as follows: We
determine first the average vector meson mass 
\begin{equation}
\bar{M}_{(\pm )}^{2}=\int \varphi ^{(\pm )}(M^{2})M^{2}dW(M^{2})/\int
\varphi ^{(\pm )}(M^{2})dW(M^{2})  \label{MAV}
\end{equation}
where $\varphi ^{(\pm )}(M^{2})$ are the kinematical parts
of the widths (\ref{GAMMA_l}) and (\ref{GAMMA_0}): 
\begin{eqnarray*}
\varphi ^{(\pm )}(M^{2}) &=&(m_{\mp }^{2}-M^{2})^{l+1/2}(m_{\pm
}^{2}-M^{2})^{l-1/2}, \\
\varphi ^{(\pm )}(M^{2}) &=&(m_{\pm }^{2}-M^{2})^{3/2}(m_{\mp
}^{2}-M^{2})^{1/2},
\end{eqnarray*}
respectively, for $l>0$ and $l=0$ resonances. The integral runs over the
Breit-Wigner distribution $dW(M^{2}).$ The sign of the vector meson decay
amplitude is given by sign of the quantity 
\begin{equation}
H_{S,\ell }^{(\pm )}=sign(\xi \xi _{\rho }^{*}\lambda _{l}^{(\pm )}){\frak H}%
_{S,\ell }^{(\pm )}  \label{VERY IMPORTANT}
\end{equation}
evaluated at $M^{2}=\bar{M}_{(\pm )}^{2}.$ The factor of $\xi _{\rho }$
brings the overall sign to the MS sign conventions \cite{Manley:1992yb}. The
phase of $\xi $ is absorbed by both the photo- and electroproduction
amplitudes and the vector meson decay amplitudes of the nucleon resonances.
The products $\xi {\frak F}_{\frac{3}{2}}^{(\pm )},$ $\xi {\frak F}_{\frac{1%
}{2}}^{(\pm )}, $ $\xi {\frak C}_{\frac{1}{2}}^{(\pm )},$ and $\xi {\frak H}%
_{S,\ell }^{(\pm )}$ are real for all partial waves \cite{Koniuk:1980vy}, so
the appearance of the complex conjugate value of $\xi $ in the vector meson
decay amplitudes does not change the relative sign, as compared with what we
have described.

For some of the resonances, as can be seen from the Figs.1-25, the amplitude
changes its sign on the interval $0<M^{2}<1$ GeV.

The plots 1-25 show the magnetic, electric, and Coulomb form factors for the
nucleon resonances with masses below $2$ GeV, listed by PDG. We give in the
space-like region the ratios between the form factors and the dipole
function 
\begin{equation}
G_{D}(t)=\frac{1}{(1-t/0.71)^{2}}  \label{DIP}
\end{equation}
where $t$ is in GeV$^{2}$ $(t=q^{2})$. The ratios between the helicity amplitudes
and the dipole function (\ref{DIP}) are also shown, also in the space-like
region. Next, we give the partial-wave amplitudes $H_{S,\ell }^{(\pm
)}$ of the vector meson emission in the time-like region. The solid curves
stand for the extended VMD model. The data for the $\pi N$ multichannel
partial-wave analysis and the quark models used in the fit at $M^{2}>0$ and
the experimental data for the transition form factors and the photo- and
electroproduction helicity amplitudes used in the fit at $M^{2}<0$ are shown
also.

In Table VII, we show the dilepton widths of the nucleon resonances.
Figs.26 and 27 show the dilepton $e^{+}e^{-}$ and $\mu ^{+}\mu ^{-}$ spectra
from decays of the nucleon resonances.

\section{Conclusion}
\setcounter{equation}{0}
$\;$
\vspace{-0.5cm}

In this work, we derived phenomenological kinematically complete relativistic 
expressions for the decay rates of nucleon resonances with arbitrary spin and 
parity into the dilepton pairs in terms of the magnetic, electric, and Coulomb 
transition form factors. The extended VMD model was used for the description of the 
transition form factors of the nucleon resonances. The quark counting rules
were taken into account to reduce the number of free parameters of the model. 
The remaining free parameters are fixed by fitting the photo- and electroproduction 
amplitudes and the decay amplitudes of the nucleon resonances into the vector mesons. 
The extended VMD model allows to treat the data both in the space- and time-like regions.
The transition form factors determined from the fit are used for the calculation 
of the dilepton widths and dilepton spectra from decays of the nucleon 
resonances with masses below $2$ GeV. 

In many cases, the experimental data we used to fix the form factors are
not stable yet. To make the study complete, we used also the quark model 
predictions. The results of our analysis can be treated as a first 
quantitative hint to the form factors and the decay amplitudes
which they determine.
The VMD predictions are useful for planning new experiments for measurements of 
the transition form factors and vector meson branchings of the nucleon resonances.

We propose thus unified description of the photo- and electroproduction data, 
the vector meson and dilepton decay amplitudes of the nucleon resonances. The 
results can be used for modeling the dilepton production in the pion-nucleon, 
nucleon-nucleon, and heavy-ion collisions. 


\vspace{1.5 cm}
\begin{center}
{\large {\bf Acknowledgments}}
\end{center}

The authors are grateful to D. M. Manley for valuable comments on the 
$\pi N$ partial-wave analysis. Two of us
(M.I.K. and B.V.M.) are indebted to the Institute for Theoretical Physics of
University of Tuebingen for kind hospitality. The work was supported by GSI
(Darmstadt) under the contract T\"{U}F\"{A}ST, by the Plesler Foundation,
and by the Deutsche Forschungsgemeinschaft under the contract 
No.~436RUS113/367/0(R).


\newpage {}

Figures captions:

Figs. 1- 25: Electromagnetic transition form factors of the magnetic, electric, 
and Coulomb types ($G_M$, $G_E$, and $G_C$), helicity amplitudes ($A_{3/2}$, 
$A_{1/2}$, and $S_{1/2}$), and partial-wave amplitudes of nucleon resonance 
decays into the $\rho $- and $\omega$-meson channels ($H_{S,\ell }^{(\pm
)} $) for nucleon
resonances listed by PDG with masses below $2$ GeV. The minimal extension 
$n=0$ of the VMD model is used for all resonances. The value $G_{D}(t)$ is the 
dipole function of Eq.(\ref{DIP}). The photo- and
electroproduction experimental data \cite
{Groom:2000in,Stein:1975yy,Bartel:1968tw,Batzner,Frolov:1999pw} are displayed.
The vector meson decay amplitudes of the nucleon resonances \cite
{Groom:2000in,Manley:1992yb,Longacre:1977ja,Koniuk:1982ej,Capstick:1994kb},
that were included into the fit, are also shown.

\newpage {}

\newpage

\begin{table}                                                                                                                                                                                                                                                       
\begin{center}                                                                                                                                                                                                                                                      
\begin{tabular}{rrrrrr}                                                                                                                                                                                                                                             
\hline                                                                                                                                                                                                                                                              
 Resonance  & $C_{10}$ & $C_{11}$ & $C_{20}$ & $C_{30}$\\                                                                                                                                                                                                           
\hline 
$N^{*}       (1535)\frac{1}{2}^-$&       0.979&            &       0.006&            \\
                                 &       1.787&            &      -0.062&            \\
$N^{*}       (1650)\frac{1}{2}^-$&       0.232&            &      -0.186&            \\
                                 &      -0.394&            &       0.157&            \\
$N^{*}       (1520)\frac{3}{2}^-$&       2.186&      -1.236&      -1.976&      -0.159\\
                                 &      -0.220&       1.899&      -0.316&      -0.249\\
$N^{*}       (1700)\frac{3}{2}^-$&       0.169&       0.410&      -0.227&      -0.454\\
                                 &       0.104&       1.666&      -0.138&      -2.281\\
$N^{*}       (1675)\frac{5}{2}^-$&      -0.205&       0.080&       0.371&      -3.812\\
                                 &       0.579&      -0.560&      -0.591&       3.357\\
$N^{*}       (1440)\frac{1}{2}^+$&       0.863&            &       1.023&            \\
                                 &       0.084&            &      -0.699&            \\
$N^{*}       (1710)\frac{1}{2}^+$&       0.287&            &      -0.185&            \\
                                 &      -0.382&            &       0.056&            \\
$N^{*}       (1720)\frac{3}{2}^+$&       0.000&       0.608&       0.187&      -5.312\\
                                 &       0.051&      -0.304&       0.194&       1.630\\
$N^{*}       (1900)\frac{3}{2}^+$&       0.024&      -0.238&       0.028&       0.991\\
                                 &      -0.054&       0.398&       0.055&      -1.777\\
$N^{*}       (1680)\frac{5}{2}^+$&       2.487&      -0.700&      -2.116&      -0.797\\
                                 &      -0.793&       4.929&       0.735&      -6.297\\
$N^{*}       (2000)\frac{5}{2}^+$&       0.201&       0.762&      -0.267&      -1.663\\
                                 &       0.049&      -0.176&      -0.029&       0.956\\
$N^{*}       (1990)\frac{7}{2}^+$&      -0.199&       0.336&       0.321&      -0.578\\
                                 &       1.028&      -2.263&      -0.796&       0.846\\
\hline                                                                                                                                                                                                                                                              
$\Delta      (1620)\frac{1}{2}^-$&      -0.155&            &      -0.081&            \\
$\Delta      (1900)\frac{1}{2}^-$&       0.123&            &      -0.025&            \\
$\Delta      (1700)\frac{3}{2}^-$&      -0.630&      -0.298&       1.080&      -0.473\\
$\Delta      (1940)\frac{3}{2}^-$&      -0.251&       0.075&       0.252&      -0.202\\
$\Delta      (1930)\frac{5}{2}^-$&       0.181&       0.845&      -0.555&      -4.531\\
$\Delta      (1750)\frac{1}{2}^+$&       0.325&            &       0.431&            \\
$\Delta      (1910)\frac{1}{2}^+$&       0.194&            &      -0.045&            \\
$\Delta      (1232)\frac{3}{2}^+$&       1.768&       0.025&      -1.096&      -0.926\\
$\Delta      (1600)\frac{3}{2}^+$&       0.086&      -0.012&      -0.238&       1.806\\
$\Delta      (1920)\frac{3}{2}^+$&      -0.120&       0.187&       0.276&      -1.386\\
$\Delta      (1905)\frac{5}{2}^+$&      -0.209&       0.090&       0.157&      -1.145\\
$\Delta      (2000)\frac{5}{2}^+$&      -0.088&      -0.388&      -0.020&       1.299\\
$\Delta      (1950)\frac{7}{2}^+$&       0.867&      -1.250&      -0.138&       1.619\\
\hline                                                                                                                                                                                                                                                              
\end{tabular}                                                                                                                                                                                                                                                       
\end{center}                                                                                                                                                                                                                                                        
\caption{Residues $C_{jk}^{(\pm )}$ of the extended VMD\ model, entering 
Eqs.(\ref
{FF_l}) and (\ref{FF_0}), in units GeV$^{-(l+1)}$ where $l=J-\frac{1}{2}.$
The $N^{*}$ residues are shown in two lines for the proton and neutron
resonances, respectively. 
}                                                                                                                                                                                                                                        
\label{table1}                                                                                                                                                                                                                                                      
\end{table}       

\newpage

\begin{table}                                                                                                                                                                                           
\begin{center}                                                                                                                                                                                          
\begin{tabular}{rrrrrrrr}                                                                                                                                                                               
\hline                                                                                                                                                                                                  
 Resonance  & $g^{\rho}_{M}$ & $g^{\rho}_{E}$ & $g^{\rho}_{C}$ & $g^{\omega}_{M}$ & $g^{\omega}_{E}$ & $g^{\omega}_{C}$ \\                                                                              
\hline
$N^{*}       (1535)\frac{1}{2}^-$&            &        2.21&        3.16&            &      -28.03&      -42.67\\
$N^{*}       (1650)\frac{1}{2}^-$&            &       -0.26&       -0.01&            &        2.01&        4.14\\
$N^{*}       (1520)\frac{3}{2}^-$&       -0.53&       -9.21&      -24.62&       -7.67&       18.16&       46.13\\
$N^{*}       (1700)\frac{3}{2}^-$&        0.02&        1.31&        2.56&       -0.17&       -1.45&       -1.45\\
$N^{*}       (1675)\frac{5}{2}^-$&       -9.79&       -4.91&      -31.08&        2.07&       -1.61&      -10.50\\
$N^{*}       (1440)\frac{1}{2}^+$&       -8.21&            &       18.16&      -14.14&            &       63.13\\
$N^{*}       (1710)\frac{1}{2}^+$&       -0.69&            &       13.35&        2.97&            &       -8.13\\
$N^{*}       (1720)\frac{3}{2}^+$&       -0.49&       -5.72&      -25.91&        0.14&       -8.27&      -37.73\\
$N^{*}       (1900)\frac{3}{2}^+$&       -1.75&        1.71&       10.66&        1.25&       -1.28&       -8.85\\
$N^{*}       (1680)\frac{5}{2}^+$&        0.00&        5.77&        8.01&       -1.34&      -11.75&       -7.98\\
$N^{*}       (2000)\frac{5}{2}^+$&        0.60&       -5.19&      -26.29&        1.72&        8.98&        5.18\\
$N^{*}       (1990)\frac{7}{2}^+$&        6.13&        1.39&        7.73&      -19.88&       -3.73&      -28.51\\
\hline                                                                                                                                                                                                  
$\Delta      (1620)\frac{1}{2}^-$&            &        1.59&        3.32&            &            &            \\
$\Delta      (1900)\frac{1}{2}^-$&            &       -0.32&       -1.31&            &            &            \\
$\Delta      (1700)\frac{3}{2}^-$&        0.05&       -5.53&      -12.08&            &            &            \\
$\Delta      (1940)\frac{3}{2}^-$&       -0.37&       -2.38&       -6.98&            &            &            \\
$\Delta      (1930)\frac{5}{2}^-$&       12.87&       -5.38&      -48.07&            &            &            \\
$\Delta      (1750)\frac{1}{2}^+$&       -6.02&            &       16.69&            &            &            \\
$\Delta      (1910)\frac{1}{2}^+$&       -0.44&            &        8.44&            &            &            \\
$\Delta      (1232)\frac{3}{2}^+$&       30.57&        0.80&        6.56&            &            &            \\
$\Delta      (1600)\frac{3}{2}^+$&        4.92&        3.31&       12.31&            &            &            \\
$\Delta      (1920)\frac{3}{2}^+$&       -1.61&       -1.38&       -9.88&            &            &            \\
$\Delta      (1905)\frac{5}{2}^+$&       -0.19&      -15.25&      -42.58&            &            &            \\
$\Delta      (2000)\frac{5}{2}^+$&       -0.54&        2.30&       20.56&            &            &            \\
$\Delta      (1950)\frac{7}{2}^+$&        5.82&        1.92&       22.90&            &            &            \\
\hline                                                                                                                                                                                                  
\end{tabular}                                                                                                                                                                                           
\end{center}                                                                                                                                                                                            
\caption{The $\rho $- and $\omega $-meson coupling constants of the magnetic,
electric, and Coulomb types with the nucleon resonances in units GeV$%
^{-(l-1)}$ where $l=J-\frac{1}{2}.$}                                                                                                                                                                            
\label{table2}                                                                                                                                                                                          
\end{table}                                                                                                                                                                                             

\newpage

{\tiny                                                                                                                                                                                                                                                              
\begin{table}                                                                                                                                                                                                                                                       
\begin{center}                                                                                                                                                                                                                                                      
\begin{tabular}{|r|l|ccc|c|ccc|c|}                                                                                                                                                                                                                                  
\hline                                                                                                                                                                                                                                                              
Resonance  & Ref.  & $N\rho$   & $N\rho  $ & $N\rho  $ & $\sqrt{\Gamma^{tot}_{N\rho}}  $                                                                                                                                                                            
                   & $N\omega$ & $N\omega$ & $N\omega$ & $\sqrt{\Gamma^{tot}_{N\omega}}$ \\                                                                                                                                                                         
\hline                                                                                                                                                                                                                                                              
\hline                                                                                                                                                                                                                                                              
                           &    & $s_{1/2}$ & $d_{3/2}$ & & & $s_{1/2}$ & $d_{3/2}$ & \\                                                                                                                                                                            
\hline                                                                                                                                                                                                                                                              
$N^{*}(1535)\frac{1}{2}^-$ &VMD &       -2.13        &       -0.25        &                    &        2.15        &        1.43        &        0.05        &                    &        1.43        \\
                           & KI & -1.7               &      -6.1          &                    & 6.3                &                    &                    &                    &              \\                                                                
                           & CR & -0.7$\pm0.3$       & 0.4$\pm0.1$        &                    & 0.8$^{+0.2}_{-0.1}$&                    &                    &                    &              \\                                                                
                           & SST&                    &                    &                    & 1.1                &                    &                    &                    &              \\                                                                
                           & MS & -1.7$\pm0.5$       &-1.3$\pm0.6$        &                    & 2.2$\pm0.6$        &                    &                    &                    &              \\                                                                
                           &PDG & -2.0$\pm0.9$       &                    &                    &  $<2.7$            &                    &                    &                    &              \\                                                                
$N^{*}(1650)\frac{1}{2}^-$ &VMD &       -1.45        &        1.04        &                    &        1.78        &       -0.97        &       -0.02        &                    &        0.97        \\
                           & KI &   -9.7             &      2.7           &                    &   10.1             &      -0.96         &       0.67         &                    &      1.2     \\                                                                
                           & CR & 0.9$^{+0.8}_{-0.6}$& 0.4$\pm0.1$        &                    & 1.0$^{+0.3}_{-0.2}$&                    &                    &              \\                                                                                     
                           & SST&                    &                    &                    &     0.6            &                    &                    &                    &              \\                                                                
                           & MS &   0.0$\pm1.6$      &      2.2$\pm0.9$   &                    & 2.2$\pm0.9$        &                    &                    &                    &              \\                                                                
                           &PDG &   $\pm1.6\pm1.2$   &      3.4$\pm1.0$   &                    & 3.6$\pm0.9$        &                    &                    &                    &              \\                                                                
\hline                                                                                                                                                                                                                                                              
                           &    & $d_{1/2}$ & $d_{3/2}$ & $s_{3/2}$ & & $d_{1/2}$ & $d_{3/2}$ & $s_{3/2}$  \\                                                                                                                                                       
\hline                                                                                                                                                                                                                                                              
$N^{*}(1520)\frac{3}{2}^-$ &VMD &       -0.37        &       -0.17        &       -5.14        &        5.16        &       -0.02        &        0.03        &        0.28        &        0.29        \\
                           & KI &      0.7           &      -1.1          &        -5.0        &       5.2          &                    &                    &                    &              \\                                                                
                           & CR &-0.1$^{+0.1}_{-0.3}$&-0.3$^{+0.2}_{-1.0}$&-2.4$^{+1.9}_{-6.4}$& 2.5$^{+6.5}_{-1.9}$&                    &                    &                    &              \\                                                                
                           & SST&     3.2            &      -2.7          &      -1.7          &       4.6          &                    &                    &                    &              \\                                                                
                           & MS &      0             &       0            &       -5.1$\pm0.6$ &  5.1$\pm0.6$       &                    &                    &                    &              \\                                                                
                           &PDG &                    &                    &       -4.9$\pm0.6$ &  4.9$\pm0.6$       &                    &                    &                    &              \\                                                                
$N^{*}(1700)\frac{3}{2}^-$ &VMD &       -0.94        &       -2.32        &       -2.22        &        3.35        &        0.21        &        0.90        &        1.39        &        1.67        \\
                           & KI &      -0.1          &     -2.7           &       -4.3         &      5.1           &      0.26          &       0.89         &        1.4         &    1.7       \\                                                                
                           & CR &       0            &-0.9$^{+0.3}_{-0.6}$&       0.0$\pm0.1$  & 0.9$^{+0.6}_{-0.4}$& 0.0$^{+0.0}_{-0.3}$& 0.0$^{+0.3}_{-0.0}$&0.0$^{+0.0}_{-16.2}$& $<16.2$      \\                                                                
                           & SST&                    &                    &                    &      3.7           &                    &                    &                    &              \\                                                                
                           & MS &       0            &       0            &  -5.6$\pm5.7$      &     5.6$\pm5.7$    &                    &                    &                    &              \\                                                                
                           &PDG &                    &                    &  $\pm2.2\pm2.1$    &      $<7.4$        &                    &                    &                    &              \\                                                                
\hline                                                                                                                                                                                                                                                              
                           &    & $d_{1/2}$ & $d_{3/2}$ & $g_{3/2}$ & & $d_{1/2}$ & $d_{3/2}$ & $g_{3/2}$  \\                                                                                                                                                       
\hline                                                                                                                                                                                                                                                              
$N^{*}(1675)\frac{5}{2}^-$ &VMD &        0.75        &       -1.70        &        0.22        &        1.87        &        0.06        &        0.00        &        0.00        &        0.06        \\
                           & KI &      -1.1          &        -0.2        &      0             &      2.3           &                    &                    &         0          &              \\                                                                
                           & CR &       0.2          &        -0.4        &      0             &      0.5           &                    &                    &                    &              \\                                                                
                           & SST&                    &                    &                    &      2.0           &                    &                    &                    &              \\                                                                
                           & MS &      0.8$\pm0.4$   &   -0.5$\pm0.5$     &      0             &     1.0$\pm0.4$    &                    &                    &                    &              \\                                                                
                           &PDG &      0.8$\pm0.4$   &   -1.7$\pm0.6$     &                    &      $<2.2$        &                    &                    &                    &              \\                                                                
\hline                                                                                                                                                                                                                                                              
\end{tabular}                                                                                                                                                                                                                                                       
\end{center}
\caption{
Predictions of the extended VMD model for the partial widths of the nucleon resonance decays into the 
$\rho $- and $\omega $-meson channels, inclusive of the sign of the amplitudes. The data quoted
by PDG [24], the results of the multichannel $\pi N$ partial-wave analysis MS [25] and LD [26], and 
predictions of the non-relativistic quark model K [27] and the quark-pair creation models CR [28] and SST 
[53,54] are given for comparison. The widths are in MeV. This table shows the $\rho$- and $\omega$-meson 
modes of the negative-parity $N^*$-resonances.
}                                                                                                                                                                                                                                        
\label{table3}                                                                                                                                                                                                                                                      
\end{table}
                                                                                                                                                                                                                                                        
\begin{table}                                                                                                                                                                                                                                                       
\begin{center}                                                                                                                                                                                                                                                      
\begin{tabular}{|r|l|ccc|c|ccc|c|}                                                                                                                                                                                                                                  
\hline                                                                                                                                                                                                                                                              
Resonance  & Ref.  & $N\rho$   & $N\rho  $ & $N\rho  $ & $\sqrt{\Gamma^{tot}_{N\rho}}  $                                                                                                                                                                            
                   & $N\omega$ & $N\omega$ & $N\omega$ & $\sqrt{\Gamma^{tot}_{N\omega}}$ \\                                                                                                                                                                         
\hline                                                                                                                                                                                                                                                              
\hline                                                                                                                                                                                                                                                              
                           &    & $p_{1/2}$ & $p_{3/2}$ & & &$p_{1/2}$ & $p_{3/2}$ & & \\                                                                                                                                                                           
\hline
$N^{*}(1440)\frac{1}{2}^+$ &VMD &       -0.29        &        0.61        &                    &        0.67        &        0.00        &        0.00        &                    &        0.00        \\
                           & KI &      0.3           &      0.1           &                    &     0.3            &                    &                    &                    &              \\                                                                
                           & CR &-0.3$^{+0.2}_{-0.3}$&-0.5$^{+0.3}_{-0.5}$&                    & 0.6$^{+0.5}_{-0.3}$&                    &                    &                    &              \\                                                                
                           & SST&                    &                    &                    &     1.5            &                    &                    &                    &              \\                                                                
                           &PDG &   $\pm3.7\pm2.2$   &                    &                    &     $<6$           &                    &                    &                    &              \\                                                                
$N^{*}(1710)\frac{1}{2}^+$ &VMD &        2.22        &        3.30        &                    &        3.97        &        0.18        &       -0.72        &                    &        0.74        \\
                           & KI &      5.5           &        2.5         &                    &       6.0          &      0.6           &      -0.7          &                    &   0.9        \\                                                                
                           & CR & 0.3$\pm0.1$        &-3.7$^{+0.9}_{-1.2}$&                    & 3.7$^{+1.2}_{-1.0}$& 0.0$^{+0.0}_{-2.3}$& 0.0$^{+0.0}_{-0.4}$&                    & $<2.3$       \\                                                                
                           & SST&                    &                    &                    &       4.1          &     0.03           &      -0.2          &                    &   0.2        \\                                                                
                           & MS & 3.9$\pm4.4$        &       0            &                    &      $<8.3$        &                    &                    &                    &              \\                                                                
                           &PDG & $\pm4.0\pm2.0$     &                    &                    &   4.3$\pm1.9$      &                    &                    &                    &              \\                                                                
\hline                                                                                                                                                                                                                                                              
                           &    & $p_{1/2}$ & $p_{3/2}$ & $f_{3/2}$ & & $p_{1/2}$ & $p_{3/2}$ & $f_{3/2}$  \\                                                                                                                                                       
\hline                                                                                                                                                                                                                                                              
$N^{*}(1720)\frac{3}{2}^+$ &VMD &       11.03        &       -2.56        &        1.02        &       11.37        &        5.29        &       -2.09        &        0.14        &        5.69        \\
                           & KI &      11.7          &     -2.6           &       -3.5         &       12.5         &       5.3          &        -2.1        &      -0.61         &    5.7       \\                                                                
                           & CR &-2.6$^{+0.7}_{-0.8}$& 1.8$^{+0.6}_{-0.5}$& 0.7$^{+0.3}_{-0.2}$& 3.3$^{+1.0}_{-0.8}$& 0.0$^{+0.0}_{-0.2}$& 0.0$^{+1.2}_{-0.0}$& 0.0$^{+0.1}_{-0.0}$&   $<1.3$     \\                                                                
                           & SST&                    &                    &                    &        5.2         &      0.1           &         0.2        &       0.1          &    0.2       \\                                                                
                           & OR &      -5.7          &    -2.5            &       -1.9         &       6.5          &                    &                    &                    &              \\                                                                
                           & MS &     18$\pm5$       &      0             &         0          &      18$\pm5$      &                    &                    &                    &              \\                                                                
                           &PDG &                    &                    &                    &      11$\pm2$      &                    &                    &                    &              \\                                                                
$N^{*}(1900)\frac{3}{2}^+$ &VMD &      -14.82        &       -1.28        &       -2.08        &       15.02        &        7.97        &       -0.49        &        1.25        &        8.09        \\
                           & KI &      -0.4          &      -1.3          &        -0.5        &        1.5         &       9.7          &         -0.4       &       -2.0         &    9.9       \\                                                                
                           & CR &-1.4$^{+0.9}_{-1.0}$&-1.0$\pm0.6$        & 0.2$^{+0.5}_{-0.2}$& 1.8$^{+1.2}_{-1.1}$&                    & 4.4$^{+1.2}_{-4.4}$& 0.6$^{+1.2}_{-0.6}$&  $<5.9$      \\                                                                
                           & SST&                    &                    &                    &        6.1         &       11.5         &         -8.2       &        6.2         &   15.4       \\                                                                
                           & MS &  -14.7$\pm2.9$     &         0          &         0          &     14.7$\pm2.9$   &       12.3$\pm1.8$ &          0         &         0          & 12.3$\pm1.8$ \\                                                                
                           &PDG &                    &                    &                    &                    &                    &                    &                    &              \\                                                                
\hline                                                                                                                                                                                                                                                              
                           &    & $f_{1/2}$ & $f_{3/2}$ & $p_{3/2}$ & & $f_{1/2}$ & $f_{3/2}$ & $p_{3/2}$  \\                                                                                                                                                       
\hline                                                                                                                                                                                                                                                              
$N^{*}(1680)\frac{5}{2}^+$ &VMD &       -1.35        &       -1.23        &       -2.62        &        3.20        &        0.09        &        0.40        &        0.58        &        0.71        \\
                           & KI &      1.6           &        -1.3        &       -4.0         &       4.5          &       0.13         &      -0.19         &      -1.2          &     1.2      \\                                                                
                           & CR &   -0.2$\pm0.0$     &-0.3$\pm0.1$        &-3.0$^{+0.4}_{-0.5}$& 3.0$^{+0.5}_{-0.4}$&                    &                    &                    &              \\                                                                
                           & SST&    3.1             &      2.7           &       -1.3         &       4.3          &                    &                    &                    &              \\                                                                
                           & MS &       0            &  -1.7$\pm0.6$      &     -2.8$\pm0.7$   &     3.3$\pm0.7$    &                    &                    &                    &              \\                                                                
                           &PDG &                    &  -2.0$\pm0.6$      &     -2.8$\pm1.4$   &     3.4$\pm1.1$    &                    &                    &                    &              \\                                                                
$N^{*}(2000)\frac{5}{2}^+$ &VMD &        2.50        &        6.99        &      -16.02        &       17.66        &        0.07        &        8.19        &        9.96        &       12.89        \\
                           & KI &      -1.7          &      -4.4          &      -6.6          &      8.1           &      4.0           &       6.7          &       10.9         &    13.4      \\                                                                
                           & CR &   -0.4$\pm0.3$     &    -0.2$\pm0.1$    &-7.8$^{+3.1}_{-0.2}$& 7.8$^{+0.2}_{-3.1}$&-0.3$^{+0.2}_{-0.3}$&-1.6$^{+1.1}_{-1.5}$& 3.1$^{+0.5}_{-0.5}$& 3.5$^{+1.3}_{-0.8}$\\                                                          
                           & MS &       0            &    8.5$\pm5.8$     &    -17.2$\pm6.2$   &     19.2$\pm6.1$   &                    &                    &                    &              \\                                                                
\hline                                                                                                                                                                                                                                                              
                           &    & $f_{1/2}$ & $f_{3/2}$ & $h_{3/2}$ & & $f_{1/2}$ & $f_{3/2}$ & $h_{3/2}$  \\                                                                                                                                                       
\hline                                                                                                                                                                                                                                                              
$N^{*}(1990)\frac{7}{2}^+$ &    &       -0.96        &        3.95        &        0.97        &        4.18        &        1.31        &       -6.90        &       -0.68        &        7.06        \\
                           & KI &      -0.8          &     4.2            &       0            &      4.3           &         1.3        &        -7.2        &       0            &   7.3        \\                                                                
                           & CR &  0.6$\pm0.3$       &-1.0$^{+0.6}_{-0.5}$&       0            & 1.2$^{+0.6}_{-0.7}$&-0.8$^{+0.4}_{-0.5}$& 1.4$^{+0.9}_{-0.7}$&       0            & 1.6$^{+1.0}_{-0.9}$\\                                                          
                           & SST&                    &                    &                    &       1.1          &         2.3        &        -2.8        &      0.7           &   14         \\                                                                
\hline                                                                                                                                                                                                                                                              
\end{tabular}                                                                                                                                                                                                                                                       
\end{center}                                                                                                                                                                                                                                                        
\caption{The $\rho$- and $\omega$-meson modes of the positive-parity $N^{*}$-resonances. The
notations are the same as in Table III.}                                                                                                                                                                                                                                        
\label{table4}                                                                                                                                                                                                                                                      
\end{table}  
    
                                                                                                                                                                                                                                                   
\begin{table}                                                                                                                                                                                                                                                       
\begin{center}                                                                                                                                                                                                                                                      
\begin{tabular}{|r|l|ccc|c|}                                                                                                                                                                                                                                        
\hline                                                                                                                                                                                                                                                              
Resonance  & Ref.  & $N\rho$   & $N\rho  $ & $N\rho  $ & $\sqrt{\Gamma^{tot}_{N\rho}}$ \\                                                                                                                                                                           
\hline                                                                                                                                                                                                                                                                                                                                                                                                                                                                                                                                 
\hline                                                                                                                                                                                                                                                              
                           &    & $s_{1/2}$ & $d_{3/2}$ &        & \\                                                                                                                                                                                               
\hline  
$\Delta(1620)\frac{1}{2}^-$&VMD &        4.05        &       -0.02        &                    &        4.05        \\
                           & KI &       7.8          &      -1.7          &                    &       8.0          \\                                                                                                                                              
                           & CR &-3.6$^{+1.3}_{-2.5}$&-0.3$^{+0.1}_{-0.2}$&                    & 3.6$^{+2.5}_{-1.3}$\\                                                                                                                                              
                           & SST&       2.5          &      -3.6          &                    &       4.4          \\                                                                                                                                              
                           & MS &   6.2$\pm0.9$      &    -2.4$\pm0.2$    &                    &    6.6$\pm0.8$     \\                                                                                                                                              
                           &PDG &   4.2$\pm1.4$      &    -2.2$\pm1.5$    &                    &    4.9$\pm1.5$     \\                                                                                                                                              
$\Delta(1900)\frac{1}{2}^-$&VMD &       -5.31        &       -1.52        &                    &        5.52        \\
                           & CR & 2.5$\pm0.6$        & 1.5$^{+0.5}_{-0.3}$&                    & 2.9$^{+0.8}_{-0.6}$\\                                                                                                                                              
                           & MS & -3.5$\pm2.7$       &  -9.3$\pm1.7$      &                    &     9.9$\pm1.9$    \\                                                                                                                                              
\hline                                                                                                                                                                                                                                                              
                           &    & $d_{1/2}$ & $d_{3/2}$ & $s_{3/2}$ & \\                                                                                                                                                                                            
\hline                                                                                                                                                                                                                                                              
$\Delta(1700)\frac{3}{2}^-$&VMD &       -1.66        &        0.66        &        6.67        &        6.91        \\
                           & KI &      4.2           &        0.9         &      16.5          &       17.0         \\                                                                                                                                              
                           & CR &-1.2$^{+0.6}_{-1.2}$& 0.5$^{+0.5}_{-0.2}$& 3.4$^{+2.2}_{-1.7}$& 3.6$^{+2.5}_{-1.8}$\\                                                                                                                                              
                           & SST&                    &                    &                    &       4.9          \\                                                                                                                                              
                           & MS &       0            &          0         &  6.8$\pm2.3$       &      6.8$\pm2.3$   \\                                                                                                                                              
                           &PDG &                    &                    & $\pm6.7\pm2.4$     &      11$\pm3$      \\                                                                                                                                              
$\Delta(1940)\frac{3}{2}^-$&VMD &       -2.70        &        1.32        &       12.83        &       13.18        \\
                           & CR &-3.8$^{+2.3}_{-2.5}$& 1.4$^{+0.9}_{-0.8}$& 1.0$\pm0.3$        & 4.2$^{+2.7}_{-2.4}$\\                                                                                                                                              
                           & MS &       0            &         0          & 12.7$\pm5.6$       &    12.7$\pm5.6$    \\                                                                                                                                              
\hline                                                                                                                                                                                                                                                              
                           &    & $d_{1/2}$ & $d_{3/2}$ & $g_{3/2}$ & \\                                                                                                                                                                                            
\hline                                                                                                                                                                                                                                                              
$\Delta(1930)\frac{5}{2}^-$&VMD &      -16.90        &       -2.62        &       -1.65        &       17.18        \\
                           & CR & 0.1$\pm0.0$        &-2.9$^{+0.5}_{-0.8}$&-0.1$^{+0.0}_{-0.1}$& 2.9$^{+0.8}_{-0.5}$\\                                                                                                                                              
                           & MS &   -20.8$\pm2.9$    &         0          &         0          &   20.8$\pm2.9$     \\                                                                                                                                              
\hline                                                                                                                                                                                                                                                              
\end{tabular}                                                                                                                                                                                                                                                       
\end{center}                                                                                                                                                                                                                                                        
\caption{The $\rho$-meson modes of the negative-parity $\Delta$-resonances. The
notations are the same as in Table III.}                                                                                                                                                                                                                                        
\label{table5}                                                                                                                                                                                                                                                      
\end{table}

\begin{table}                                                                                                                                                                                                                                                       
\begin{center}                                                                                                                                                                                                                                                      
\begin{tabular}{|r|l|ccc|c|}                                                                                                                                                                                                                                        
\hline                                                                                                                                                                                                                                                              
Resonance  & Ref.  & $N\rho$   & $N\rho  $ & $N\rho  $ & $\sqrt{\Gamma^{tot}_{N\rho}}  $\\                                                                                                                                                                          
\hline                                                                                                                                                                                                                                                                                                                                                                                                                                                                                                                                  
\hline                                                                                                                                                                                                                                                              
                           &    & $p_{1/2}$ & $p_{3/2}$ &  \\                                                                                                                                                                                                       
\hline
$\Delta(1750)\frac{1}{2}^+$&VMD &        2.20        &       -7.65        &                    &        7.97        \\
                           & KI &       2.2          &      -7.6          &                    &     7.9            \\                                                                                                                                              
                           & CR &-6.5$^{+4.6}_{-4.1}$& 4.7$^{+3.1}_{-3.3}$&                    & 8.0$^{+5.1}_{-5.7}$\\                                                                                                                                              
                           & SST&                    &                    &                    &    17.1            \\                                                                                                                                              
$\Delta(1910)\frac{1}{2}^+$&VMD &       -2.75        &       -5.44        &                    &        6.10        \\
                           & KI &       -3.7         &      -4.9          &                    &     6.1            \\                                                                                                                                              
                           & CR & 5.6$^{+0.9}_{-0.4}$& 2.6$^{+0.4}_{-0.2}$&                    & 6.1$^{+1.0}_{-0.5}$\\                                                                                                                                              
                           & SST&                    &                    &                    &     6.9            \\                                                                                                                                              
                           & MS &                    &                    &                    &    4.9$\pm1.1$     \\                                                                                                                                              
\hline                                                                                                                                                                                                                                                              
                           &    & $p_{1/2}$ & $p_{3/2}$ & $f_{3/2}$ & \\                                                                                                                                                                                            
\hline                                                                                                                                                                                                                                                              
$\Delta(1600)\frac{3}{2}^+$&VMD &        0.56        &       -1.30        &        0.13        &        1.42        \\
                           & KI &      -1.3          &      -5.5          &      -0.4          &        5.7         \\                                                                                                                                              
                           & CR & 0.4$^{+0.7}_{-0.3}$&-0.9$^{+0.6}_{-1.4}$&        0           & 1.0$^{+1.6}_{-0.6}$\\                                                                                                                                              
                           & SST&                    &                    &                    &        2.9         \\                                                                                                                                              
                           & L  &       4.5          &       4.5          &        0           &        6.4         \\                                                                                                                                              
                           &PDG &                    &                    &                    &        $<11$       \\                                                                                                                                              
$\Delta(1920)\frac{3}{2}^+$&VMD &       -6.19        &        6.75        &       -2.05        &        9.39        \\
                           & KI &      -8.1          &       6.2          &       5.5          &       11.6         \\                                                                                                                                              
                           & CR & 5.3$^{+1.3}_{-0.5}$& 6.6$^{+1.6}_{-0.7}$&-0.7$^{+0.2}_{-0.4}$& 8.5$^{+2.0}_{-0.8}$\\                                                                                                                                              
                           & SST&                    &                    &                    &        5.2         \\                                                                                                                                              
\hline                                                                                                                                                                                                                                                              
                           &    & $f_{1/2}$ & $f_{3/2}$ & $p_{3/2}$ & \\                                                                                                                                                                                            
\hline                                                                                                                                                                                                                                                              
$\Delta(1905)\frac{5}{2}^+$&VMD &       -1.40        &       -0.46        &       17.46        &       17.53        \\
                           & KI &     -0.1           &       -6.4         &      -2.1          &       6.7          \\                                                                                                                                              
                           & CR & -0.7$\pm0.2$       &-0.7$^{+0.1}_{-0.2}$& 6.3$^{+0.8}_{-0.4}$& 6.4$^{+0.8}_{-0.4}$\\                                                                                                                                              
                           & SST&                    &                    &                    &       5.1          \\                                                                                                                                              
                           & OR &       0.3          &        1.3         &       -6.6         &       6.7          \\                                                                                                                                              
                           & MS &       0            &         0          &     16.8$\pm1.3$   &   16.8$\pm1.3$     \\                                                                                                                                              
                           &PDG &                    &                    &     20$\pm6$       &    $>17$           \\                                                                                                                                              
$\Delta(2000)\frac{5}{2}^+$&VMD &        2.43        &        5.20        &       -6.73        &        8.84        \\
                           & KI &     7.2            &       4.6          &     17.8           &     19.7           \\                                                                                                                                              
                           & CR & 2.6$^{+2.8}_{-2.1}$&-3.1$^{+2.4}_{-3.2}$&  3.1$\pm1.2$       & 5.1$^{+4.2}_{-3.0}$\\                                                                                                                                              
                           & SST&                    &                    &                    &      8.9           \\                                                                                                                                              
                           & MS &        0           &         0          &   -6.7$\pm2.4$     &    6.7$\pm2.4$     \\                                                                                                                                              
\hline                                                                                                                                                                                                                                                              
                           &    & $f_{1/2}$ & $f_{3/2}$ & $h_{3/2}$ & \\                                                                                                                                                                                            
\hline                                                                                                                                                                                                                                                              
$\Delta(1950)\frac{7}{2}^+$&VMD &        1.28        &       -2.38        &        0.28        &        2.72        \\
                           & KI &    -4.7            &      -8.2          &       0            &     9.4            \\                                                                                                                                              
                           & CR & 1.3$\pm0.1$        &-2.3$\pm0.2$        &       0            & 2.6$\pm0.2$        \\                                                                                                                                              
                           & SST&                    &                    &                    &      4.5           \\                                                                                                                                              
                           & OR &     1.1            &       1.9          &        0           &      2.2           \\                                                                                                                                              
                           & MS &       0            &   11.4$\pm0.5$     &        0           &    11.4$\pm0.5$    \\                                                                                                                                              
                           &PDG &                    &                    &                    &      $<6$          \\                                                                                                                                              
\hline                                                                                                                                                                                                                                                              
\end{tabular}                                                                                                                                                                                                                                                       
\end{center}                                                                                                                                                                                                                                                        
\caption{The $\rho$-meson modes of the positive-parity $\Delta$-resonances. The
notations are the same as in Table III.}                                                                                                                                                                                                                                        
\label{table6}                                                                                                                                                                                                                                                      
\end{table}                                                                                                                                                                                                                                                         
}                                                                                                                                                                                                                                                                   

\newpage

\begin{table}                                                                                                                                                                                                                                                       
\begin{center}                                                                                                                                                                                                                                                      
\begin{tabular}{rrr}                                                                                                                                                                                                                                                
\hline                                                                                                                                                                                                                                                              
Resonance$\;\;$& $\Gamma_{e^+e^-}$ & $\Gamma_{\mu^+\mu^-}$ \\                                                                                                                                                                                                          
             &    KeV $\;\;$       &      KeV $\;\;$         \\                                                                                                                                                                                                          
\hline
$N^{*}       (1535)\frac{1}{2}^-$&        2.01&        1.87\\
                                 &        5.30&        4.85\\
$N^{*}       (1650)\frac{1}{2}^-$&        3.23&        0.79\\
                                 &        2.00&        0.31\\
$N^{*}       (1520)\frac{3}{2}^-$&        6.02&        0.73\\
                                 &        4.42&        0.41\\
$N^{*}       (1700)\frac{3}{2}^-$&        0.41&        0.32\\
                                 &        2.86&        2.64\\
$N^{*}       (1675)\frac{5}{2}^-$&        0.21&        0.10\\
                                 &        1.09&        0.22\\
$N^{*}       (1440)\frac{1}{2}^+$&        1.40&        0.22\\
                                 &        0.56&        0.05\\
$N^{*}       (1710)\frac{1}{2}^+$&        0.58&        0.31\\
                                 &        0.60&        0.56\\
$N^{*}       (1720)\frac{3}{2}^+$&        7.93&        7.77\\
                                 &        3.14&        2.77\\
$N^{*}       (1900)\frac{3}{2}^+$&        4.62&        4.54\\
                                 &       12.22&       11.91\\
$N^{*}       (1680)\frac{5}{2}^+$&        2.58&        0.43\\
                                 &        1.47&        1.13\\
$N^{*}       (2000)\frac{5}{2}^+$&       14.17&       13.99\\
                                 &       20.89&       21.56\\
$N^{*}       (1990)\frac{7}{2}^+$&        3.09&        2.97\\
                                 &        8.24&        4.78\\
\hline                                                                                                                                                                                                                                                              
$\Delta      (1620)\frac{1}{2}^-$&        1.33&        0.88\\
$\Delta      (1900)\frac{1}{2}^-$&        1.19&        1.09\\
$\Delta      (1700)\frac{3}{2}^-$&        6.10&        1.65\\
$\Delta      (1940)\frac{3}{2}^-$&        6.57&        6.20\\
$\Delta      (1930)\frac{5}{2}^-$&        9.63&        9.16\\
$\Delta      (1750)\frac{1}{2}^+$&        4.61&        2.77\\
$\Delta      (1910)\frac{1}{2}^+$&        1.33&        1.27\\
$\Delta      (1232)\frac{3}{2}^+$&        5.02&        0.04\\
$\Delta      (1600)\frac{3}{2}^+$&        0.24&        0.13\\
$\Delta      (1920)\frac{3}{2}^+$&        4.22&        3.50\\
$\Delta      (1905)\frac{5}{2}^+$&       10.51&       10.36\\
$\Delta      (2000)\frac{5}{2}^+$&        3.25&        3.00\\
$\Delta      (1950)\frac{7}{2}^+$&        3.18&        0.81\\
\hline                                                                                                                                                                                                                                                              
\end{tabular}                                                                                                                                                                                                                                                       
\end{center}                                                                                                                                                                                                                                                        
\caption{The decay widths of the nucleon resonances into the $e^{+}e^{-}$
and $\mu ^{+}\mu ^{-}$ pairs. The first line of the $N^{*}$-resonances
with $I=1/2$
refers to the proton, the second one to the neutron resonances.}                                                                                                                                                                                                                                        
\label{table7}                                                                                                                                                                                                                                                      
\end{table}                                                                                                                                                                                                                                                         

\widetext

%
%

\clearpage


\begin{figure}[tbp]
\begin{center}
\leavevmode
\epsfxsize = 12cm 
\epsffile[20 20 440 810]{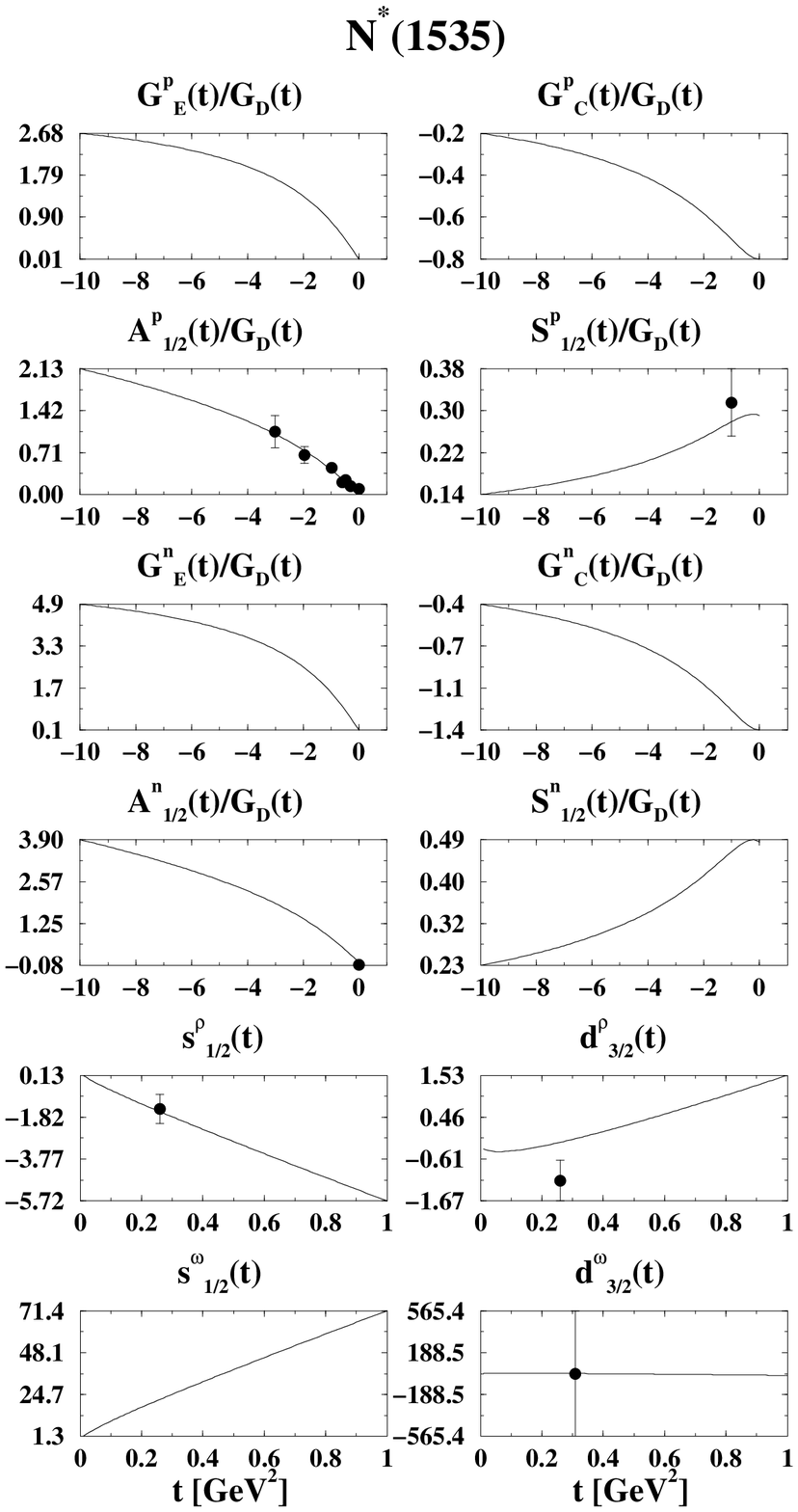}
\end{center}
\caption{ }
\label{fig1}
\end{figure}

\begin{figure}[tbp]
\begin{center}
\leavevmode
\epsfxsize = 12cm 
\epsffile[20 20 440 810]{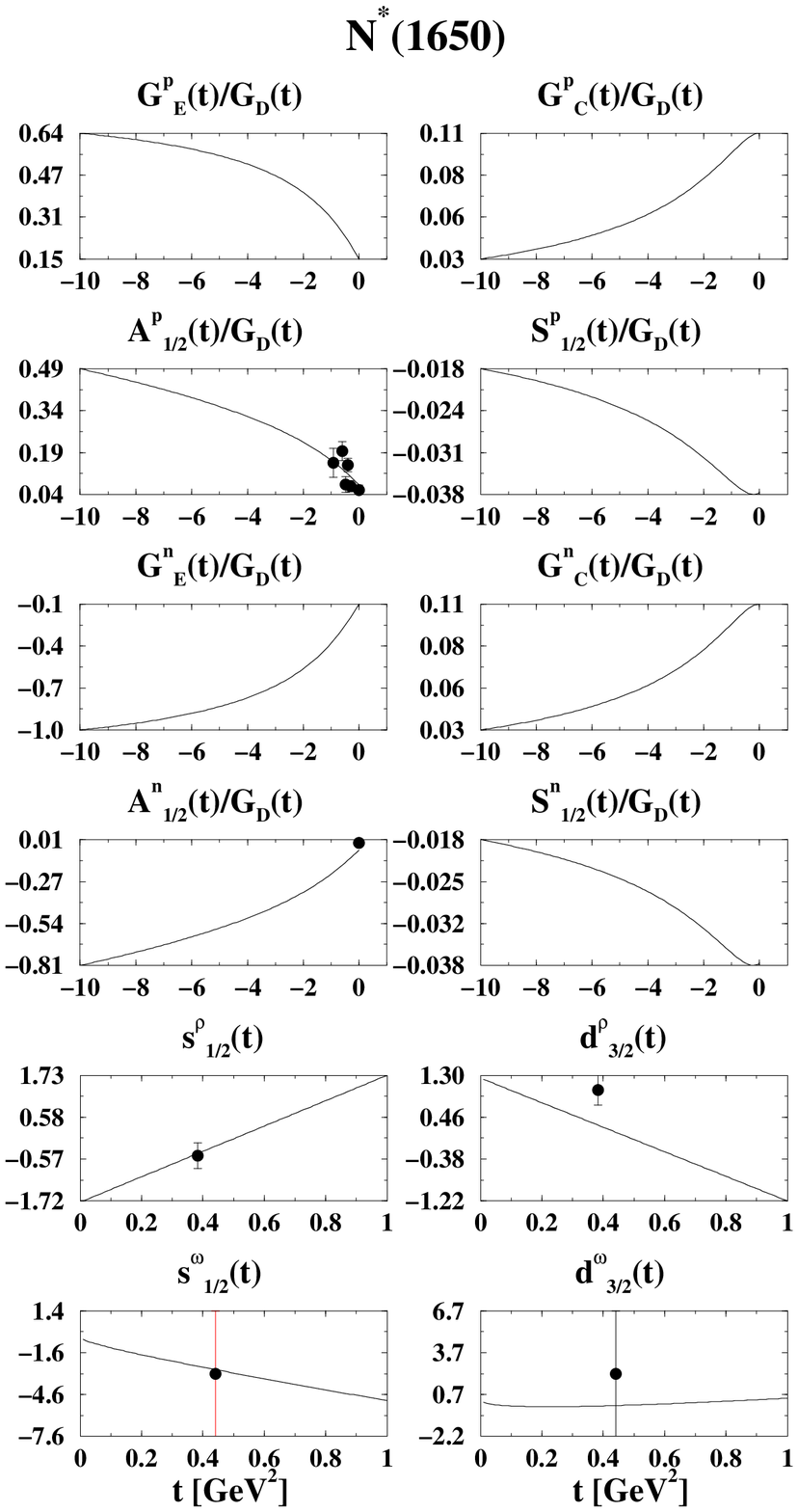}
\end{center}
\caption{ }
\label{fig2}
\end{figure}

\begin{figure}[tbp]
\begin{center}
\leavevmode
\epsfxsize = 12cm 
\epsffile[10 30 590 740]{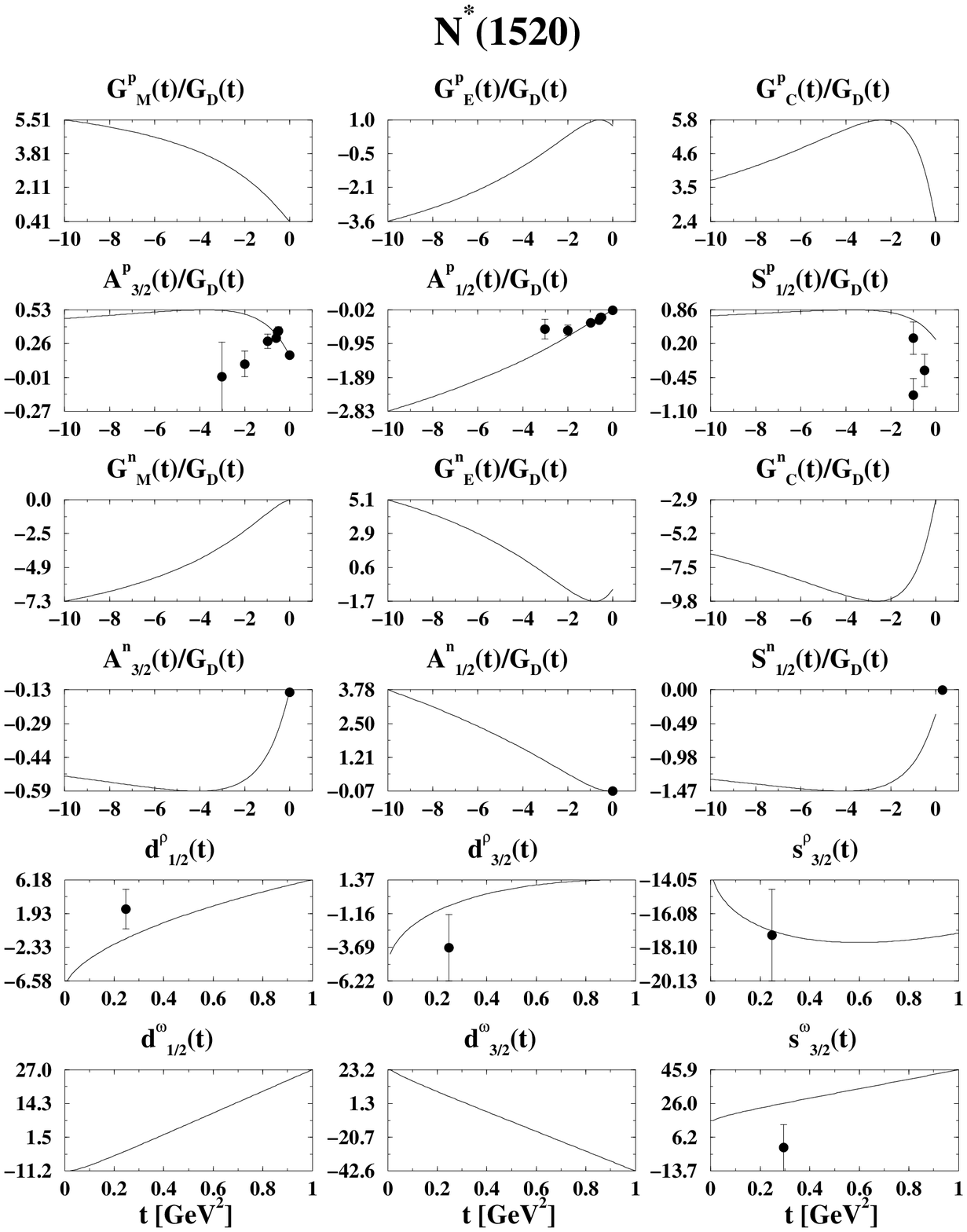}
\end{center}
\caption{ }
\label{fig3}
\end{figure}

\begin{figure}[tbp]
\begin{center}
\leavevmode
\epsfxsize = 12cm 
\epsffile[10 30 590 740]{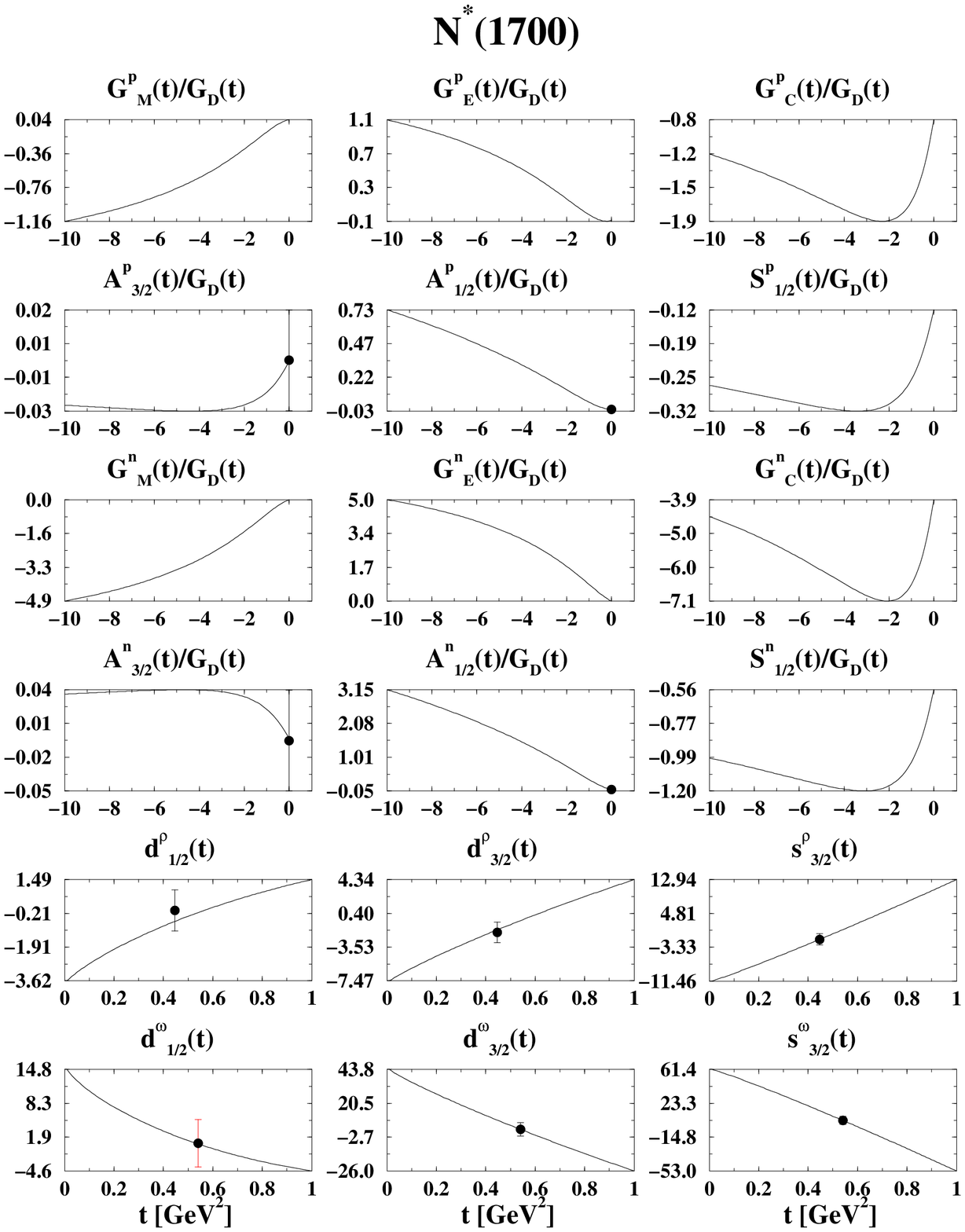}
\end{center}
\caption{ }
\label{fig4}
\end{figure}

\begin{figure}[tbp]
\begin{center}
\leavevmode
\epsfxsize = 12cm 
\epsffile[10 30 590 740]{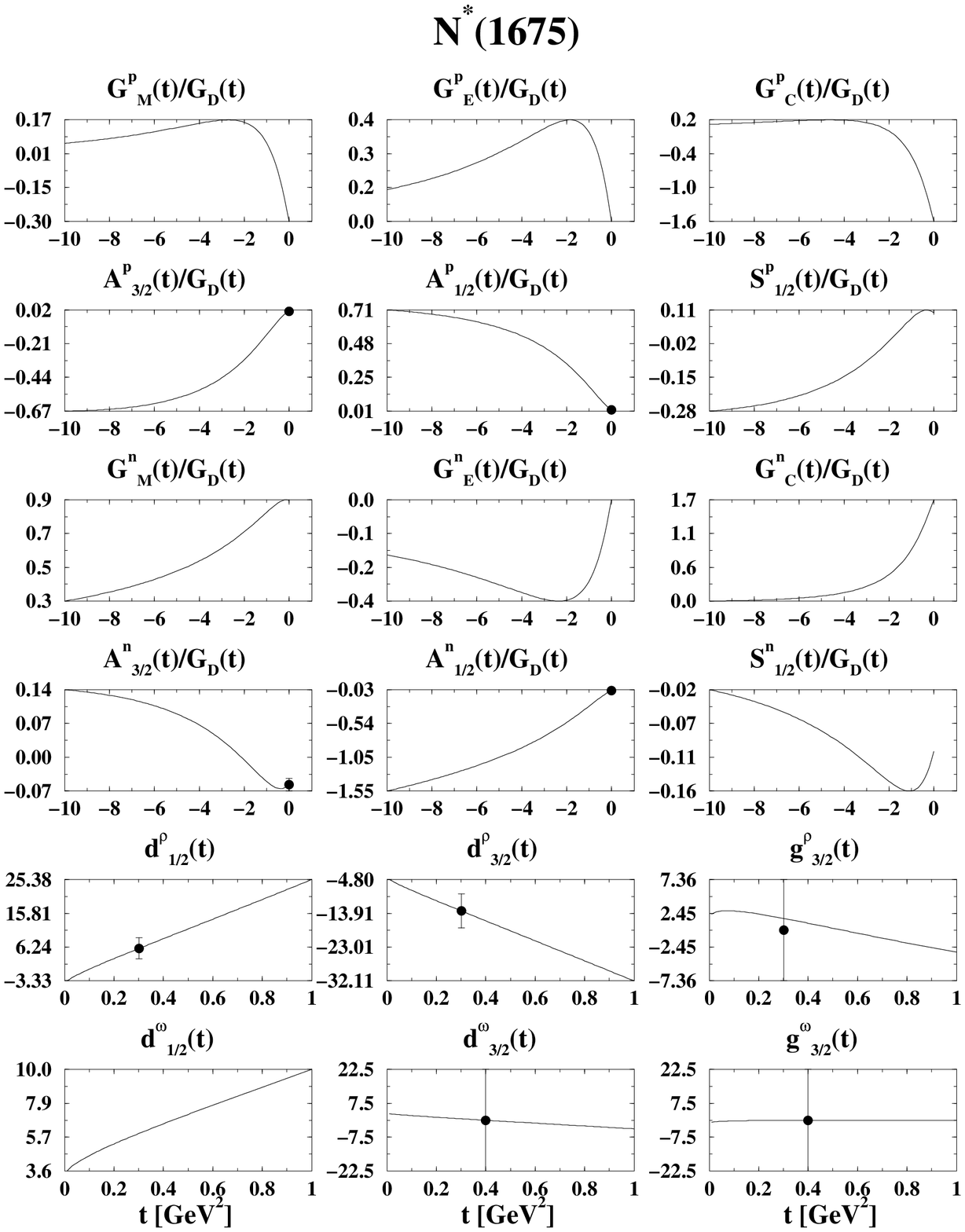}
\end{center}
\caption{ }
\label{fig5}
\end{figure}

\begin{figure}[tbp]
\begin{center}
\leavevmode
\epsfxsize = 12cm 
\epsffile[20 20 440 810]{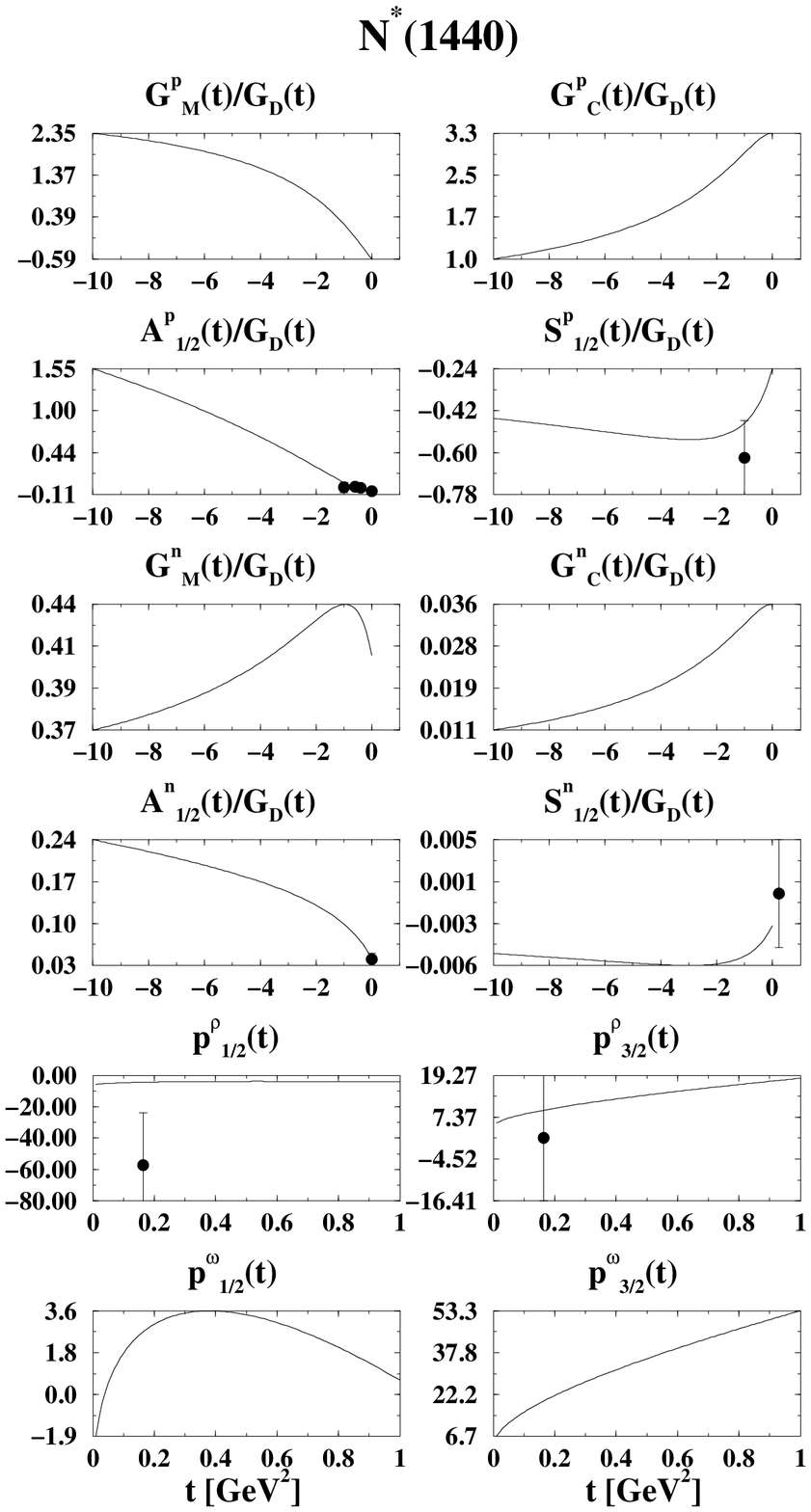}
\end{center}
\caption{ }
\label{fig6}
\end{figure}

\begin{figure}[tbp]
\begin{center}
\leavevmode
\epsfxsize = 12cm 
\epsffile[20 20 440 810]{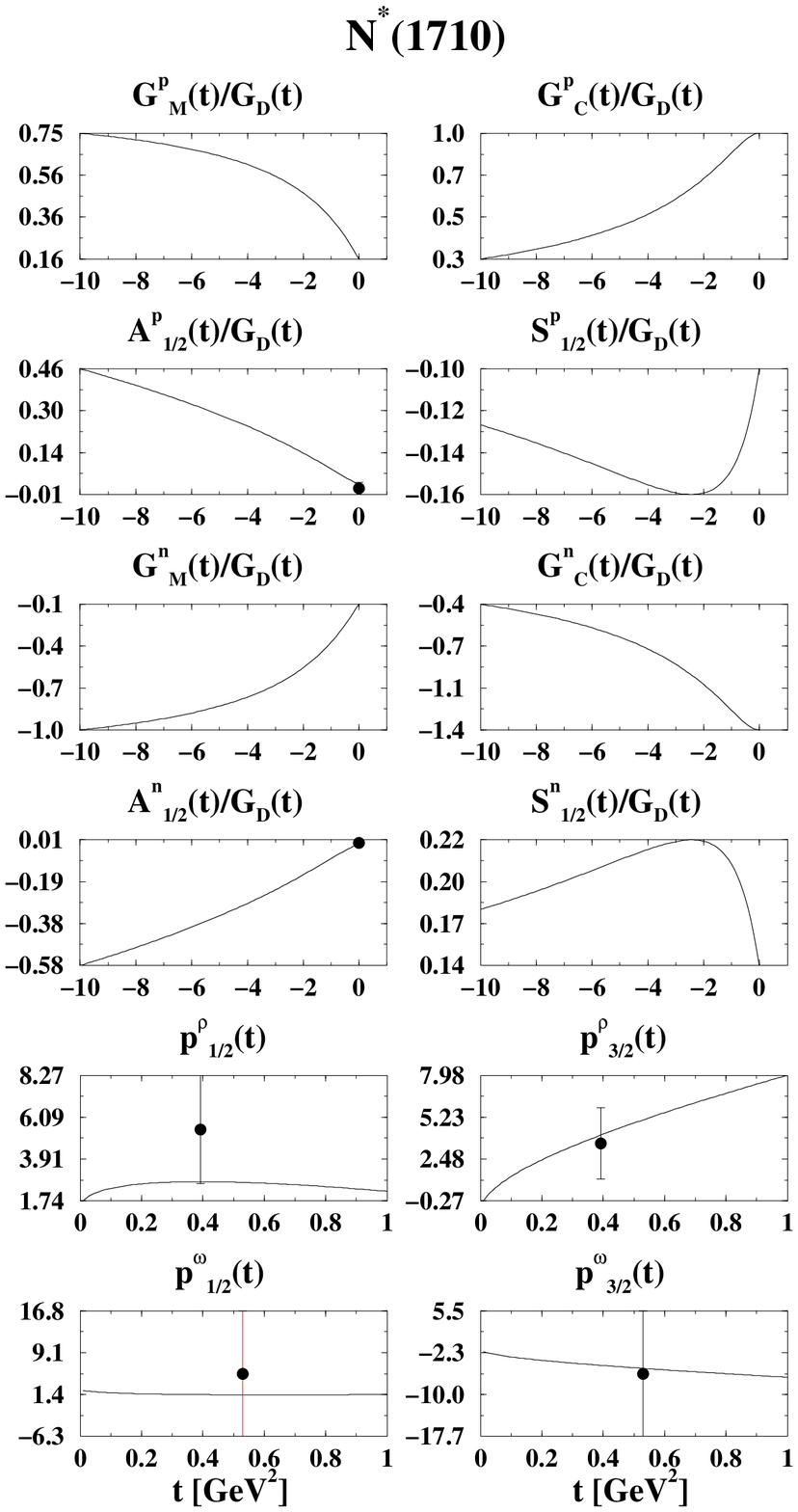}
\end{center}
\caption{ }
\label{fig7}
\end{figure}

\begin{figure}[tbp]
\begin{center}
\leavevmode
\epsfxsize = 12cm 
\epsffile[10 30 590 740]{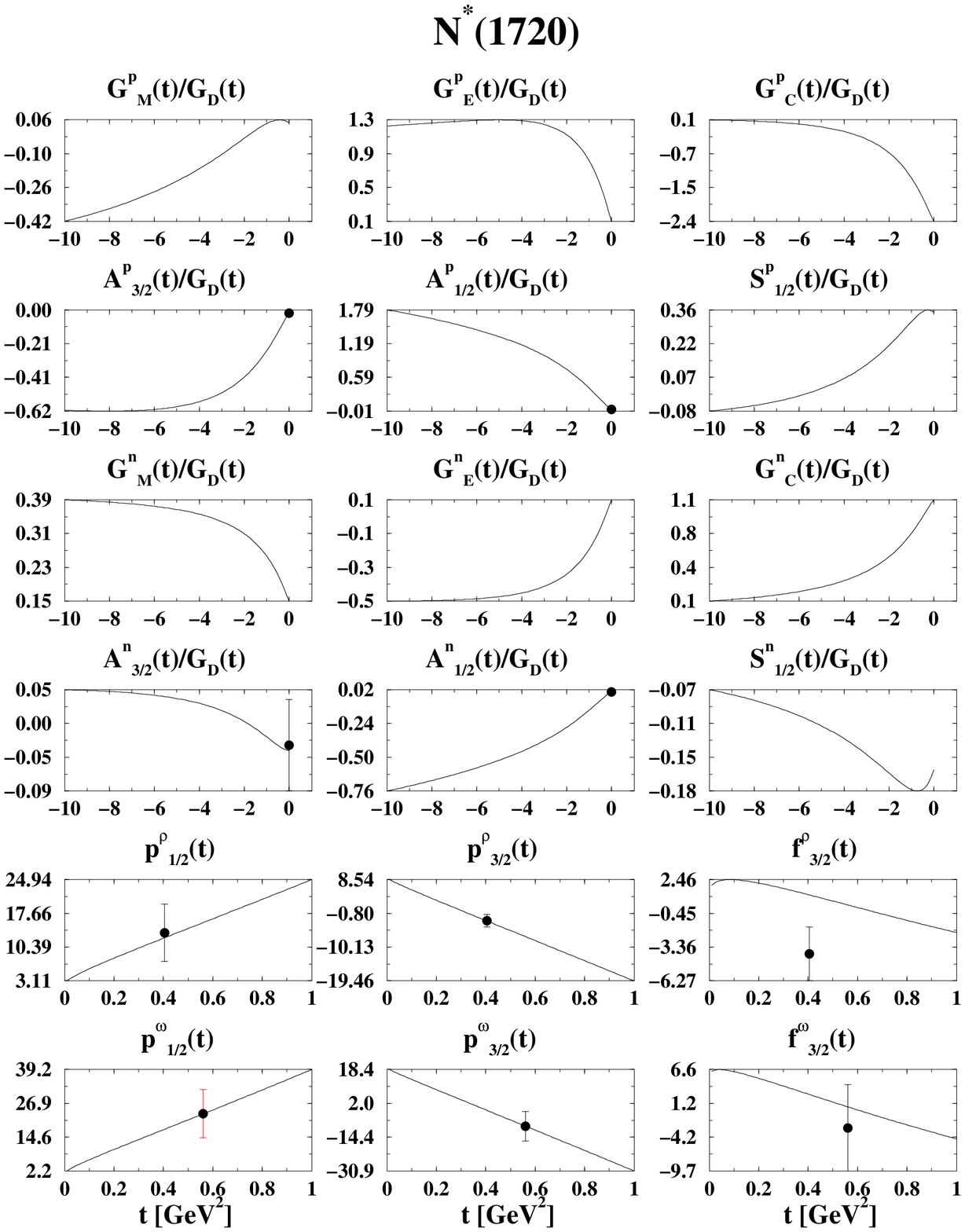}
\end{center}
\caption{ }
\label{fig8}
\end{figure}

\begin{figure}[tbp]
\begin{center}
\leavevmode
\epsfxsize = 12cm 
\epsffile[10 30 590 740]{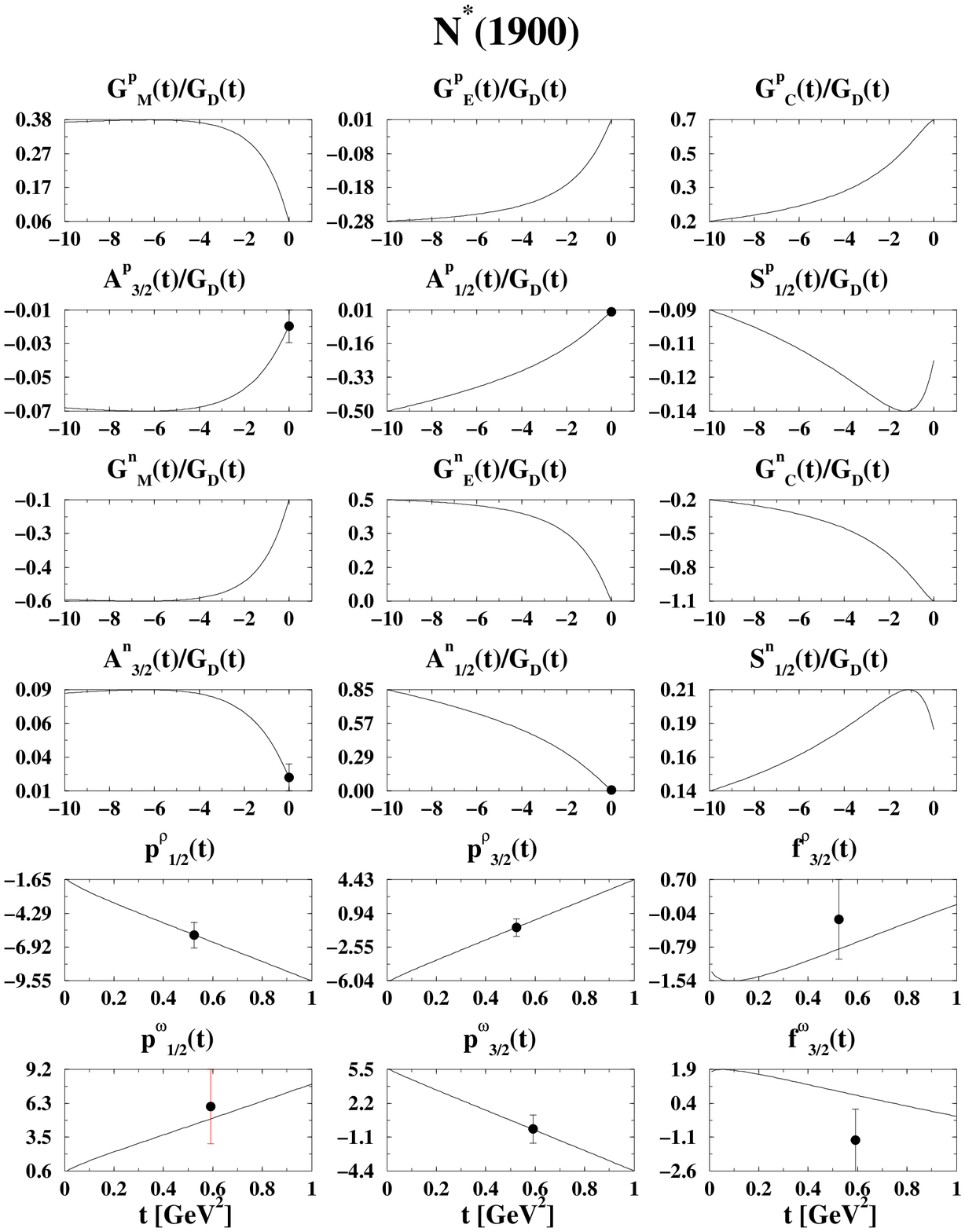}
\end{center}
\caption{ }
\label{fig9}
\end{figure}

\begin{figure}[tbp]
\begin{center}
\leavevmode
\epsfxsize = 12cm 
\epsffile[10 30 590 740]{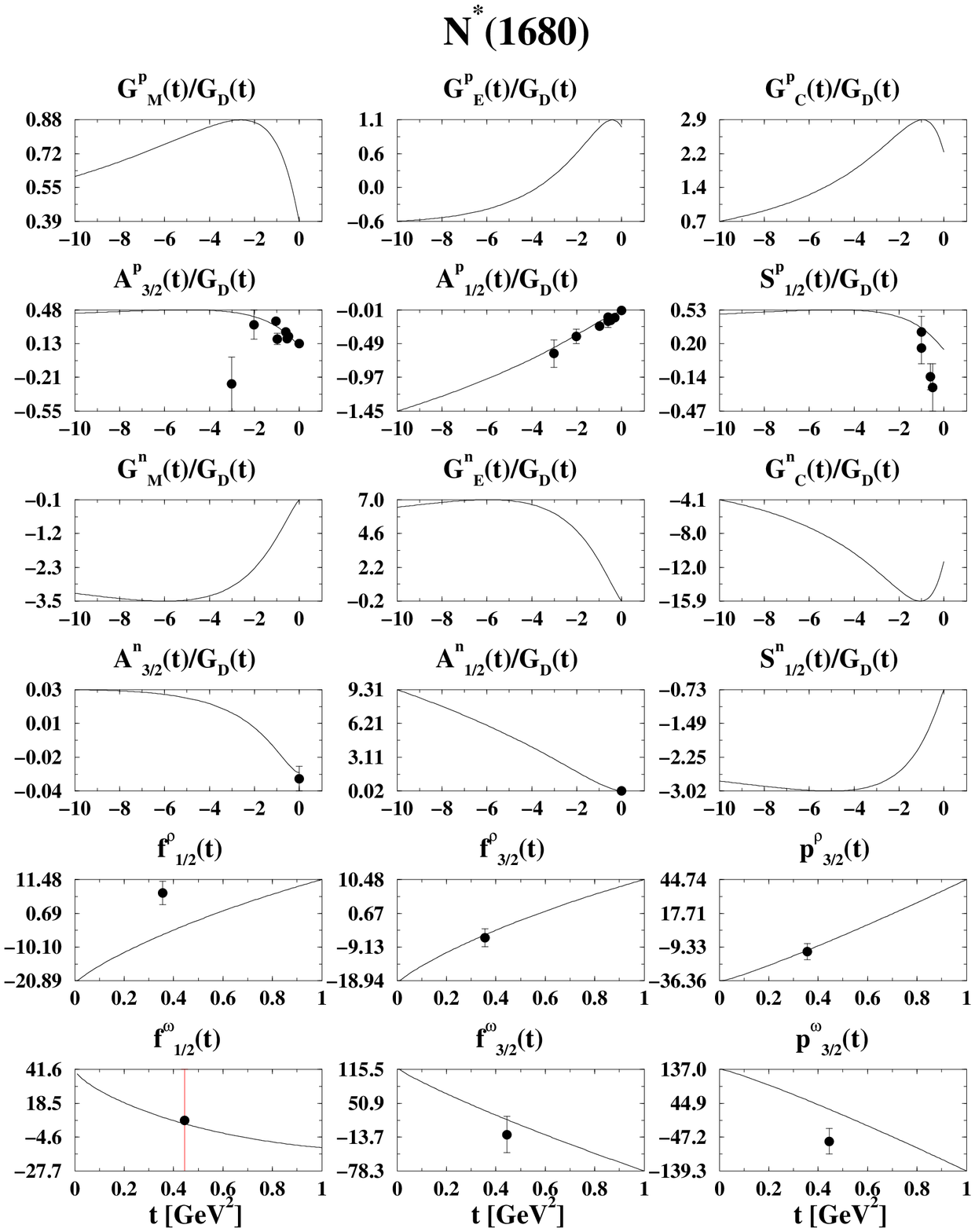}
\end{center}
\caption{ }
\label{fig10}
\end{figure}

\begin{figure}[tbp]
\begin{center}
\leavevmode
\epsfxsize = 12cm 
\epsffile[10 30 590 740]{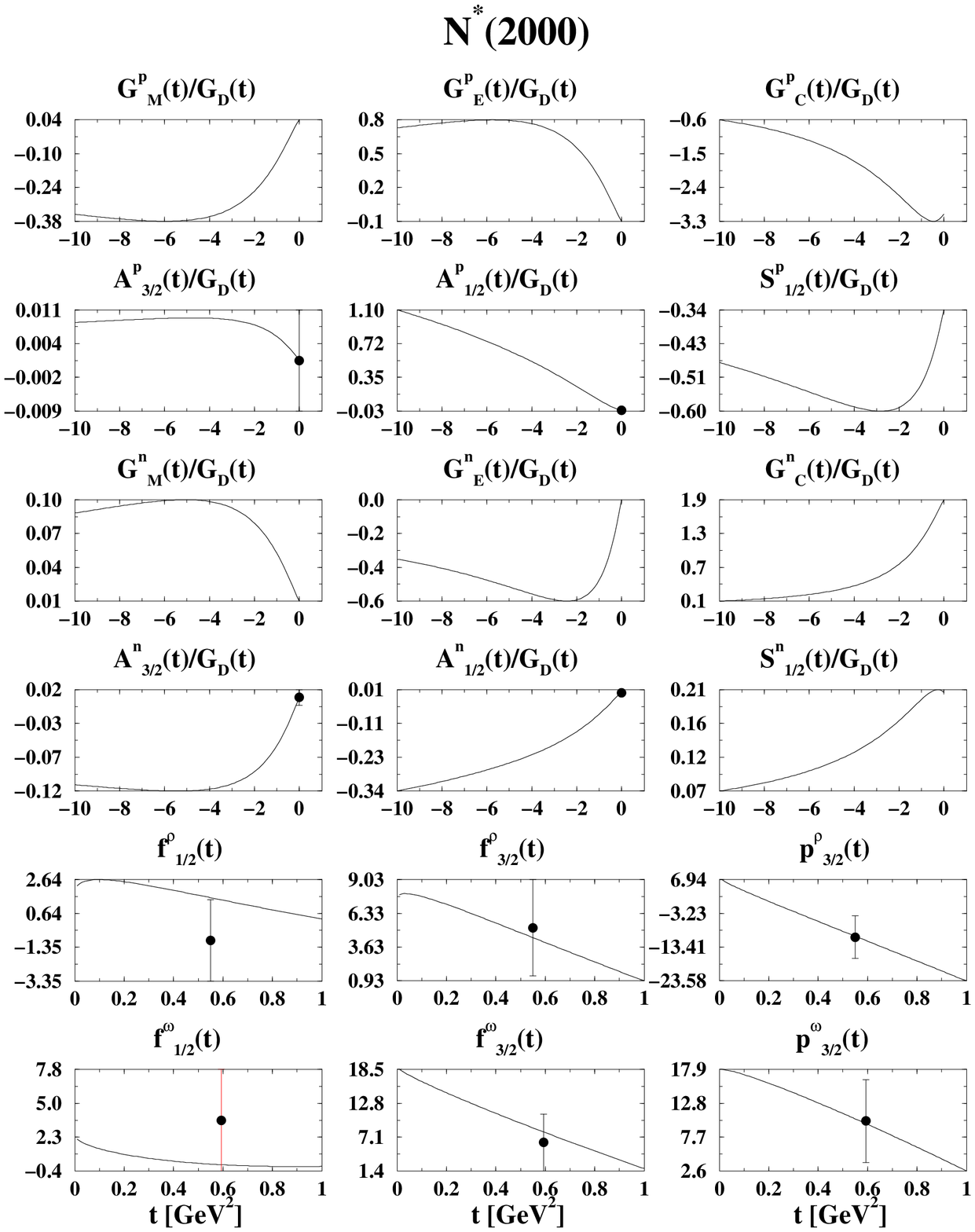}
\end{center}
\caption{ }
\label{fig11}
\end{figure}

\begin{figure}[tbp]
\begin{center}
\leavevmode
\epsfxsize = 12cm 
\epsffile[10 30 590 740]{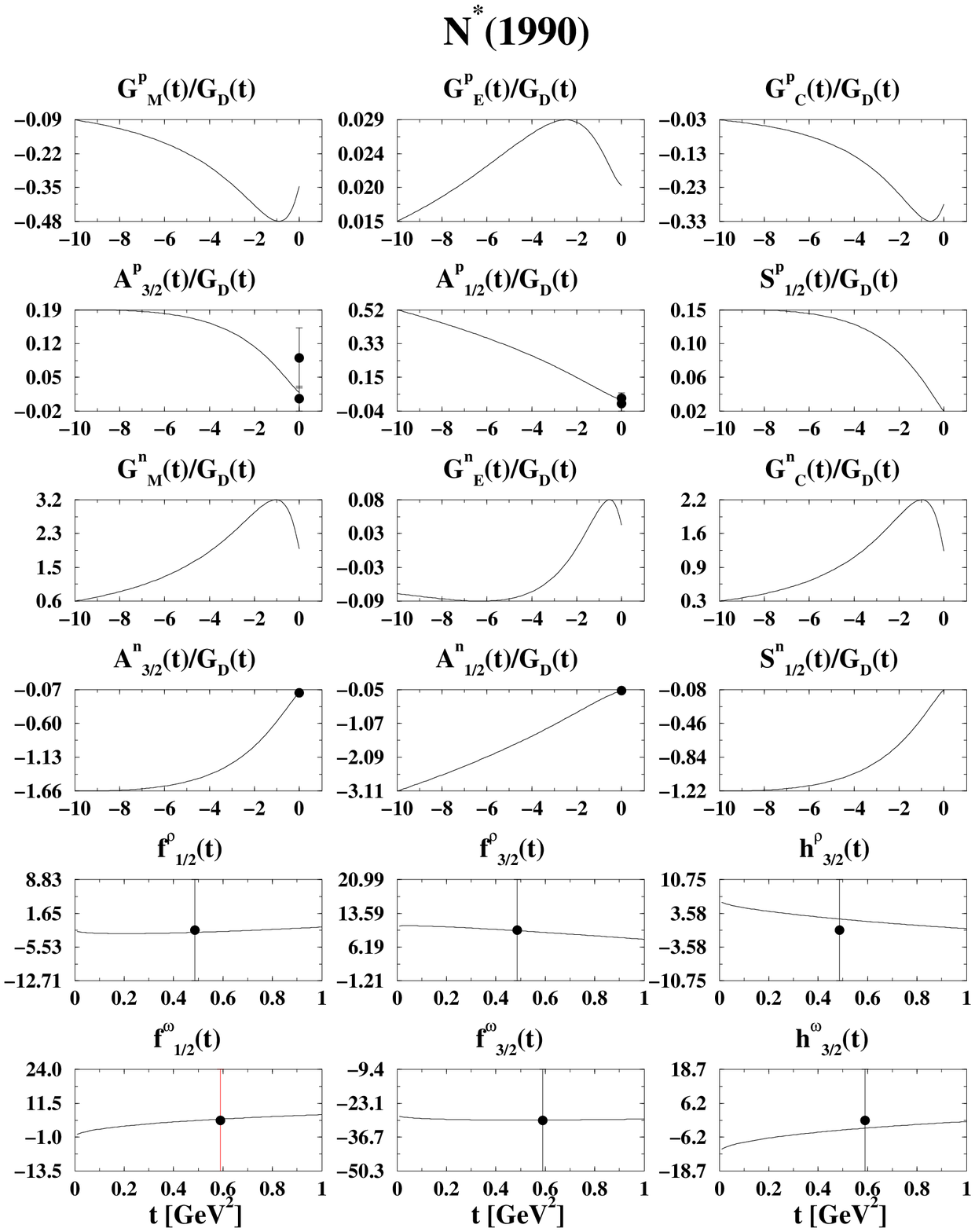}
\end{center}
\caption{ }
\label{fig12}
\end{figure}


\begin{figure}[tbp]
\begin{center}
\leavevmode
\epsfxsize = 14cm 
\epsffile[30 340 420 740]{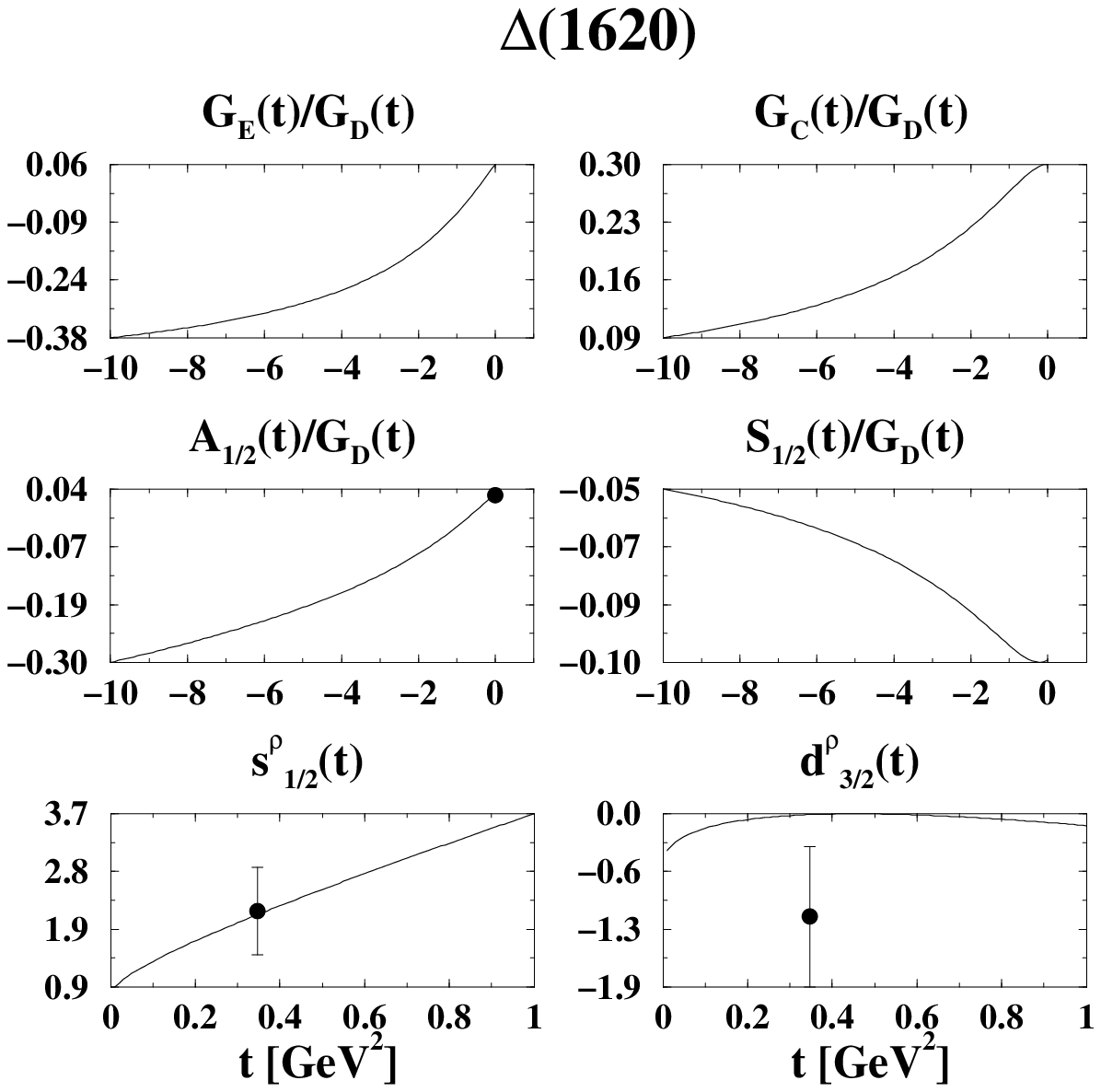}
\end{center}
\caption{ }
\label{fig13}
\end{figure}

\begin{figure}[tbp]
\begin{center}
\leavevmode
\epsfxsize = 14cm 
\epsffile[30 340 420 740]{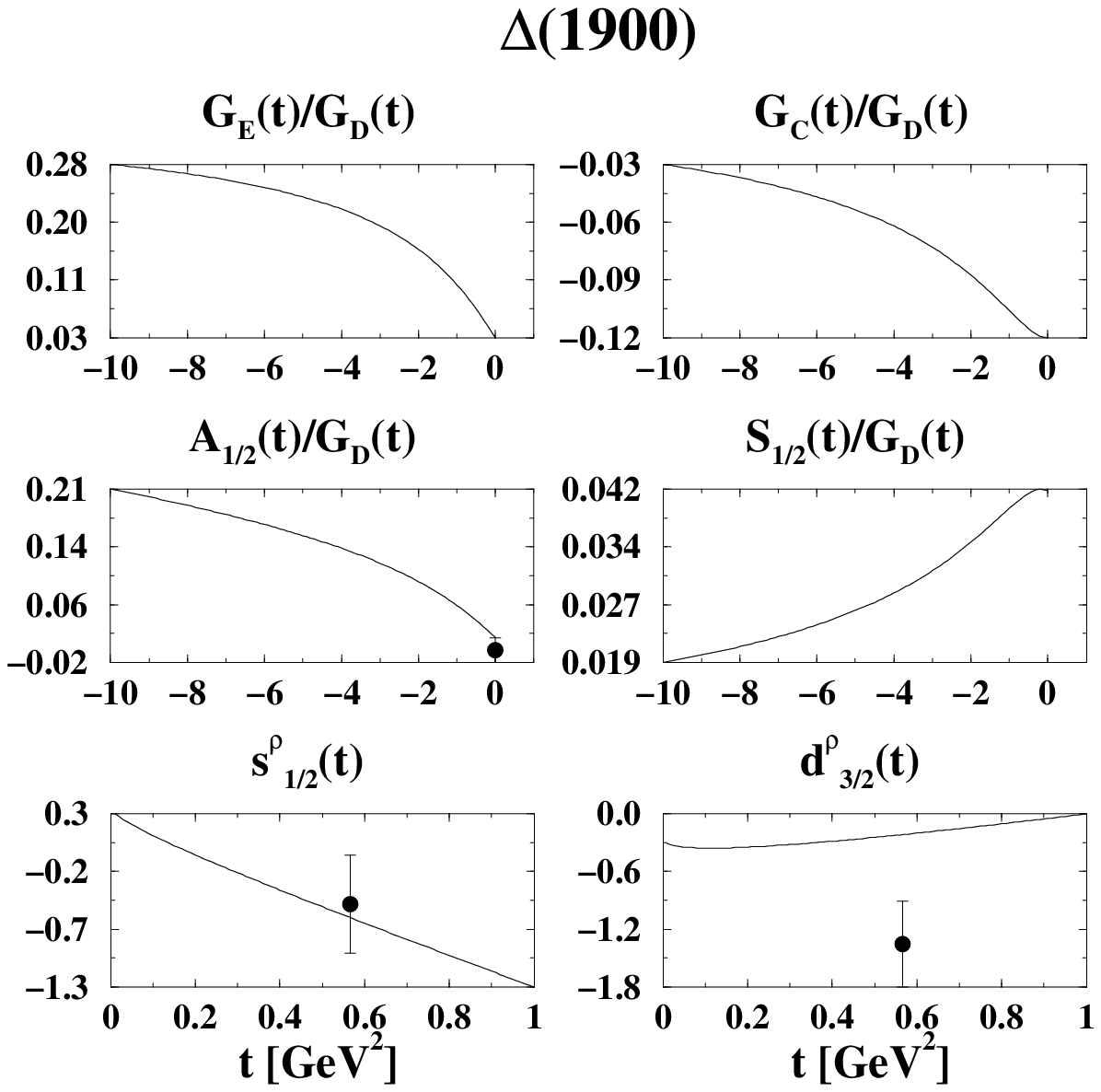}
\end{center}
\caption{ }
\label{fig14}
\end{figure}


\begin{figure}[tbp]
\begin{center}
\leavevmode
\epsfxsize = 18cm 
\epsffile[20 330 590 750]{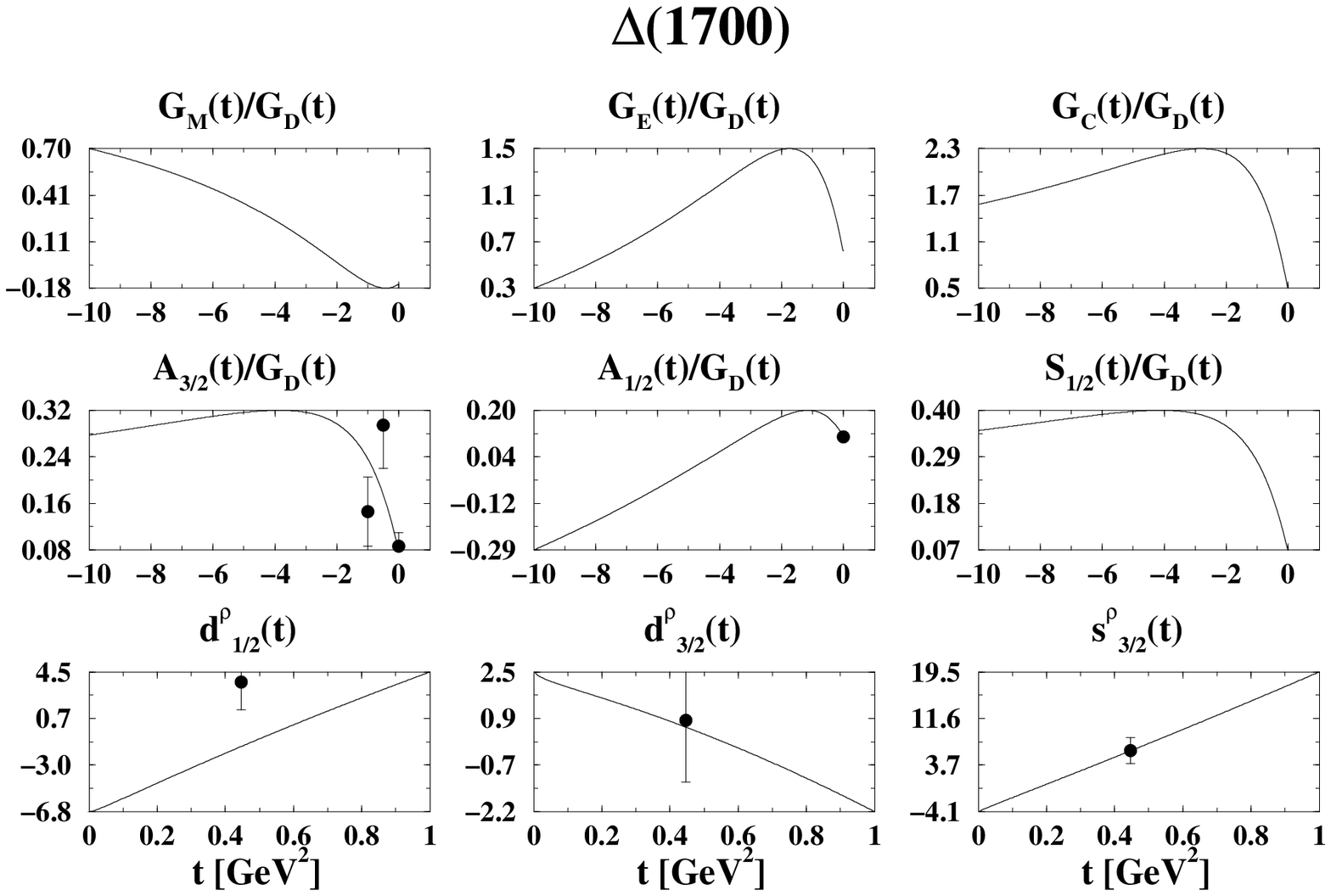}
\end{center}
\caption{ }
\label{fig15}
\end{figure}

\begin{figure}[tbp]
\begin{center}
\leavevmode
\epsfxsize = 18cm 
\epsffile[20 330 590 750]{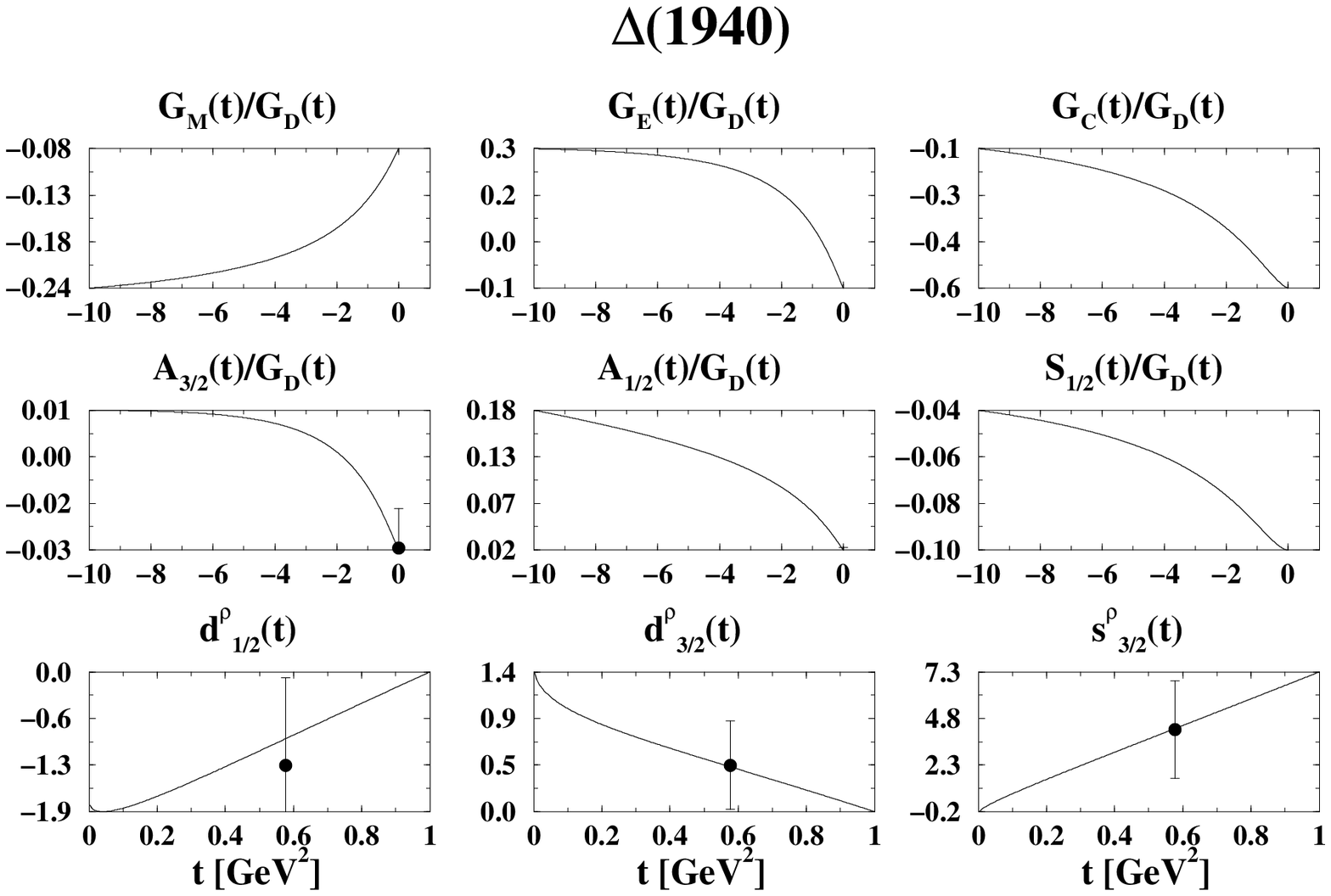}
\end{center}
\caption{ }
\label{fig16}
\end{figure}

\begin{figure}[tbp]
\begin{center}
\leavevmode
\epsfxsize = 18cm 
\epsffile[20 330 590 750]{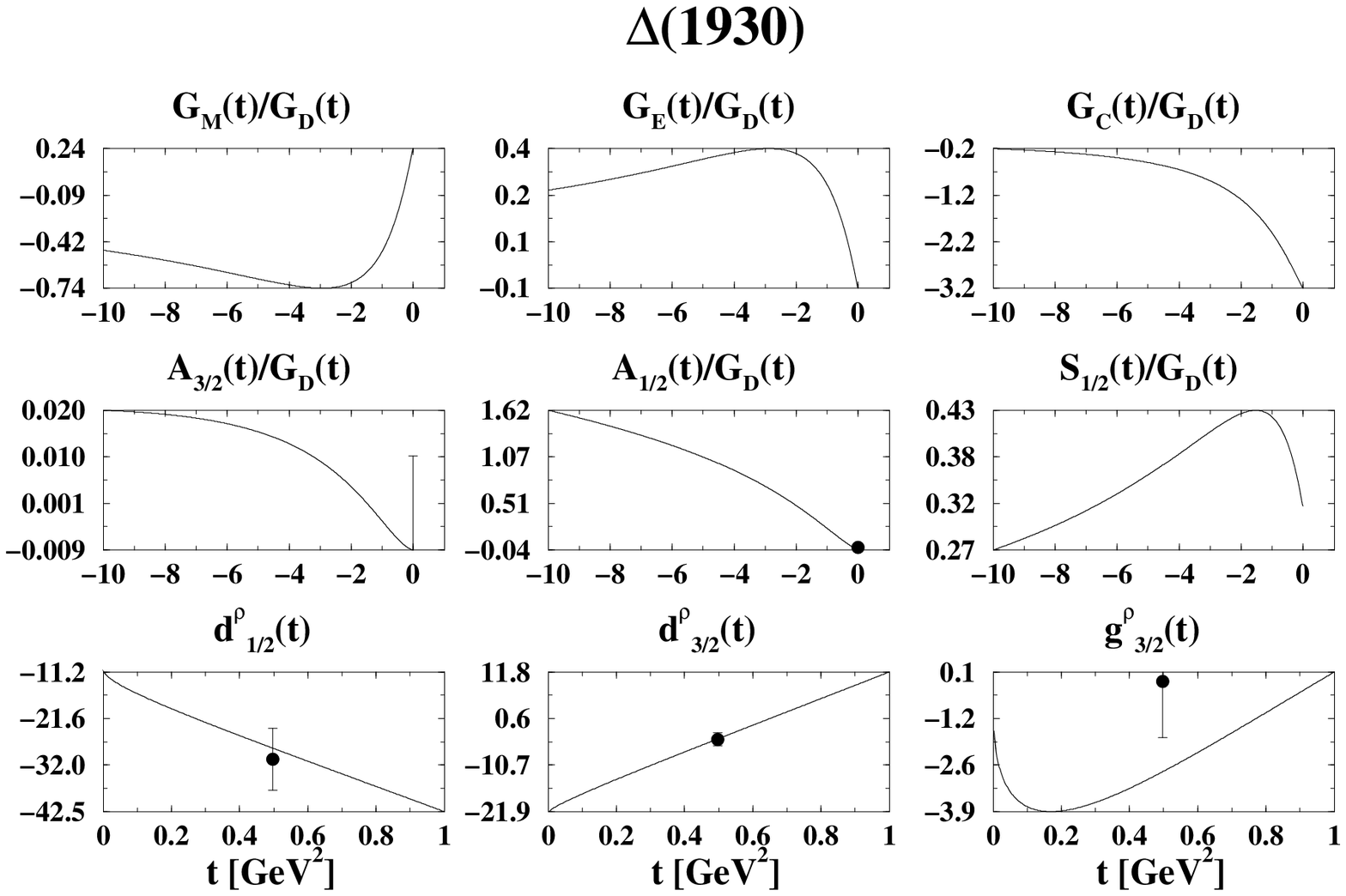}
\end{center}
\caption{ }
\label{fig17}
\end{figure}

\begin{figure}[tbp]
\begin{center}
\leavevmode
\epsfxsize = 14cm 
\epsffile[30 340 420 740]{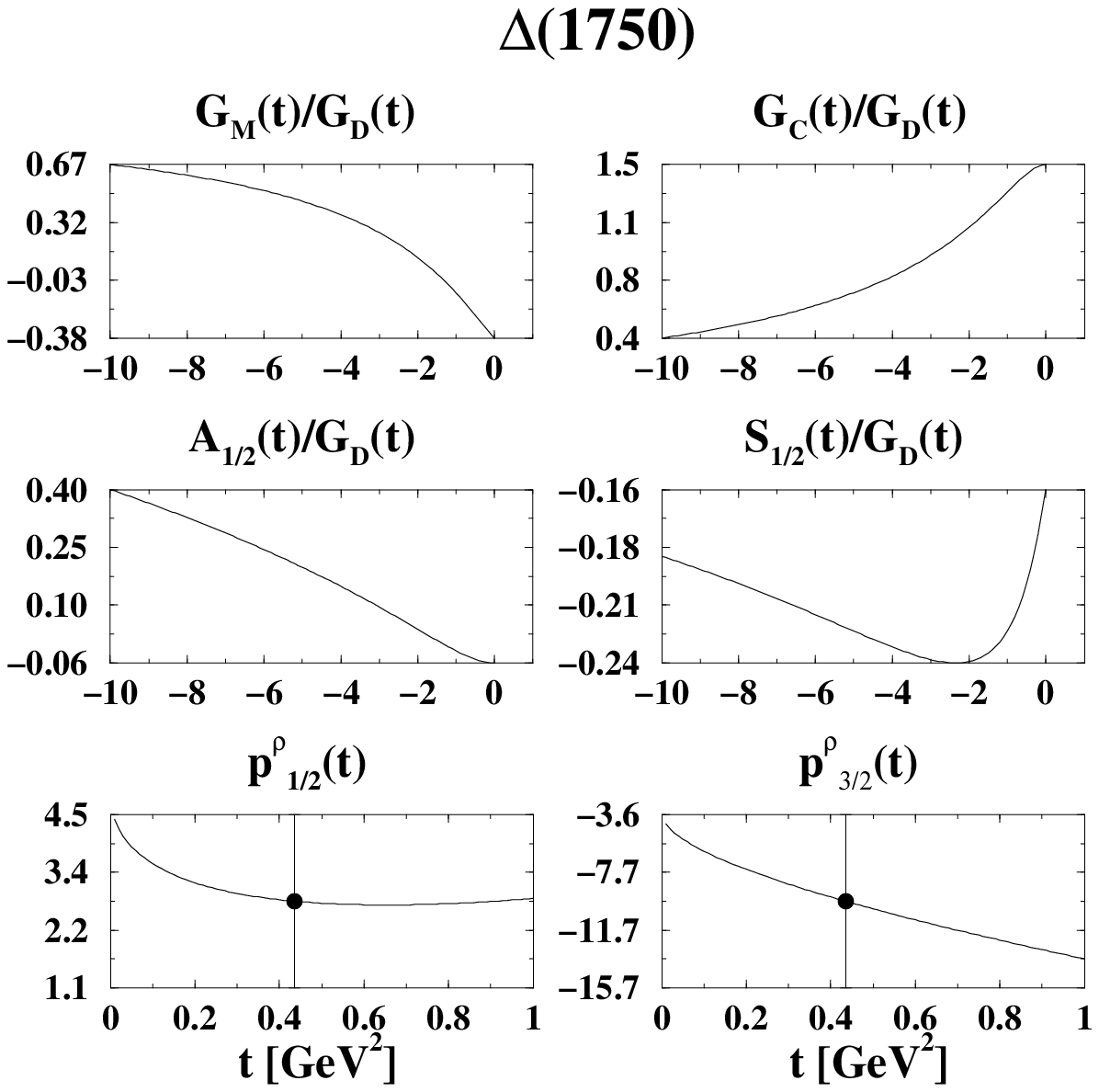}
\end{center}
\caption{ }
\label{fig18}
\end{figure}

\begin{figure}[tbp]
\begin{center}
\leavevmode
\epsfxsize = 14cm 
\epsffile[30 340 420 740]{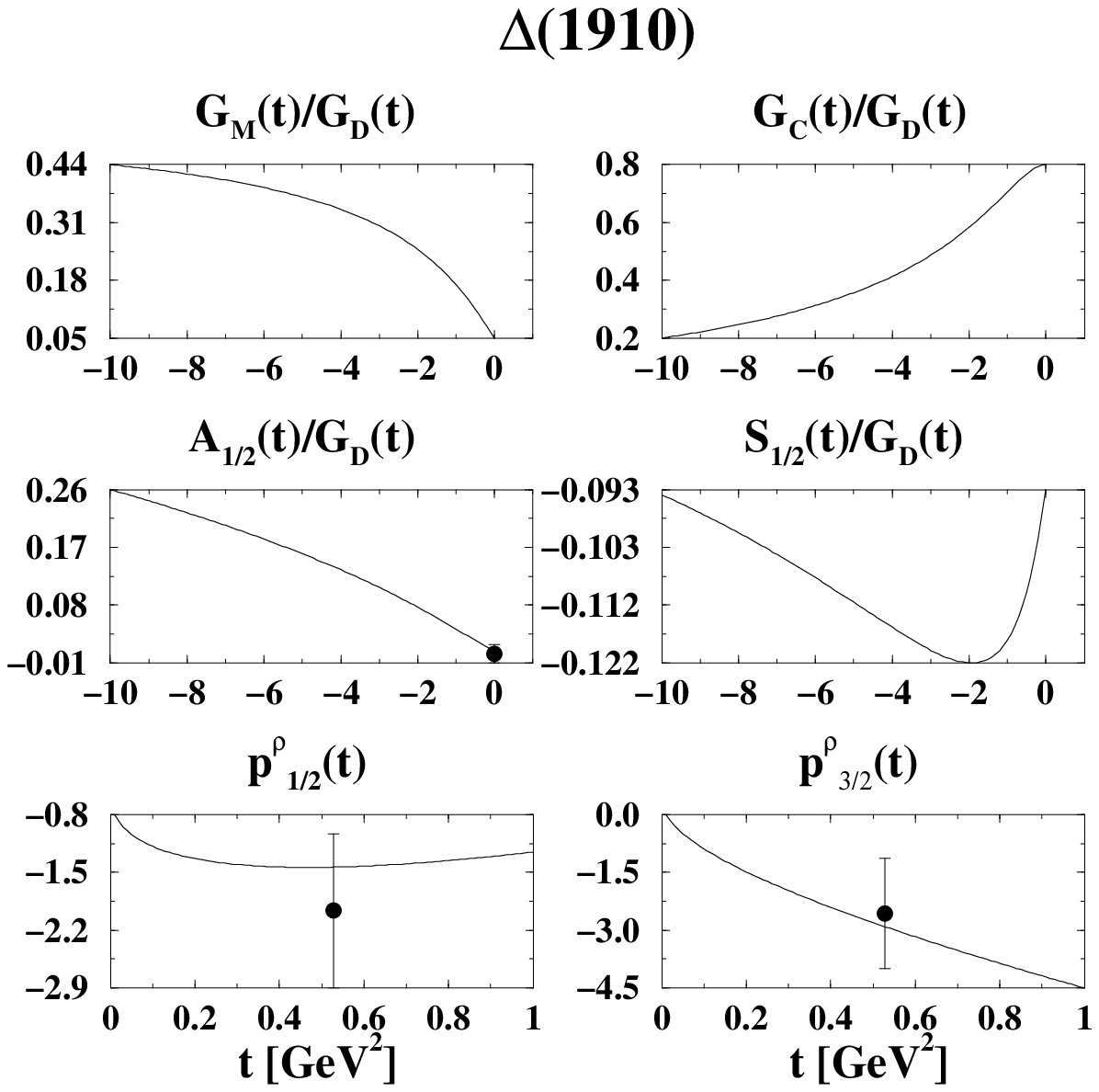}
\end{center}
\caption{ }
\label{fig19}
\end{figure}

\begin{figure}[tbp]
\begin{center}
\leavevmode
\epsfxsize = 18cm 
\epsffile[20 330 590 750]{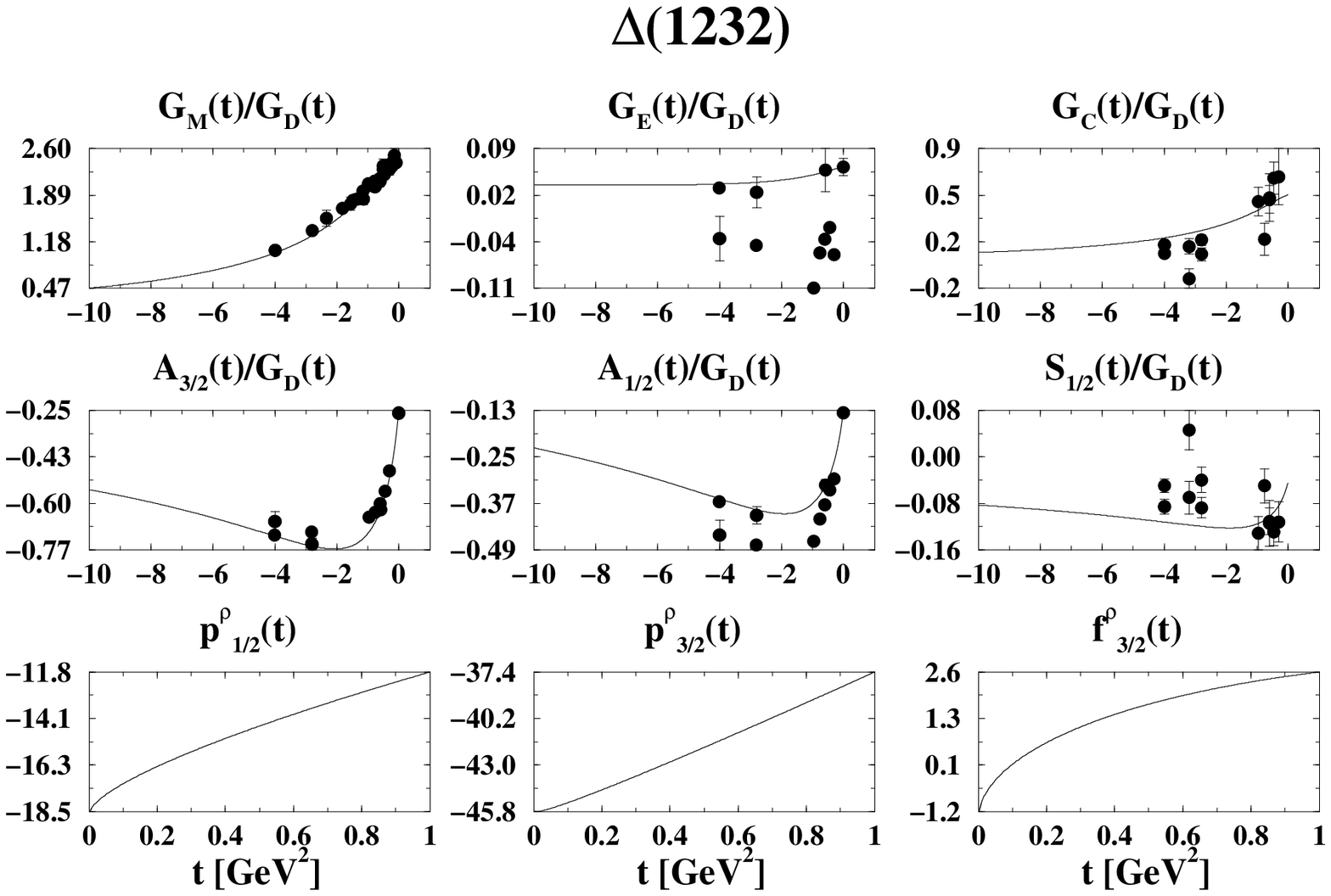}
\end{center}
\caption{ }
\label{fig20}
\end{figure}

\begin{figure}[tbp]
\begin{center}
\leavevmode
\epsfxsize = 18cm 
\epsffile[20 330 590 750]{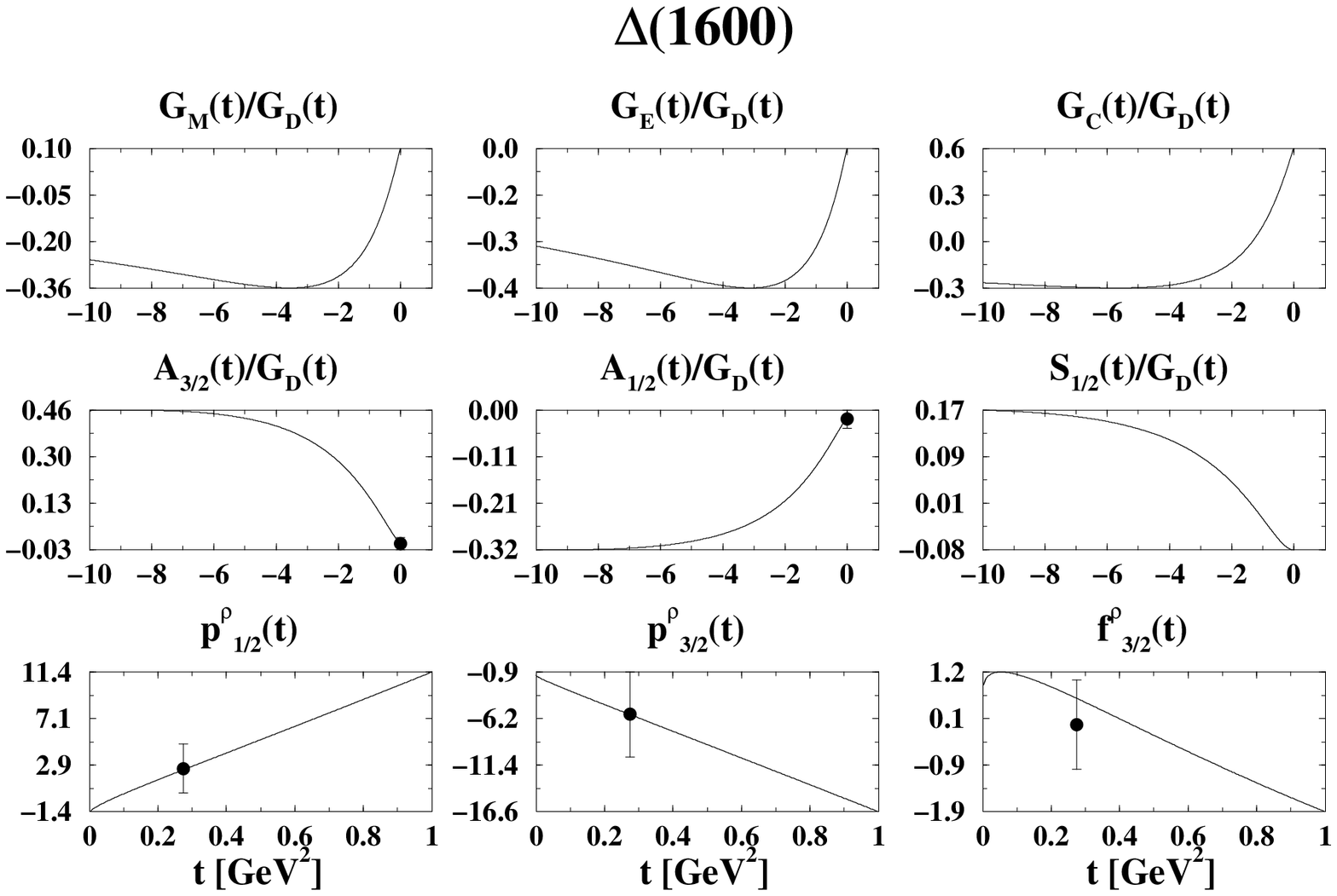}
\end{center}
\caption{ }
\label{fig21}
\end{figure}

\begin{figure}[tbp]
\begin{center}
\leavevmode
\epsfxsize = 18cm 
\epsffile[20 330 590 750]{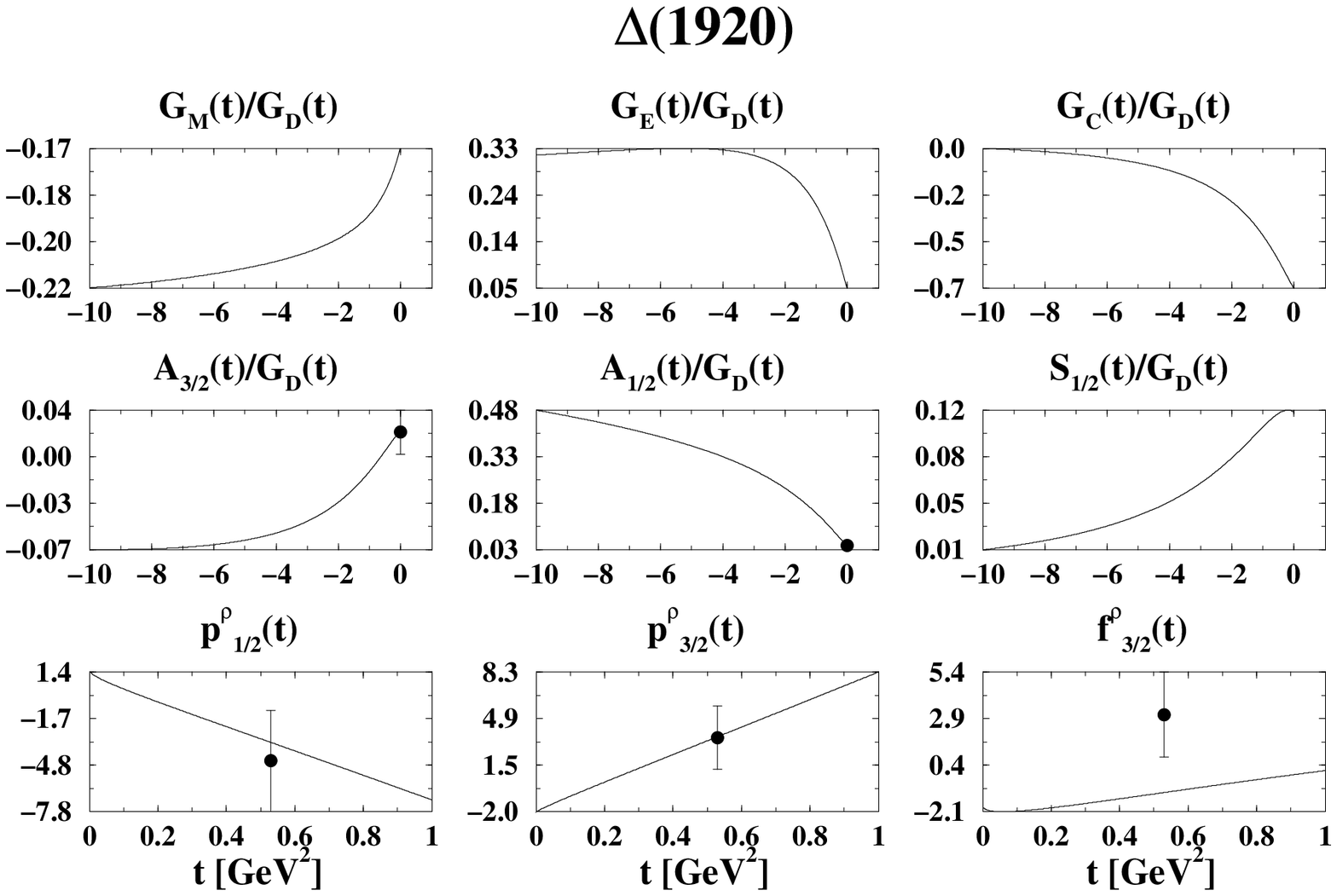}
\end{center}
\caption{ }
\label{fig22}
\end{figure}

\begin{figure}[tbp]
\begin{center}
\leavevmode
\epsfxsize = 18cm 
\epsffile[20 330 590 750]{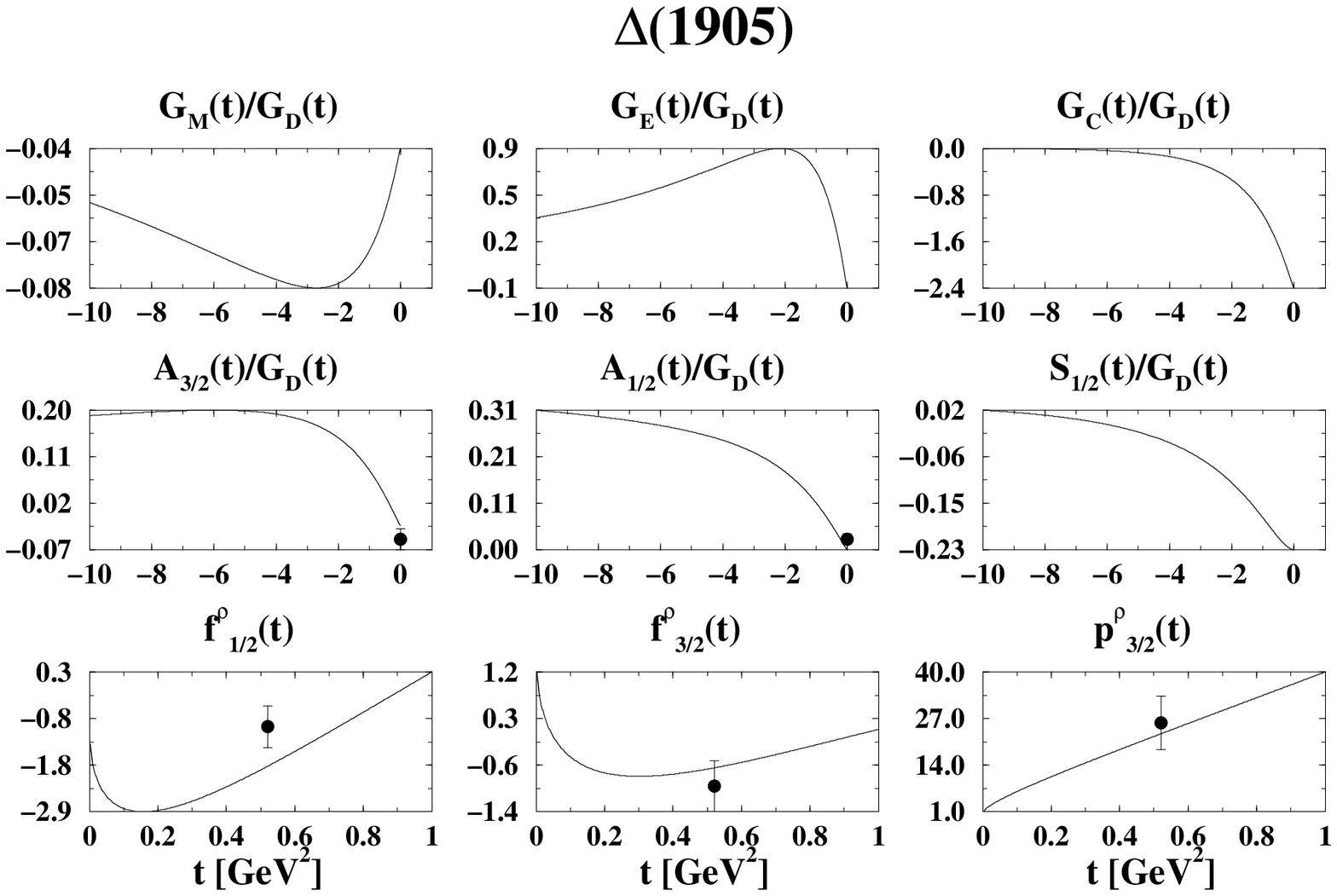}
\end{center}
\caption{ }
\label{fig23}
\end{figure}

\begin{figure}[tbp]
\begin{center}
\leavevmode
\epsfxsize = 18cm 
\epsffile[20 330 590 750]{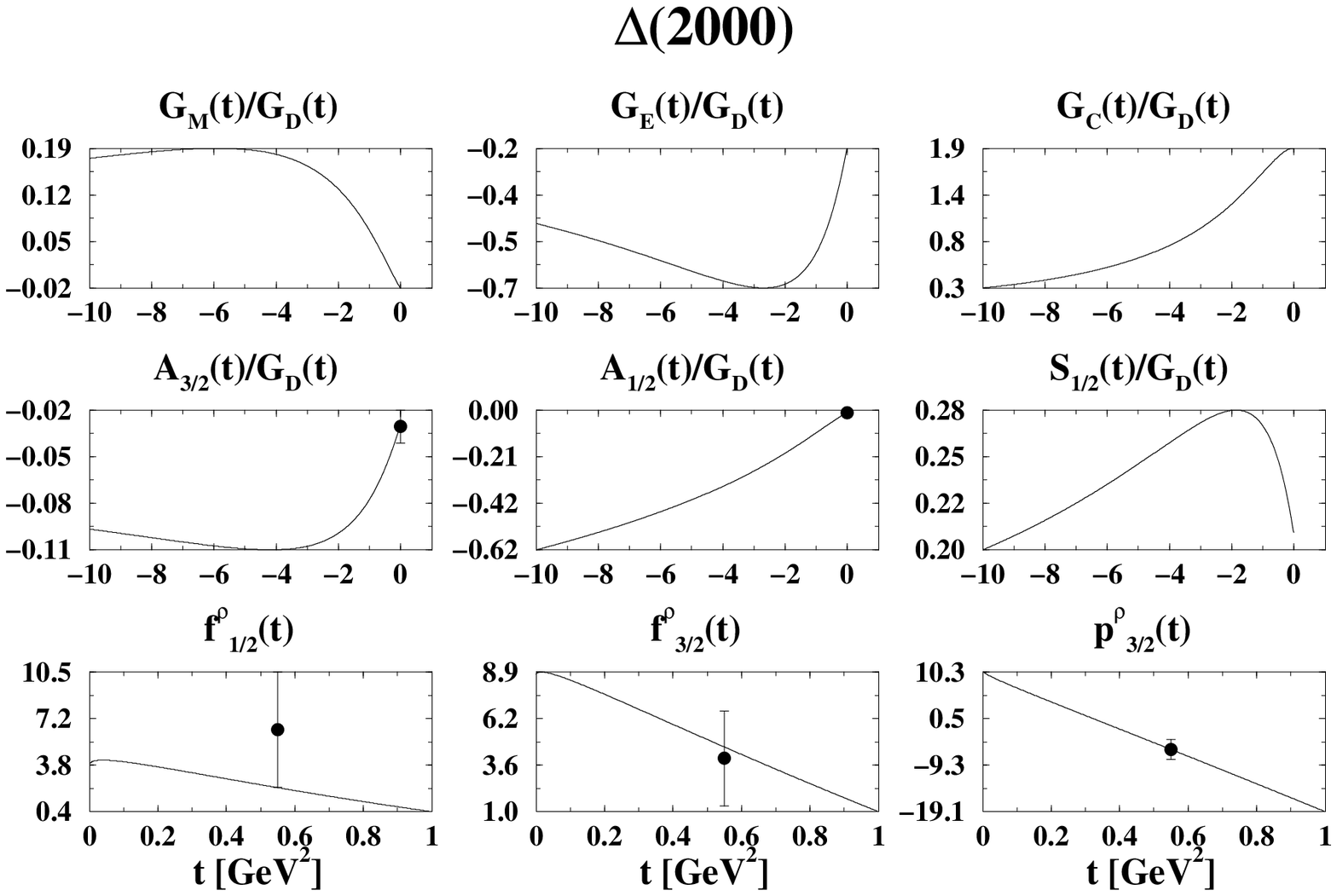}
\end{center}
\caption{ }
\label{fig24}
\end{figure}

\begin{figure}[tbp]
\begin{center}
\leavevmode
\epsfxsize = 18cm 
\epsffile[20 330 590 750]{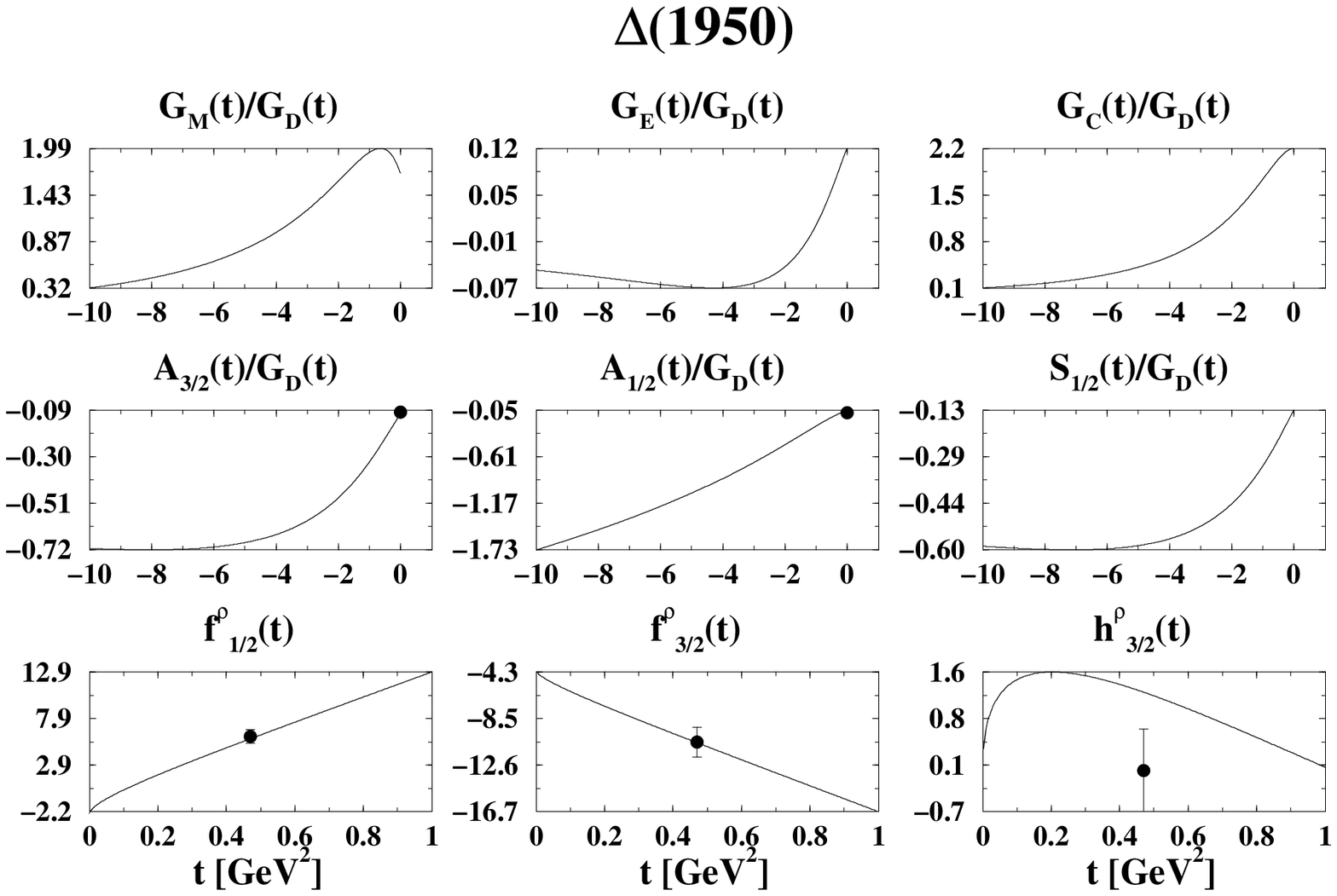}
\end{center}
\caption{ }
\label{fig25}
\end{figure}


\begin{figure}[tbp]
\begin{center}
\leavevmode
\epsfxsize = 16cm 
\epsffile[20 200 574 700]{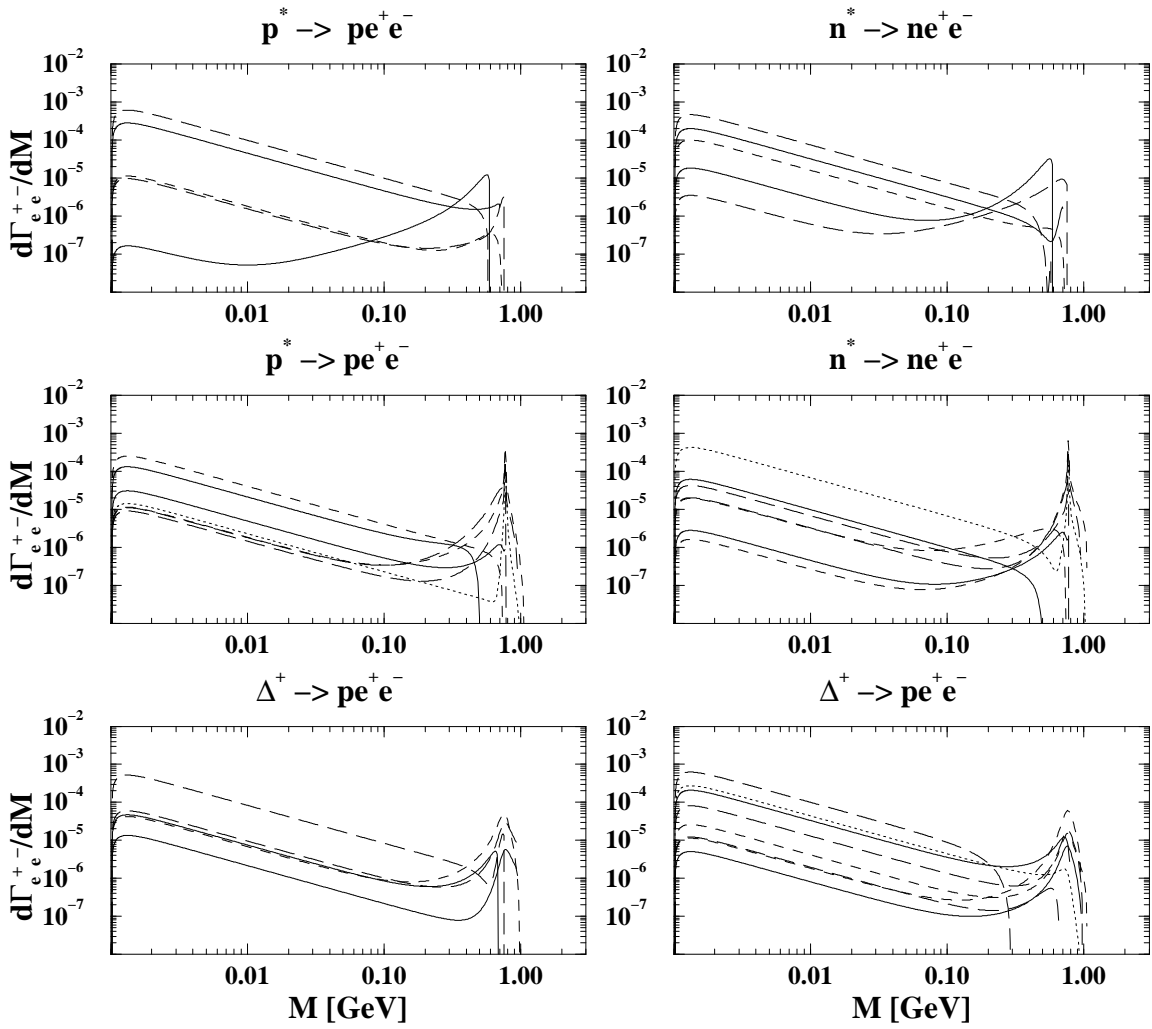}
\end{center}
\caption{The $e^{+}e^{-}$ spectra from decays of the nucleon resonances
listed in Tables I - III. The bold lines, long-dashed lines, short-dashed
lines, and the dotted lines stand, respectively, for the $J=\frac{1}{2},%
\frac{3}{2},\frac{5}{2},$ and $\frac{7}{2}$ resonances. The two upper plots
show the results for the negative-parity $N^{*}$-resonances (protons and neutrons), 
the two next plots show the results for the positive-parity $N^{*}$-resonances,
the two lower plots show the spectra from decays of the $\Delta$-resonances,
the left one for the negative-parity and the right one for the positive-parity resonances. 
The peaks are due to the $\rho$- and $\omega$-mesons.}
\label{fig26}
\end{figure}

\begin{figure}[tbp]
\begin{center}
\leavevmode
\epsfxsize = 16cm 
\epsffile[20 200 574 700]{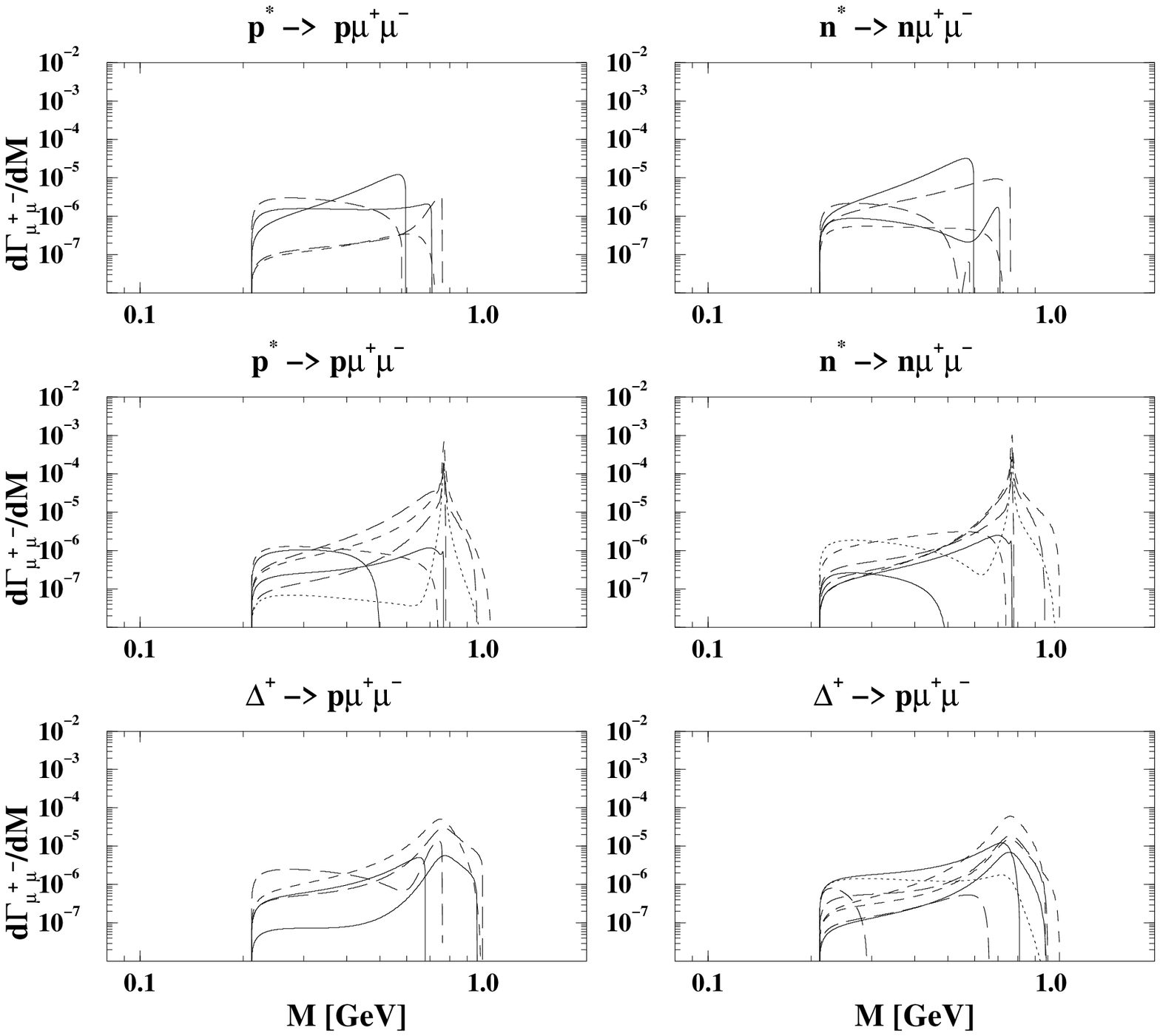}
\end{center}
\caption{ The $\mu ^{+}\mu ^{-}$ spectra from decays of the nucleon
resonances listed in Tables I - III. The bold lines, long-dashed lines,
short-dashed lines, and the dotted lines stand, respectively, for the $J=%
\frac{1}{2},\frac{3}{2},\frac{5}{2},$ and $\frac{7}{2}$ resonances.
The two upper plots
show the dimuon spectra for the negative-parity proton and neutron resonances ($I=1/2$), 
the two next plots show the results for the positive-parity proton and neutron resonances,
the two lower plots show the dimuon spectra from decays of the $\Delta$-resonances,
the left one for the negative-parity and the right one for the positive-parity $\Delta$-resonances. 
The peaks are due to the $\rho$- and $\omega$-mesons.}
\label{fig27}
\end{figure}


\begin{thebibliography}{999}

\bibitem{Walecka:1974qa}  J.~D.~Walecka, 
Annals Phys.\ {\bf 83}, 491 (1974). 


\bibitem{Chin:1977iz}  S.~A.~Chin, 
Annals Phys.\ {\bf 108}, 301 (1977). 


\bibitem{Drukarev:1988ib}  E.~G.~Drukarev and E.~M.~Levin, 
JETP Lett.\ {\bf 48}, 338 (1988) [Pisma Zh.\ Eksp.\ Teor.\ Fiz.\ {\bf 48},
307 (1988)]; 
E.~G.~Drukarev and E.~M.~Levin, 
Nucl.\ Phys.\ A {\bf 511}, 679 (1990) [Erratum-ibid.\ A {\bf 516}, 715
(1990)]. 


\bibitem{Tsushima:1991fe}  K.~Tsushima, T.~Maruyama and A.~Faessler, 
Nucl.\ Phys.\ A {\bf 535}, 497 (1991). 
T.~Maruyama, K.~Tsushima and A.~Faessler, 
Nucl.\ Phys.\ A {\bf 537}, 303 (1992). 


\bibitem{Brown:1991kk}  G.~E.~Brown and M.~Rho, 
Phys.\ Rev.\ Lett.\ {\bf 66}, 2720 (1991). 


\bibitem{Adami:1993tp}  C.~Adami and G.~E.~Brown, 
Phys.\ Rept.\ {\bf 234}, 1 (1993). 
T.~Hatsuda, H.~Shiomi and H.~Kuwabara, 
Prog.\ Theor.\ Phys.\ {\bf 95}, 1009 (1996) [nucl-th/9603043]. 


\bibitem{Bianchi:1993nh}  N.~Bianchi {\it et al.}, 
Phys.\ Lett.\ B {\bf 309}, 5 (1993). 
N.~Bianchi {\it et al.}, 
Phys.\ Lett.\ B {\bf 325}, 333 (1994). 


\bibitem{Kondratyuk:1994ah}  L.~A.~Kondratyuk, M.~I.~Krivoruchenko,
N.~Bianchi, E.~D.~Sanctis and V.~Muccifora, 
Nucl.\ Phys.\ A {\bf 579}, 453 (1994). 

\bibitem{Wei}  V. Weisskopf, Physikalische Zeitschrift {\bf 34}, 1 (1933).


\bibitem{Agakishiev:1995xb}  G.~Agakishiev {\it et al.} [CERES
Collaboration], 
Phys.\ Rev.\ Lett.\ {\bf 75}, 1272 (1995). 
A.~Drees, 
Nucl.\ Phys.\ A {\bf 610}, 536C (1996). 


\bibitem{Masera:1995ck}  M.~Masera [HELIOS Collaboration], 
Nucl.\ Phys.\ A {\bf 590}, 93C (1995). 


\bibitem{Porter:1997rc}  R.~J.~Porter {\it et al.} [DLS Collaboration], 
Phys.\ Rev.\ Lett.\ {\bf 79}, 1229 (1997) [nucl-ex/9703001]. 


\bibitem{Bratkovskaya:1999pr}  E.~L.~Bratkovskaya and C.~M.~Ko, 
Phys.\ Lett.\ B {\bf 445}, 265 (1999) [nucl-th/9809056]. 


\bibitem{Friese:1999qm}  J.~Friese [HADES Collaboration], 
Prog.\ Part.\ Nucl.\ Phys.\ {\bf 42}, 235 (1999). 


\bibitem{Matveev:1973ra}  V.~A.~Matveev, R.~M.~Muradian and
A.~N.~Tavkhelidze, 
Lett.\ Nuovo Cim.\ {\bf 7}, 719 (1973); 
S.~J.~Brodsky and G.~R.~Farrar, 
in {\it C73-09-06.4} Phys.\ Rev.\ Lett.\ {\bf 31}, 1153 (1973); 
S.~J.~Brodsky and G.~R.~Farrar, 
Phys.\ Rev.\ D {\bf 11}, 1309 (1975); 
A.~I.~Vainstein and V.~I.~Zakharov, Phys. Lett. B {\bf 72}, 368 (1978).


\bibitem{Hohler:1976ax}  G.~Hohler, E.~Pietarinen, I.~Sabba Stefanescu,
F.~Borkowski, G.~G.~Simon, V.~H.~Walther and R.~D.~Wendling, 
Nucl.\ Phys.\ B {\bf 114}, 505 (1976). 


\bibitem{Krivoruchenko:1994qb}  M.~I.~Krivoruchenko, 
ITEP-41-94 {\it Talk given at International Conference on Mesons and Nuclei
at Intermediate Energies, Dubna, Russia, 3-8 May 1994}.


\bibitem{Mergell:1996bf}  P.~Mergell, U.~G.~Meissner and D.~Drechsel, 
Nucl.\ Phys.\ A {\bf 596}, 367 (1996) [hep-ph/9506375]. 


\bibitem{Dubnicka:1996sp}  S.~Dubnicka, 
Acta Phys.\ Polon.\ B {\bf 27}, 2525 (1996). 


\bibitem{Faessler:2000md}  A.~Faessler, C.~Fuchs, M.~I.~Krivoruchenko and
B.~V.~Martemyanov, 
nucl-th/0010056. 


\bibitem{Faessler:2000de}  A.~Faessler, C.~Fuchs and M.~I.~Krivoruchenko, 
Phys.\ Rev.\ C {\bf 61}, 035206 (2000) [nucl-th/9904024]. 


\bibitem{Akhmetshin:2001bw}
R.~R.~Akhmetshin {\it et al.}  [CMD-2 Collaboration],
Phys.\ Lett.\ B {\bf 501}, 191 (2001)
[arXiv:hep-ex/0012039].

\bibitem{JS}  H. F. Jones and M. D. Scadron, Ann. Phys. (N.Y.) {\bf 81}, 1
(1973).

\bibitem{DEK}  R. C. E. Devenish, T. S. Eisenschitz and J. G. K\"orner, Phys.
Rev. D {\bf 14}, 3063 (1976).


\bibitem{Krivoruchenko:2001hs}  M.~I.~Krivoruchenko and A.~Faessler, 
Phys. Rev. D, in press [nucl-th/0104045]. 


\bibitem{Groom:2000in}  D.~E.~Groom {\it et al.} [Particle Data Group
Collaboration], 
Eur.\ Phys.\ J.\ C {\bf 15} (2000) 1. 


\bibitem{Manley:1992yb}  D.~M.~Manley and E.~M.~Saleski, 
Phys.\ Rev.\ D {\bf 45}, 4002 (1992). 


\bibitem{Longacre:1977ja}  R.~S.~Longacre and J.~Dolbeau, 
Nucl.\ Phys.\ B {\bf 122}, 493 (1977); 
R.~Longacre, A.~H.~Rosenfeld, T.~A.~Lasinski, G.~Smadja, R.~J.~Cashmore and
D.~W.~Leith, 
Phys.\ Lett.\ B {\bf 55}, 415 (1975). 


\bibitem{Koniuk:1982ej}  R.~Koniuk, 
Nucl.\ Phys.\ B {\bf 195}, 452 (1982). 


\bibitem{Capstick:1994kb}  S.~Capstick and W.~Roberts, 
Phys.\ Rev.\ D {\bf 49}, 4570 (1994) [nucl-th/9310030]. 

\bibitem{BW}  J. D. Bjorken and J. D. Walecka, Ann. Phys. (N.Y.) {\bf 38},
35 (1966).

\bibitem{LAN}  W. B. Berestetzki, E. M. Lifschitz and L. P. Pitaevwski,
{\it Relativistische Quantumtheorie}, Academie-Verlag, Berlin (1980).

\bibitem{Trueman:1969wn}  T.~L.~Trueman, 
Phys.\ Rev.\ {\bf 182}, 1469 (1969); 
W. R. Theis and P. Hertel, Nuovo. Cim. {\bf 66}, 152 (1970); 
F.~E.~Close and W.~N.~Cottingham, 
Nucl.\ Phys.\ B {\bf 99}, 61 (1975); 
W. A. Bardin and Wu-Ki Tung, Phys. Rev. {\bf 173}, 1423 (1968); 
R.~Tarrach, 
Nuovo Cim.\ A {\bf 28}, 409 (1975). 

\bibitem{Rose}  M. E. Rose, {\it Elementary Theory of Angular Momentum},
John Wiley \& Sons, Inc., New York e. a., (1957).


\bibitem{Manley:1984jz}  D.~M.~Manley, R.~A.~Arndt, Y.~Goradia and
V.~L.~Teplitz, 
Phys.\ Rev.\ D {\bf 30}, 904 (1984). 

\bibitem{JW} M. Jacob and G. C. Wick, Ann. Phys. (N.Y.) {\bf 7}, 404 (1959).

\bibitem{Frazer:1959gy}  W.~R.~Frazer and J.~R.~Fulco, 
Phys.\ Rev.\ Lett.\ {\bf 2}, 365 (1959); 
W.~R.~Frazer and J.~R.~Fulco, Phys. Rev. {\bf 117}, 1603 (1960);
W.~R.~Frazer and J.~R.~Fulco, Phys. Rev. {\bf 117}, 1609 (1960).

\bibitem{LLME}  A. I. Lendel, V. I. Lendyel, V. A. Meshcheryakov and B. M.
Ernst, Yad. Fiz. {\bf 3}, 1093 (1966).

\bibitem{SSM}  D. V. Shirkov, V. V. Serebryakov and V. A. Meshcheryakov, 
{\it Dispersion Theories of Strong Interactions at Low Energies}, Moscow,
Nauka (1967).


\bibitem{Hohler:1975ht}  G.~Hohler and E.~Pietarinen, 
Nucl.\ Phys.\ B {\bf 95}, 210 (1975). 

\bibitem{HBOOK}  G. Hohler, {\it Pion-Nucleon Scattering}, Vol.I/9b, ed.
H.Schopper, Springer, Landolt-B\"{o}rnstein (1983).

\bibitem{FUR}  S. Furuichi and K. Watanabe, Progr. Th. Phys. {\bf 92}, 339
(1994).


\bibitem{Krivoruchenko:1995cv}  M.~I.~Krivoruchenko, 
Phys.\ Atom.\ Nucl.\ {\bf 58}, 1738 (1995) [Yad.\ Fiz.\ {\bf 58N10}, 1839
(1995)]; 
PiN Newslett.\ {\bf 10}, 156 (1995), [hep-ph/9611323]; 
M.~I.~Krivoruchenko, 
nucl-th/9710072. 

\bibitem{Gounaris:1968mw}  G.~J.~Gounaris and J.~J.~Sakurai, 
Phys.\ Rev.\ Lett.\ {\bf 21}, 244 (1968). 

\bibitem{BD}  J. D. Bjorken and S. D. Drell, {\it Relativistic Quantum Fields%
}, McGraw-Hill, New-York (1965).


\bibitem{Krivoruchenko:1993vd}  M.~I.~Krivoruchenko, 
JETP Lett.\ {\bf 58} (1993) 6 [Pisma Zh.\ Eksp.\ Teor.\ Fiz.\ {\bf 58}
(1993) 7]. 
M.~I.~Krivoruchenko, 
Z.\ Phys.\ A {\bf 350}, 343 (1995). 

\bibitem{SAK}  J. J. Sakurai, {\it Currents and Mesons}, University of
Chicago Press, Chicago (1969).


\bibitem{Koniuk:1980vy}  R.~Koniuk and N.~Isgur, 
Phys.\ Rev.\ D {\bf 21}, 1868 (1980) [Erratum-ibid.\ D {\bf 23}, 818
(1980)]. 


\bibitem{Warns:1990ic}  M.~Warns, H.~Schroder, W.~Pfeil and H.~Rollnik, 
Z.\ Phys.\ C {\bf 45}, 613 (1990). 


\bibitem{Li:1990qu}  Z.~Li and F.~E.~Close, 
Phys.\ Rev.\ D {\bf 42}, 2207 (1990). 


\bibitem{Bijker:1994yr}  R.~Bijker, F.~Iachello and A.~Leviatan, 
Annals Phys.\ {\bf 236}, 69 (1994) [nucl-th/9402012]. 


\bibitem{Capstick:1992uc}  S.~Capstick, 
Phys.\ Rev.\ D {\bf 46}, 2864 (1992). 

\bibitem{note} The Koniuk model \cite{Koniuk:1982ej} has two coupling constants, $g$ and $h$%
, of the electric and magnetic types. The polarization operator $V\gamma $
is transversal provided that the anomalous quark magnetic moments with
respect to the vector field $V$ are zero. It gives $g=h.$ The coupling
constants $g$ and $h$ have in the Koniuk model the same sign, but a factor
of two different from each other. We consider it as an artefact of the
model and speak everywhere in terms of one constant ($g$).



\bibitem{LeYaouanc:1975mr}  A.~Le Yaouanc, L.~Oliver, O.~Pene and
J.~C.~Raynal, 
Phys.\ Rev.\ D {\bf 11}, 1272 (1975). 


\bibitem{Stassart:1990zt}  P.~Stassart and F.~Stancu, 
Phys.\ Rev.\ D {\bf 42}, 1521 (1990). 


\bibitem{Stancu:1993xz}  F.~Stancu and P.~Stassart, 
Phys.\ Rev.\ D {\bf 47}, 2140 (1993). 


\bibitem{Bijker:1997tr}  R.~Bijker, F.~Iachello and A.~Leviatan, 
Phys.\ Rev.\ D {\bf 55}, 2862 (1997) [nucl-th/9608057]. 


\bibitem{Godfrey:1985xj}  S.~Godfrey and N.~Isgur, 
Phys.\ Rev.\ D {\bf 32}, 189 (1985). 


\bibitem{Bartel:1968tw}  W.~Bartel {\it et al.}, 
Phys.\ Lett.\ B {\bf 28}, 148 (1968). 


\bibitem{Stein:1975yy}  S.~Stein {\it et al.}, 
Phys.\ Rev.\ D {\bf 12}, 1884 (1975). 

\bibitem{Batzner}  K. Batzner et.al Phys. Lett. B {\bf 39}, 575 (1972).

\bibitem{Siddle:1971ug}
R.~Siddle {\it et al.},
Nucl.\ Phys.\ B {\bf 35}, 93 (1971).

\bibitem{Burkert} V. D. Burkert and L. Elouadrhiri, Phys. Rev. Lett. {\bf 75}, 3614 (1995).

\bibitem{Beck:1997ew}
R.~Beck {\it et al.},
Phys.\ Rev.\ Lett.\  {\bf 78}, 606 (1997).


\bibitem{Frolov:1999pw}  V.~V.~Frolov {\it et al.}, 
Phys.\ Rev.\ Lett.\ {\bf 82}, 45 (1999) [hep-ex/9808024]. 

\bibitem{AS} I. G. Aznauryan and S. G. Stepanyan, Phys. Rev. D {\bf 59}, 054009 (1999).


\bibitem{ALDER} J. C. Alder {\it et al.}, Nucl. Phys. B {\bf 46}, 573 (1972).

\bibitem{Burk} V. Burkert,  Proc. of the Topical
Workshop on "Excited Baryons 1988", Troy, New York (1988), eds.
G. Adams, N. Mukhopadhyay and P. Stoler, World Scientific, Singapore (1989), p. 122.  


\bibitem{foster} F. Foster and G. Huges, Rep. Progr. Phys. {\bf 46}, 1445 (1983).

\bibitem{Gerhardt:1980yg}
C.~Gerhardt,
Z.\ Phys.\ C {\bf 4}, 311 (1980).
                           
\end{thebibliography}
\end{document}